\documentclass[12pt]{iopart}
\pdfoutput=1
\usepackage{amsfonts}
\usepackage{graphicx}

\usepackage{xcolor}
\usepackage{subfigure}
\usepackage{textcomp}
\usepackage{cite}
\usepackage{float}
\usepackage{soul}
\usepackage{stackrel}

\definecolor{dgreen}{rgb}{0,0.7,0}

\expandafter\let\csname equation*\endcsname\relax
\expandafter\let\csname endequation*\endcsname\relax
\usepackage{amsmath,amssymb}

\usepackage{esint}

\begin{document}

\title[]{Run-and-Tumble particle in inhomogeneous media in one dimension}

\author{Prashant Singh}
\address{International Centre for Theoretical Sciences, TIFR, Bengaluru 560089, India}
\author{Sanjib Sabhapandit}
\address{Raman Research Institute, Bangalore, India}
\author{Anupam Kundu}
\address{International Centre for Theoretical Sciences, TIFR, Bengaluru 560089, India}
\ead{prashant.singh@icts.res.in}

%\begin{indented}
%\item[]August 2017
%\end{indented}
\date{\today}
\begin{abstract}
\noindent
We investigate the run and tumble particle (RTP), also known as persistent Brownian motion, in one dimension. A telegraphic noise $\sigma(t)$ drives the particle which changes between $\pm 1$ values with some rates. Denoting the rate of flip from $1$ to $-1$ as $R_1$ and the converse rate as $R_2$, we consider the position and direction dependent rates of the form $R_1(x)=\left(\frac{\mid x \mid}{l}\right) ^{\alpha}\left[\gamma_1~\theta(x)+\gamma_2 ~\theta (-x)\right]$ and $R_2(x)=\left(\frac{\mid x \mid}{l}\right) ^{\alpha}\left[\gamma_2~\theta(x)+\gamma_1 ~\theta (-x)\right]$ with $\alpha \geq 0$. For $\gamma_1 >\gamma_2$, we find that the particle exhibits a steady-state probability distriution even in an infinite line whose exact form depends on $\alpha$. For $\alpha =0$ and $1$, we solve the master equations exactly for arbitrary $\gamma_1$ and $\gamma_2$ at large $t$. From our  explicit expression for time-dependent probability distribution $P(x,t)$ we find that it exponentially relaxes to the steady-state distribution for $\gamma_1 > \gamma_2$. On the other hand, for $\gamma_1<\gamma_2$, the large $t$ behaviour of $P(x,t)$ is drastically different than $\gamma_1=\gamma_2$ case where the distribution decays as $t^{-\frac{1}{2}}$. Contrary to the latter, detailed balance is not obeyed by the particle even at large $t$ in the former case. 
For general $\alpha$, we argue that the approach to the steady state in $\gamma_1>\gamma_2$ case is exponential which we numerically demonstrate. On the other hand for $\gamma_1\leq \gamma_2$, the distribution $P(x,t)$ does not reach a steady state, however posseses certain scaling  behaviour. For $\gamma_1=\gamma_2$ we derive this scaling behaviour as well as the scaling function rigorously whereas for $\gamma_1< \gamma_2$ we provide heuristic arguments for the scaling behaviour and the corresponding scaling functions. We also study the dynamics in semi-infinite line with an absorbing barrier at the origin. We analytically compute the survival probabilities and first-passage time distributions for $\alpha =0$ and $1$. For general $\alpha \geq 0$, once again we compute the value of survival probability at large $t$ and approach to it.  Finally, we consider RTP in an  finite interval $[0,M]$ and compute the associated exit probability from that interval for all $\alpha$. All our analytic results match with the numerical simulation of the same. \\
\end{abstract}

%\maketitle

\section{Introduction}
\label{intro}
\noindent
Active matter is a class of non-equilibrium systems that can transduce the supplied energy to a systematic movement through some internal mechanisms \cite{Ramaswamy 2010,Romanczuk 2012,Marchetti 2013,Ramaswamy 2017,Schweitzer 2003,Bechinger 2016}. The dynamics of these systems does not respect time-reversal symmetry and thus break the detailed balance. A plethora of interesting phenomena like - motility induced phase transition \cite{Cates 2015,Gonnella 2015,Partridge 2019}, flocking \cite{Ballerini 2008, Katz 2011}, clustering \cite{Redner 2013, Bricard 2013}, non existence of equation of state in terms of pressure \cite{Solon 2015} etc, has been observed and studied in these systems which arise due to the activity and interaction among the particles. At the level of single particle also,  such systems exhibit interesting behaviours like accumulation at the boundaries inside confinement \cite{Li 2009, Elgeti 2015, Kanaya_2018}, non-Botlzmann stationary distribution \cite{Erdmann 2002,Tailleur_2008,Tailleur_2009,Das 2018, Maggi 2014, Maggi 2015}, anomalous behaviours \cite{Ebeling 2000,Basu 2018} which are remarkably different than their passive counterparts. "Run-and-tumble" particle (RTP) and Active Brownian particle (ABP) are two paradigmatic models of the dynamics of active particles which have extensively been studied in the recent few years. A single ABP, free or in harmonic trap, shows rich features like anomalous first passage distributions \cite{Basu 2018}, re-entrant phase transition \cite{Kanaya2019}, position distribution \cite{Pototsky2012, UrnaB2019} and many more. These particles have also been used as models of microscopic constituents in many theoretical studies of active matter or collective behaviour of many active agents \cite{Tailleur_2008, Solon_2015}.

%In this paper, we will consider the other paradigmatic model of Run-and-tumble particle (RTP).

The run-and-tumble mechanism describes the stochastic dynamics of a particle which moves in a straight line for some time $t_{run}$ and undergo tumble, a state of rest, which lasts for another time $t_{tum}$. The particle then chooses the direction randomly for the next run. For example E. Coli bacteria runs for some time along a straight line and then tumbles to randomly choose a new direction of run \cite{Berg2003, Berg1972}. For a bacteria such times scales are of thee order of $t_{run} \sim 1$ sec and $t_{tum} \sim 0.1$ sec, respectively \cite{Berg1972, Block1982}.   In RTP model, such tumble events often considered to occur instantaneously and  after each tumble the direction of motion is changed. The time for which the particle runs is taken from exponential distribution with some rate. In one dimension the particle tumbles between the positive and the negative direction and its equation of motion  is given by 
\begin{align}
\frac{dx}{dt} = v \sigma (t),
\label{langevin}
\end{align}
where $x(t)$ is the position of the particle at time $t$, $v (>0)$ is it's speed and $\sigma(t)$ represents it's instantaneous direction of motion governed by the telegraphic or dichotomous noise. This 
noise $\sigma (t)$ switches between $\pm 1$ with rate $R$ and consequently it's values at different times are  correlated exponentially as $\langle \sigma (t_1) \sigma (t_2) \rangle= 2Re^{-2R \mid t_1-t_2\mid}$ which makes the evolution of the position $x(t)$ non-Markovian. The RTP model is in some sense an amalgamation of ballistic motion and Brownian motion as by tuning the parameters $R$ and $v$, one can go from from a pure ballistic particle ($R \to 0$) to pure Brownian particle ($R \to \infty$ and $v\to \infty$ keeping $v^2/R$ fixed). In the physics literature the telegraphic noise and the RTP process have been studied in various settings starting from persistent Brownian motion, electromagnetic theory, optics, Lorentz gas to polymers \cite{Behn1989,Masoliver1993,Masoliver1995,Weiss2002,Kawai,Weissbook, Dhar2002, Masoliver2017,Shee2020}. The model has gained renewed interest in the recent years due to its applicability in mimicking the movement of E-Coli \cite{Berg2003}. Also the exact solvability of this model makes it a quintessential candidate for study of the rich and remarkably different behaviours of the active systems. The model has been extensively studied and a variety of its properties are known. Some examples are - joint distribution of maximum and minimum of the position \cite{Masoliver1993},  distributions of first-passage times and exit times from an interval \cite{Kanaya_2018,Angelani2014,Scacchi2017,Mori2019}, behaviour under resetting \cite{Evans2018}, large deviation forms \cite{Gradenigo2019, Banerjee2019, Ion2020}, convex hull \cite{Hartmann2019}, distributions in harmonic trap and other confining potentials \cite{Dhar2019, Urna2020}, behaviour in inhomogeneous force field \cite{Doussal2020}. Recently the authors have also investigated the "Generalised" Arcsine laws for this model and found some interesting features in comparison to pure Brownian particle \cite{Singh2019}. There has also been reasonable amount of study of the microscopic dynamics of multiple interacting RTPs on  continuous as well as lattice space\cite{Slowman2016,Slowman2017,Slowman2018,Das2019,Doussal2019,Put2019}.

\begin{figure}[t]
\includegraphics[scale=0.31]{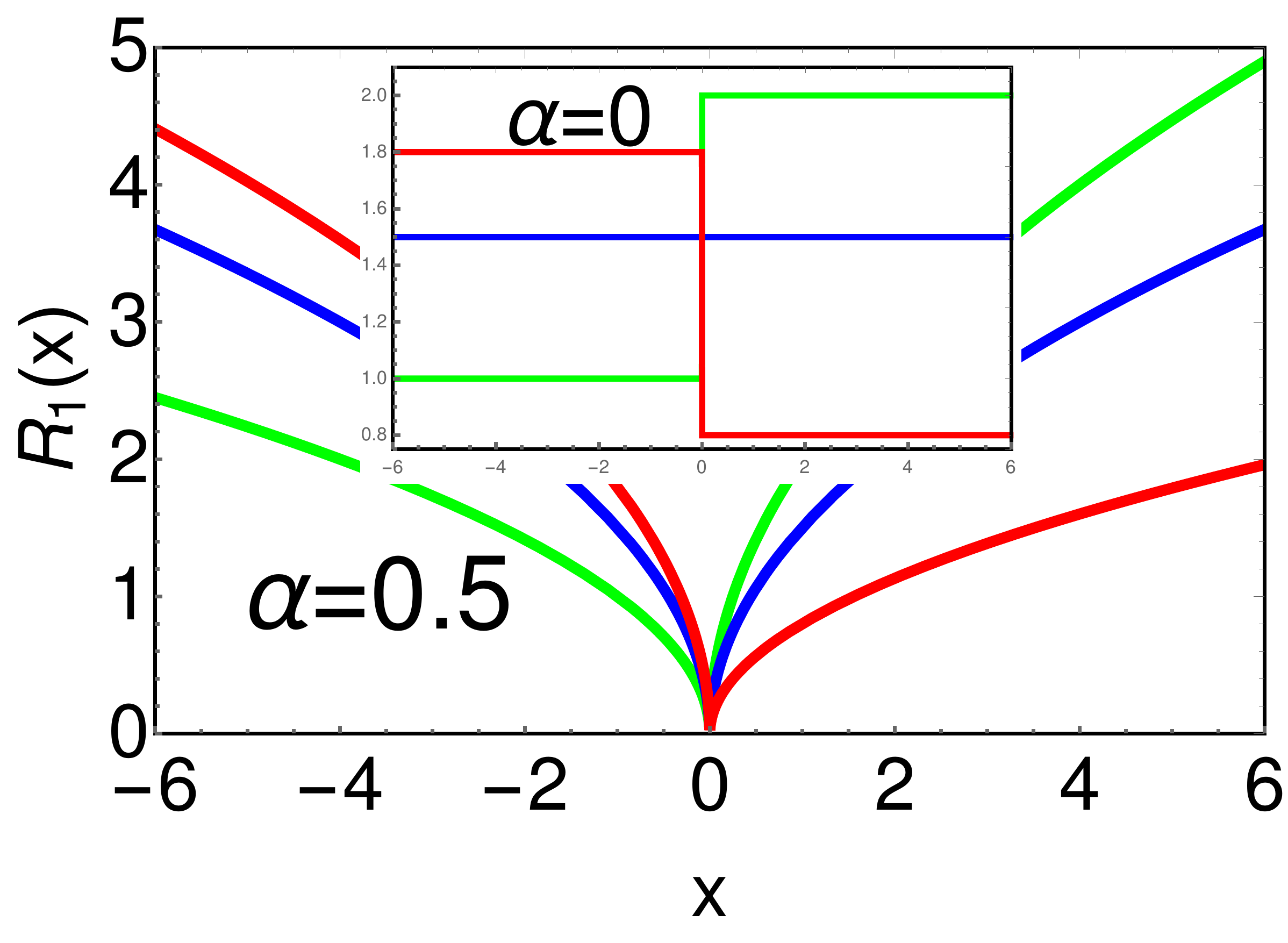}
\includegraphics[scale=0.33]{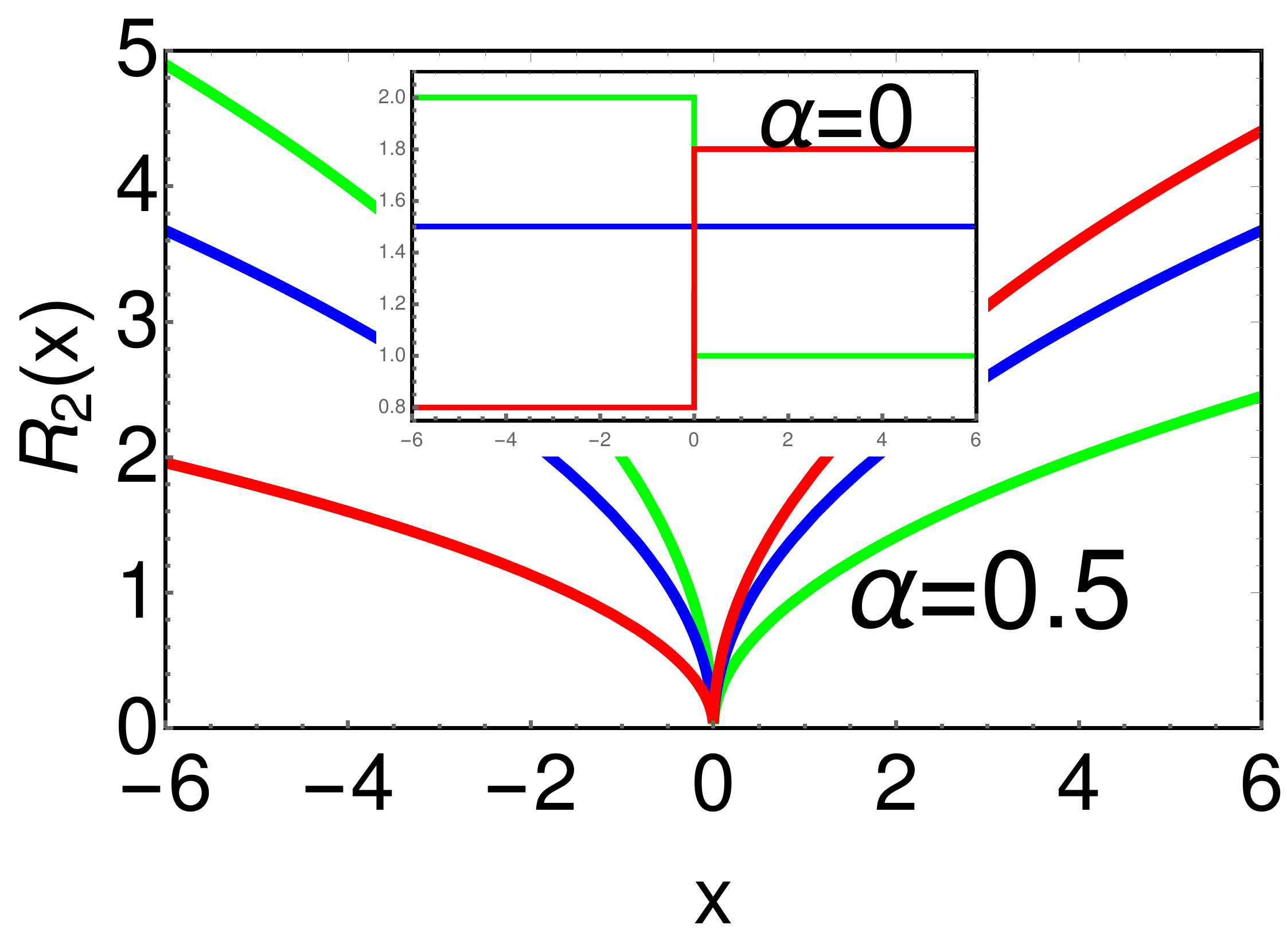}
\centering
\caption{Plot of rates defined in Eqs. \eqref{rates} for $\alpha = 0.5$ and various signatures of $\Delta$. In the left panel, we have plotted $R_1(x)$ vs $x$ for (i) $\gamma_1 =2, \gamma_2 =1$ (green), (ii) $\gamma_1 = \gamma_2 =1.5$ (blue) and (iii) $\gamma_1 =0.8, \gamma_2 =1.8$ (red). \textbf{Inset:} Shows the same plot for $\alpha =0$. In right panel, we have plotted $R_2(x)$ vs $x$ for the same choice of parameters and colour. For all plots $l=1$.}
\label{ratepic1}
\end{figure}

It is imperative to emphasise that the telegraphic noise considered in the above settings is characterised only by the constant rate $R$. The flip from $+1$ to $-1$ $(1 \to -1 )$ occurs with the same rate as from $-1$ to $+1$ $(-1 \to 1 )$. This consideration, however, is a clich\'e specially when the particle is exposed to some chemoattractants or chemorepellents. For example it is seen experimentally (and used theoretically) \cite{Berg1972,Turner2000,Degennes2004,Adler J 1973,Sakuntala2011} that in E-Coli the run-time depends strongly on the concentration of the nutrients and the nutrient-gradient. In \cite{Berg1972}, the run duration for E-Coli is found to depend on whether the bacteria are moving towards or away from the chemo-attractants although the distribution for the times is still exponential. This observation is suggestive to generalise the telegraphic noise in Eq.~\eqref{langevin} whereby the flips from $1 \leftrightarrow-1$ occurs with position dependent rates. 
In this paper we consider the dynamics of a single  RTP particle in one dimension with generalised telegraphic noise which is characterised by position and direction dependent rates $R_1(x)$ and $R_2(x)$ given by 
% More specifically we will take the rates of the form,
\begin{align}
\begin{split}
R_1(x)=\left(\frac{\mid x \mid}{l}\right) ^{\alpha}\left[\gamma_1~\theta(x)+\gamma_2 ~\theta (-x)\right],\\
R_2(x)=\left(\frac{\mid x \mid}{l}\right) ^{\alpha}\left[\gamma_2~\theta(x)+\gamma_1 ~\theta (-x)\right],
\label{rates}
\end{split}
\end{align}
where $\alpha \geq 0$, $\theta (x)$ is the Heaviside function and, $\gamma_1$ and $\gamma_2$ (both positive) are position independent rates (see Fig.(\ref{ratepic1})). Here $l$ is a length scale over which the rate functions are varying. Similar generalisations of RTP motions  have been considered in various settings such as Markovian robots \cite{Luis2018}, active diffusion \cite{Maes2018}, response to stochastic input \cite{Sakuntala2019}, chemotaxis \cite{Tailleur_2008,Mercedes1989,Schnitzer1993}, quorum sensing \cite{Farrell2012, ASolon2018} and motion with space dependent speed $v(x)$ \cite{Angelani2019}. These studies mostly deal with either steady-state behaviours or hydrodynamic descriptions. 
%\redw{Also the RTP model with space depedent speed has been considered in \cite{Angelani2019}.} 
In this paper, we study the occupation probability, the survival problem and the exit problem, going beyond the steady state properties of non-interacting RTPs with flipping rates depending on both position and orientation. We find that, in addition to being proximal to realistic situations, this model of the dynamics of RTP also exhibits interesting features like the existence of steady-state and non-trivial and richer large time properties of the occupation probability as well as survival probability compared to pure Brownian particles which are otherwise absent in RTPs with constant rate. We find that for this generalised RTP the survival probability $S(t)$ for large time decays as $S(t) \sim t^{-\theta}$ with a persistent exponent $\theta$ which strongly depends on $\alpha$ and the rates $\gamma_1$ and $\gamma_2$. Note that for a pure Brownian particle as well as for a RTP with constant rate, $\theta=1/2$.

The paper is organised as follows. We start with the computation of the occupation probability $P(x,t)$ in sec.~\ref{prob}  for different $\alpha$ and $\Delta= (\gamma_1-\gamma_2)/2$. We perform computations for $\alpha=0$ in sec.~\ref{alpha=0}, for $\alpha=1$ in sec.~\ref{alpha=1} and for general $\alpha$ in sec.~\ref{gen-alpha}. For each choices of $\alpha$, three different cases 
of $\Delta =0$, $\Delta >0$ and $\Delta <0$ are discussed in subsequent sections. After studying occupation probability $P(x,t)$ we study, survival probability of the RTP in inhomogeneous media with an absorbing site at the origin in sec.~\ref{surv_prob}. In this case also we perform computations for different $\alpha$ separately in secs.~\ref{surv-alpha=0} (for  $\alpha=0$), sec.~\ref{surv-alpha=1} (for $\alpha=1$) and in sec.~\ref{surv-gen-alpha} (for general $\alpha$). Finally we study exit probability of the RTP from a finite box in sec.~\ref{exit_prob_label} which is followed by our conclusion in sec.~\ref{conc}.

%Contrary to the latter, the large $t$ behaviour of the former still does not have the detailed balance for $\gamma_1 \neq \gamma_2$. These rich structures and experimental findings motivate us to consider 

%For many stochastic processes, the survival probability $S(t)$ decays as $S(t) \sim t^{-\theta}$ where $\theta$ is called the persistence exponent. The exponent $\theta$ is computed for a class of processes and is non-trivial for non-markovian processes \cite{Majumdar1999}. For example for a RTP with constant rate, $\theta$ is equal to $\frac{1}{2}$ \cite{Kanaya 2018}. The other aim of this paper is to address these questions in the context of RTP with generalised telegraphic noise and see how $\theta$ gets modified for position and direction dependent rates. 

%We emphasise that in literature the generalised telegraphic noise has been studied in various settings like Markovian robots \cite{Luis2018}, active diffusion \cite{Maes2018}, response to stocastic input \cite{Sakuntala2019}, chemotaxis \cite{Mercedes1989,Cates2008,Schnitzer1993} and so on which mostly deal with either steady-state behaviours or hydrodynamic descriptions. In contrast our work is oriented more towards understanding the behaviour from a microscopic perspective and go beyond the steady-states in some cases. Our work also addresses questions in presence of absorbing barriers which arise naturally in various contexts and not covered in \cite{Luis2018,Maes2018,Sakuntala2019,Mercedes1989,Cates2008,Schnitzer1993}. 

\section{The occupation probability density}
\label{prob}
%We start by computing the probability distribution of the RTP moving in an infinite line. 
Let $P_{\sigma}(x,t)$ denote the probability distribution for the RTP, starting at the origin with orientations $\pm 1$ (chosen with probabilities $a_{\pm}$  such that $a_++a_-=1$), to be at position $x$ in time $t$ with velocity direction $\sigma \in \{+, -\}$. Starting from the Langevin equation \eqref{langevin}, it is easy to show that the distributions $P_{\pm}(x,t)$ satisfy the following master equations \cite{Kanaya_2018}
\begin{align}
\begin{split}
  \partial_t P_+(x,t) &= -v \partial_x P_+(x,t) -R_1(x) P_+(x,t) + R_2(x) P_-(x,t), \\
   \partial_t P_-(x,t) &=~~~v \partial_x P_-(x,t) +R_1(x) P_+(x,t) - R_2(x) P _-(x,t),
 \end{split}
\label{fokker}
\end{align} 
where $R_1(x)$ and $R_2 (x)$ are the position and direction dependent rates defined in Eq.~\eqref{rates}. To solve these equations we need to specify the initial as well as the boundary conditions. The initial conditions of the problem are $P_{\pm}(x,0)=a_{\pm} \delta (x)$. Note that for a given finite time $t$, the particle can at most travel a distance $\pm vt$ depending on the initial velocity direction which implies the boundary conditions $P_{\pm}(x \to \pm \infty,t)=0$. Throughout the paper, we will work with the symmetric initial condition $a_{+}=a_{-}=\frac{1}{2}$ for which the particle starts with $\pm v$ velocity with equal probability. It is interesting to note that the choice of rates $R_1(x)$ and $R_2(x)$ are such that the timescale over which the particle moves towards  the origin is $\sim~\frac{1}{\gamma_2}$. Similarly, the time scale to go away from the origin is $\frac{1}{\gamma_1}$. For $\gamma_1 > \gamma_2$, the motion is drifted on an average  towards the origin which suggests one to anticipate a stationary state distribution at large times even when the particle is moving on an infinite line. On the other hand for $\gamma_1\leq \gamma_2$, the probability distribution of the particle never reaches a steady state. 

To solve the master equations in Eq.~\eqref{fokker}, we first take Laplace transformation of the distributions with respect to time $t$, defined as 
\begin{align}
\bar{P}_{\pm}(x,s)=L_{t\to s}[P_\pm(x,t)]=\int_{0}^{\infty} dt e^{-st} P_{\pm}(x,t), \label{LP}
\end{align} 
on both sides and  get the following ordinary but coupled differential equations for $\bar{P}_{\pm}(x,s)$ as
\begin{align}
\big(v \partial_x +R_1(x)+s\big)\bar{P}_+ &= R_2(x)\bar{P}_-+\frac{1}{2} \delta (x),\label{fin_eq13}\\
\big(-v\partial_x +R_2(x)+s\big)\bar{P}_-&=R_1(x)\bar{P}_++\frac{1}{2} \delta (x). \label{fin_eq14}
\end{align}
Defining,
\begin{align}
\bar{P}(x,s)&=\bar{P}_+(x,s)+\bar{P}_-(x,s), \label{def1}\\
\bar{Q}(x,s)&=\bar{P}_+(x,s)-\bar{P}_-(x,s), \label{def2}
\end{align}
we rewrite the above equations as 
\begin{align}
&v \partial _x \bar{P}+ s\bar{Q} +  \frac{2~\text{sgn}(x) \Delta \mid x \mid^{\alpha}}{l^{\alpha}}  \bar{P}+\frac{2 \gamma \mid x \mid^{\alpha}}{l^{\alpha}}  \bar{Q}=0, \label{main_eq1}\\
&~~~~~~~~~~~~s \bar{P} +v \partial _x \bar{Q}=\delta(x), \label{main_eq2}
\end{align}
with $2 \Delta=\gamma_1-\gamma_2$ and $2 \gamma=\gamma_1+\gamma_2$. The signum function $\text{sgn}(x)$ takes values $1$ for $x>0$, $0$ for $x=0$ and $-1$ for $x<0$. Substituting $\bar{P}(x,s)$ from Eq.~\eqref{main_eq2} in Eq.~\eqref{main_eq1}, one can eliminate $\bar{P}(x,s)$ and get a second order differential equation of  $\bar{Q}(x,s)$ valid for $x \neq 0$ as,
\begin{align}
\partial_x^2 \bar{Q}+\frac{2~\text{sgn}(x) \Delta \mid x \mid^{\alpha}}{v~l^{\alpha}}\partial_x \bar{Q}- \left(\frac{2 \gamma~s \mid x \mid^{\alpha}}{v^2~l^{\alpha}}+\frac{s^2}{v^2} \right)\bar{Q}=0.\label{main_eq3}
\end{align}
Similarly eliminating $\bar{Q}(x,s)$, one can get the following equation for $\bar{P}(x,s)$
\begin{align}
s\bar{P}(x,s)-\delta(x)=\partial_x \left(\frac{v^2}{s+ \frac{2\gamma~ |x|^\alpha}{l^\alpha}}\right) 
\left[\partial_x \bar{P}(x,s) 
+ \frac{2 \Delta ~\text{sgn}(x)|x|^\alpha}{v l^\alpha}\bar{P}(x,s) \right] .
\label{eq-bar-P}
\end{align}
To obtain $P(x,t)$ one can in principle directly solve this equation, however, as we will see it turns out convenient to first solve Eq.~\eqref{main_eq3} first and then obtain $\bar{P}(x,s)$ from Eq.~\eqref{main_eq2}.
To get rid of the first order derivative term (second) in L.H.S of Eq.~\eqref{main_eq3} we define
\begin{align}
\bar{Q}(x,s)=e^{-\frac{\Delta \mid x \mid^{\alpha+1}}{v (\alpha+1) l^{\alpha}}} G(x,s), \label{main_eq4}
\end{align}
substituting which in Eq.~\eqref{main_eq3} and simplifying we get
\begin{align}
\partial_x^2 G-\left[\frac{\Delta \alpha \mid x \mid ^{\alpha-1}}{v~ l^{\alpha}}+\frac{2 \gamma s\mid x \mid ^{\alpha}}{v^2~ l^{\alpha}}+\frac{\Delta ^2 \mid x \mid ^{2\alpha}}{v^2~ l^{2\alpha}} +\frac{s^2}{v^2}\right] G=0. \label{main-eq-G}
\end{align}
We solve this equation with boundary conditions that $G(x,s)$ should not diverge at $x \to \pm \infty$ for arbitrary $\alpha$ and $\Delta$. This turns out to be difficult job except for $\alpha=0$ and $\alpha=1$ for which one can obtain explicit results. For general $\alpha$, it is however possible to derive some general results. For example,  the probability distribution $P_{\pm}(x,t)$ reaches a stationary state at large times for $\Delta>0$  with arbitrary $\alpha \geq 0$. For this case ($\gamma_1>\gamma_2$) particle tumbles  from $+1$ to $-1$ more frequently if it is on the positive side and from $-1$ to $+1$ more frequently if it is on the negative side. As a result there is an overall effective bias on the particle towards the origin which makes the RTP to reach a stationary state, form of which depends on the value of alpha. On the other hand for $\Delta \leq 0$ the distribution never reaches a stationary state where some properties of the time dependent distribution $P(x,t)=P_+(x,t)+P_-(x,t)$ can be obtained in the asymptotically large time limit. In this limit we demonstrate that the dynamics of the RTP can be described by an effective Langevin equation of a particle diffusing in an inhomogeneous medium with position dependent drift and diffusion constant.

%We will see below that Eq.~\eqref{main-eq-G} can be solved exactly for $\alpha=0$ and approximately for $\alpha=1$ for all $\Delta$. For general $\alpha$, we show that the probability distribution attains a stationary state for $\Delta>0$ (equivalently $\gamma_1>\gamma_2$) whose form is explicitly derived. On the other hand for $\Delta \leq 0$, at large time $t$ one finds that our model effectively becomes a particle diffusing in an inhomogenous medium with position dependent drift and diffusion constant. We write an effective Langevin equation for the particle and obtain the correct scaling of $x$ with time $t$ for $\Delta =0$. 
In what follows, we first consider $\alpha=0$ and $\alpha=1$ cases separately and consider the general $\alpha$ case in the subsequent section. For each values of $\alpha$, we discuss the three cases (i) $\Delta >0$ (ii) $\Delta = 0$ and (iii) $\Delta <0$ separately.

\begin{figure}[t]
\includegraphics[scale=0.3]{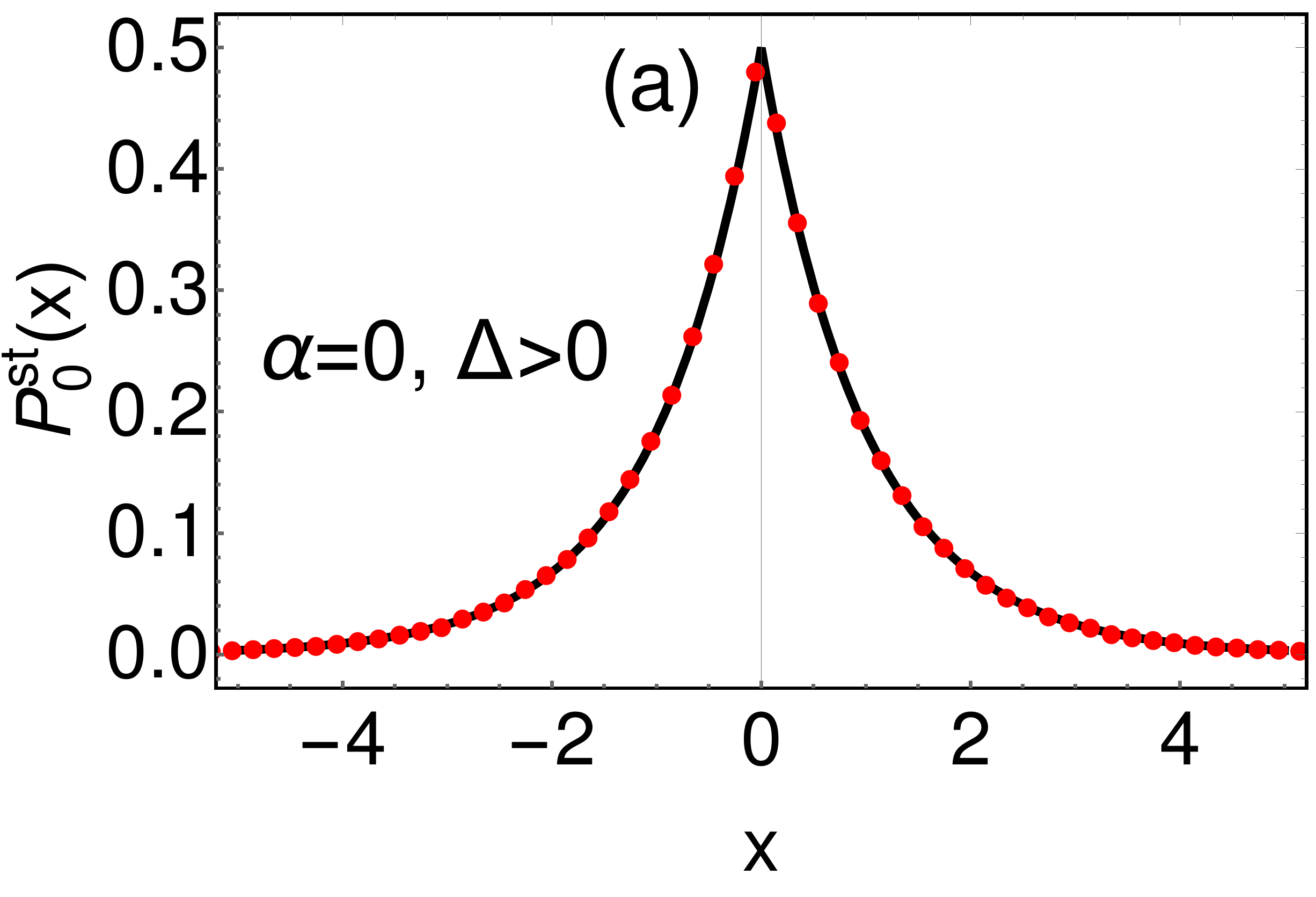}
\includegraphics[scale=0.25]{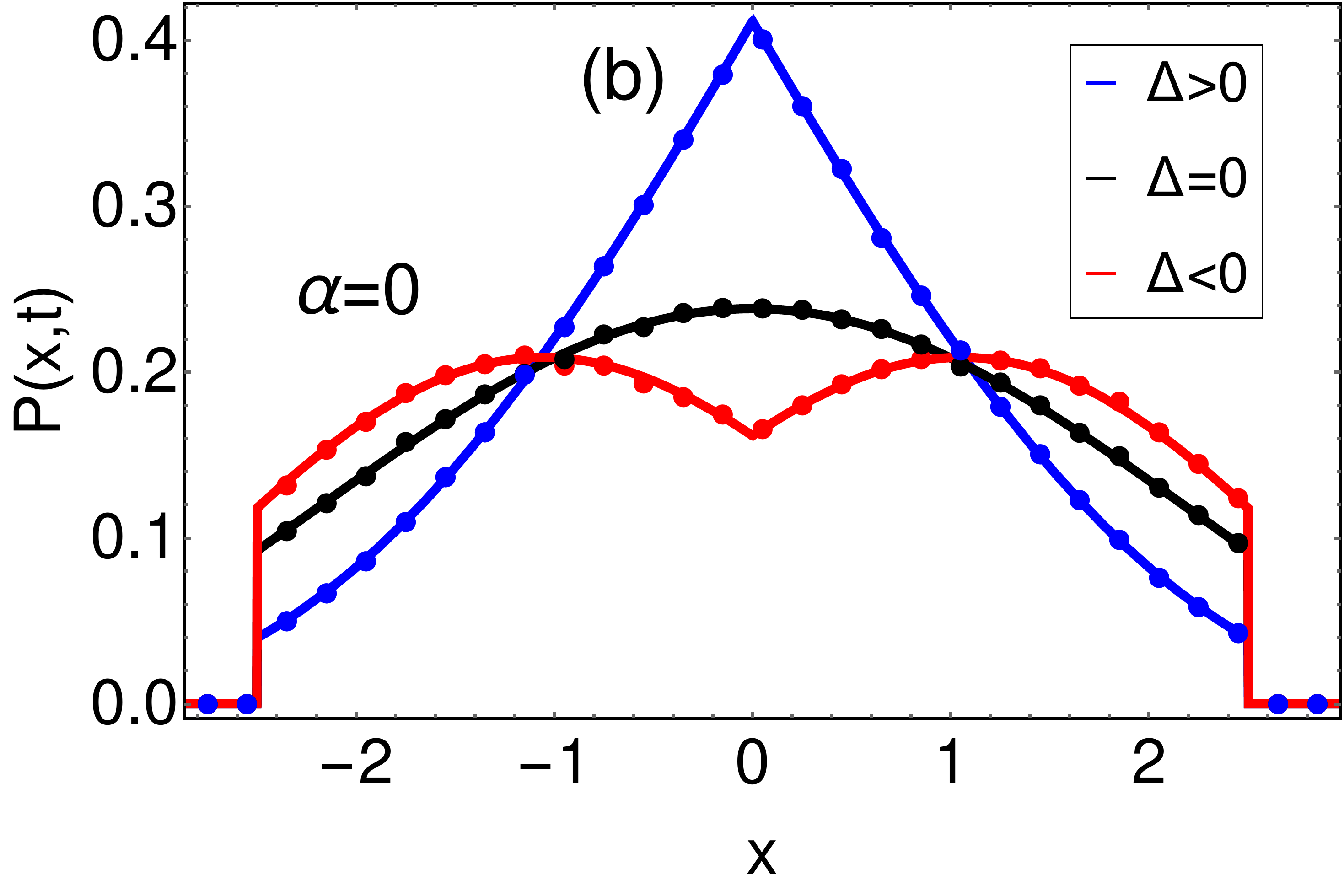}
\centering
\caption{(a) Comparison of the stationary distribution $P^{st}_{0}(x)$ in Eq.\eqref{Pss-a-0} for $\alpha =0$ with the same obtained from simulation of the microscopic dynamics (shown by filled circles). The parameters chosen are $v=1, \gamma_1 = 2 \text{ and }\gamma_2 = 1$. The histogram is constructed for $10^6$ realisations at time $t=20$. (b) The time dependent distribution $P(x,t)$ in Eq.~\eqref{main_eq14} of the position of the RTP for $\alpha=0$ case ( solid lines) has been shown in comparison with simulation data (fillled circles) for $t=2.5$. The blue,  black and red corresponds to the $\Delta>0$, $\Delta=0$ and $\Delta<0$. The explicit values for the parametrs for the three curves are given as follows: (i) $\gamma_1=1.5, \gamma_2=1$ (Blue) (ii) $\gamma_1=\gamma_2=1$ (Black) and (iii) $\gamma_1=1, \gamma_2=1.5$ (Red).  For all plots we have taken $v=1$.}
\label{new_steadypic1}
\end{figure}

\begin{figure}[t]
\includegraphics[scale=0.25]{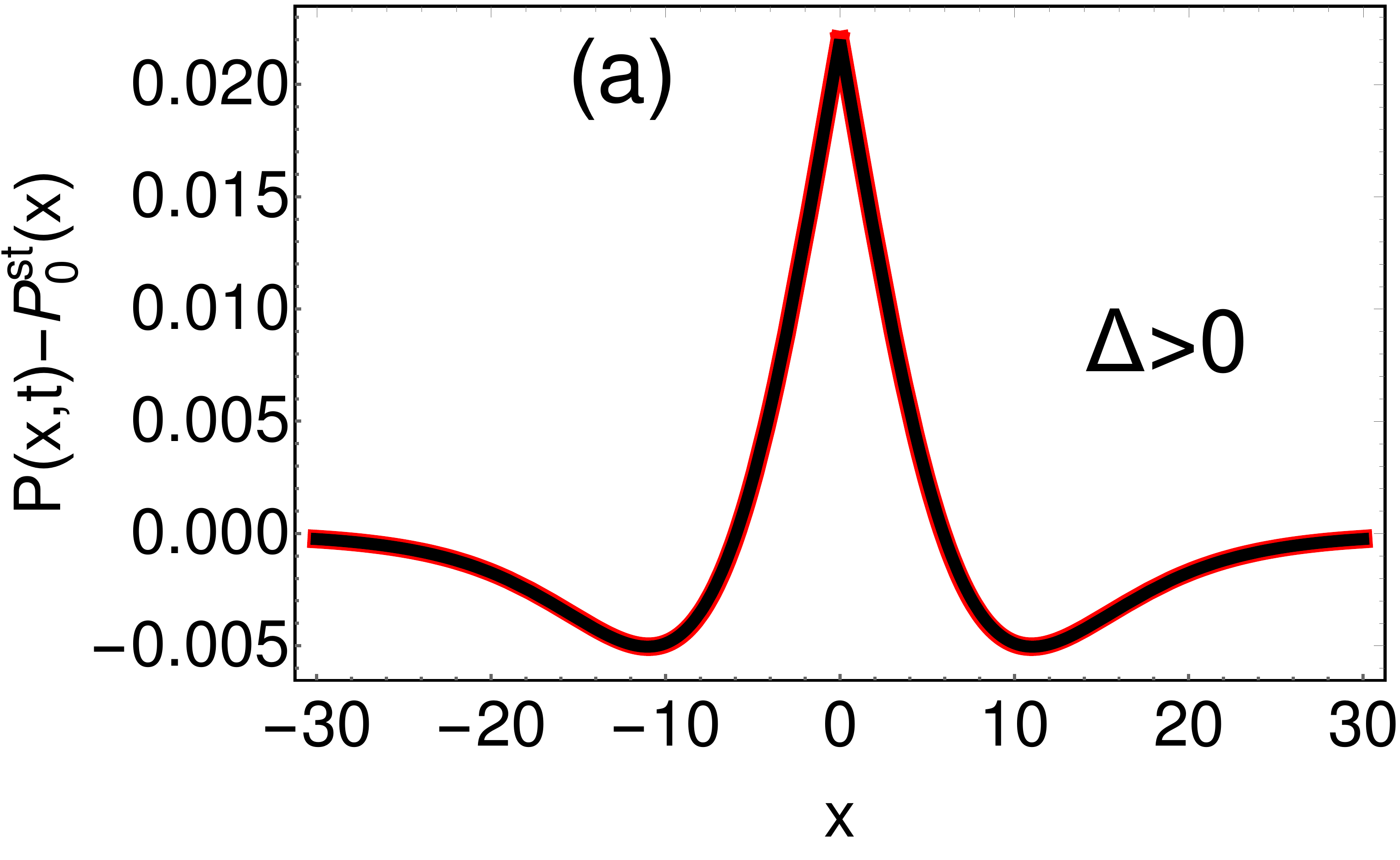}
\includegraphics[scale=0.33]{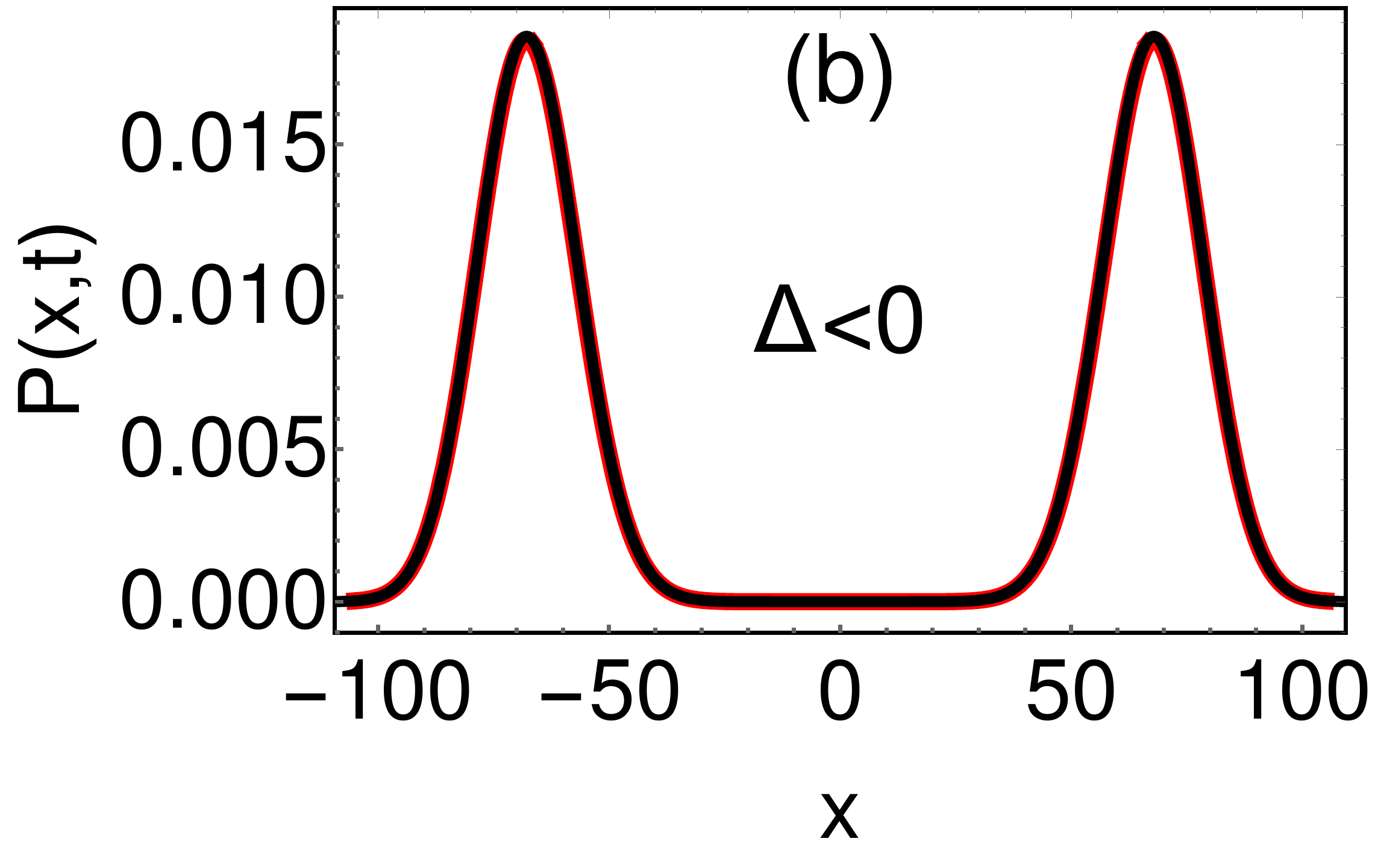}
\centering
\caption{Comparison of the approximate expressions (red) of $P(x,t)$ given in Eq. \eqref{main_eq15} with the exact result (black) in Eq.~\eqref{main_eq14} for $\Delta \neq 0$.  In Figure (a), we have plotted $P(x,t)-P_0^{st}(x)$ vs $x$ for $\gamma _1 = 1.2$, $\gamma_2 =1$ and $t = 50$ while in Figure (b), we plot $P(x,t)$ vs $x$ for $\gamma _1 = 1$, $\gamma_2 =2$ and $t = 200$. For both plots $v=1$.}
\label{asymptotic-alph-0}
\end{figure}

\subsection{Case I: $\alpha=0$}
\label{alpha=0}
For this case the rates $R_1(x)$ and $R_2(x)$ are independent of the magnitude of $x$ but depends on the sign of $x$. In this case, Eq.~\eqref{main-eq-G} reduces to 
\begin{align}
\partial_x^2 G-\lambda^2 G=0, \label{main_eq6}
\end{align}
where $\lambda (s)=\frac{1}{v} \sqrt{\Delta^2 + 2 \gamma s +s^2 }$. We solve this equation with the boundary conditions $G(x \to \pm \infty,s)=0$ and using the solution in Eqs. \eqref{main_eq2} and \eqref{main_eq4} we finally  get $\bar{Q}(x,s)$ and $\bar{P}(x,s)$. For clarity and compactness of the presentation, we have relegated the details of calculation of $G(x,s)$ to \ref{new_stdy_appen}. We here instead present the final expression of $\bar{P}(x,s)$ which reads as 
 \begin{align}
\bar{P}(x,s)=\frac{1}{2s}\left(\lambda(s)+\frac{\Delta}{v} \right)e^{-\left(\lambda (s)+\frac{\Delta}{v}\right) \mid x \mid}. \label{main_eq12}
\end{align}
%To get the probability distribution $P(x,t)$, one has to perform the Laplace inversion of $\bar{P}(x,s)$. 
%Before inverting Eq.~\eqref{main_eq12}, r
Recall that for $\Delta >0$ ({\it i.e.} $\gamma_1>\gamma_2$) one anticipates a stationary state distribution at late times given by
%$P^{st}_{0}(x)=\lim_{s \to 0}\left[ s \bar{P}(x,s)\right]$ (where $0$ in the subscript of $P^{st}_{0}(x)$ represents $\alpha=0$) which for $\Delta >0$ goes to
\begin{align}
P^{st}_{0}(x)=\lim_{s \to 0}\left[ s \bar{P}(x,s)\right]=\frac{\Delta}{v}~e^{-\frac{2\Delta}{v } \mid x \mid}. 
\label{Pss-a-0}
\end{align} 
which is an exponential distribution decaying over length scale $l_d=\frac{v}{2 \Delta}$. In Fig.\ref{new_steadypic1}(a), we have plotted our analytic result of $P^{st}_{0}(x)$ in Eq.~\eqref{Pss-a-0} with the numerical simulation of the same and find excellent agreement. Note that the decay length $l_d$ diverges as $\Delta \to 0$ which indicates that there is no stationary state for $\Delta =0$. 
For $\Delta \leq 0$, the $\lim_{s \to 0}\left[ s \bar{P}(x,s)\right]=0$ again implies that there is no stationary state for this case either. 

To get the distribution in time domain, one has to perform the inverse Laplace transform over $s$  (which for Laplace transform $\tilde{f}(s)$ of a function $f(t)$ is denoted by $f(t)=L_{s \to t}^{-1}[\bar{f}(s)]$). 
%Looking at the expressin of $\bar{P}(x,s)$, we need the following inverse Laplace transform (see \ref{laplace-appen} for derivation).
%\begin{align}
%&L_{s\to t} \left[e^{- \lambda(s) y}\right]=-v \frac{d}{dy} \left[ e^{-\gamma t}   I_0\left(\sqrt{\gamma_1 \gamma_2 \left(t^2-\frac{y^2}{v^2}\right)} \right)\Theta(vt-y)\right].
%\label{laplace-eq1}
%\end{align}
%Using this one can now perform inverse Laplace transform over $s$ in Eq.~\eqref{main_eq12}. 
The details of inversion of $\bar{P}(x,s)$ is relegated to \ref{laplace-appen} and we provide only the final result here. 
%Performing inverse Laplace transform over $s$ in Eq.~\eqref{main_eq12}, 
%(defined for a function $\tilde{f}(s)$ as $f(t)=L_{s \to t}^{-1}[\bar{f}(s)]$) 
%we get the following expression for the distribution $P(x,t)$
%symbolically represents inverse Laplace transform of $\bar{f}(s)$ with respect to $s$

\begin{align}
P(x,t)=&\frac{1}{2} e^{-\gamma_1 t} \delta (\mid x \mid-vt)+\frac{\gamma_1}{2 v}\left( 1+\frac{\gamma_2 \mid x \mid}{2 v}\right)e^{-\frac{\gamma_1 \mid x \mid}{2 v}}\Theta \left(v t -\mid x \mid \right)\nonumber \\
&~~~~~~~~ -\frac{\sqrt{\gamma_1 \gamma_2}}{2v} \int_{0}^{t} d\tau~e^{-\gamma \tau}~ \frac{d ~\mathcal{I}(\mid x \mid, \tau)}{d \mid x \mid}\Theta \left(v \tau -\mid x \mid \right),
\label{main_eq14}
\end{align}
where $\mathcal{I}(x,t)=\frac{x e^{-\frac{\Delta x}{v}}}{v}\frac{I_{1}\left( \sqrt{\gamma_1 \gamma_2(t^2-\frac{x^2}{v^2})}\right)}{\sqrt{t^2-\frac{x^2}{v^2}}}$ with $I_1$ being the modified Bessel function of first kind. Note that the distribution $P(x,t)$ contains $\delta$-function terms at $x=\pm vt$. They arise from those trajectories in which the particle has not changed its velocity direction till time $t$ starting from $x=0$ with equal probability for $\pm v$. In Fig.\ref{new_steadypic1}(b), we plot the above result for $P(x,t)$ for three cases and compare them against the direct simulation of the Langevin equation \eqref{langevin}. It is interesting to note that $P(x,t)$ has a dip at $x=0$ for $\gamma_1<\gamma_2$ and a peak for $\gamma_1>\gamma_2$. Appearance of this behaviour can be understood from the fact that for $\gamma_1 <\gamma_2$, the particle is drifted away from the origin while for $\gamma_1>\gamma_2$the drift is towards the origin. Another interesting point to note is the derivative of $P(x,t)$ has discontinuity at $x=0$ for $\gamma_1 \neq \gamma_2$ while it is continous for $\gamma_1 = \gamma_2$. For $\alpha =0$ and $\gamma_1 \neq \gamma_2$, the rates $R_1(x)$ and $R_2(x)$ in Eq.~\eqref{rates} have discontinuty at $x=0$ which amounts to the discontinuity in the derivative of $P(x,t)$. Although the theoretical result in Eq.~\eqref{main_eq14} is exact but less explicit and illuminating. For that it is instructive to get more explicit but approximate expression of the distribution $P(x,t)$ for both $\gamma_1>\gamma_2$ and $\gamma_1\leq \gamma_2$ cases in the large time $t$ limit. After some algebra ( presented in \ref{assy-del-gt-0} and \ref{assy-del-lt-0} with details) we find the following approximate  expressions valid for large $t$:
%In principle one can take the large $t$ limit directly in the expression of $P(x,t)$ given by Eq.~\eqref{main_eq14}. However it seems more convenient to take small $s$ limit in $\bar{P}(x,s)$ given by Eq.~\eqref{main_eq12} and then perform the inversion of corresponding $\bar{P(x,s)}$. 
%\redw{The details of the calculation of asymptotics are given in \ref{assy-del-gt-0} and \ref{assy-del-lt-0}. The asymtotics finally read as, }
%defining $\mathcal{P}(x,t)=P(x,t)-P(x,t \to \infty)$ with $P(x,t \to \infty)$ is $0$ for $\gamma_1 \leq \gamma_2$ and $P^{st}_{0}(x)$ for $\gamma_1 >\gamma_2$ 
\begin{align}
{P}(x,t)\simeq
    \begin{cases}
     &  P^{st}_{0}(x)+\frac{e^{-t\left[\gamma - \sqrt{\gamma_1 \gamma_2} \left(1-\frac{\bar{x}^2}{2} \right)+\Delta \bar{x} \right]}}{4 |x|}\left[\sqrt{\frac{8 t  \sqrt{\gamma_1 \gamma_2 }}{\pi}}|\bar{x}| -t e^{t \rho _{-}^2} \text{Erfc}[\sqrt{t} \rho _-]\left( \Delta \bar{x} + \frac{\rho _+ ^2-\rho _ -^2}{2}\right) \right.\\
 &      \left.  ~~~~~~~~~~~~~~  +t e^{t \rho _{+}^2}~ \text{Erfc}[\sqrt{t} \rho _+]\left( \Delta \bar{x} - \frac{\rho _+ ^2-\rho _ -^2}{2}\right)\right]
   , ~~~~~~~~~~~~~~~~~~~~~~~~~ \text{if}\ \gamma_1>\gamma_2 \\
 & \\  
 &    \frac{1}{\sqrt{4 \pi D_0 t}} e^{-\frac{x^2}{4 D_0 t}}~~~~~~~~~~~~~~~~~~~~~~~~~~~~~~~~~~~~~~~~~~~~~~~~~~~~~~~~~~~~~~~~~~~~~~~ \text{if}\ \gamma_1 =  \gamma_2 \\ 
 & \\
&   \sqrt{\frac{\sqrt{\gamma ^2-\Delta ^2}}{2 \pi t}} \frac{\bar{x}}{2 v (1-\bar{x}^2)^{\frac{3}{4}}} \frac{\bar{x} \sqrt{\gamma ^2-\Delta ^2}+\Delta \sqrt{1-\bar{x}^2}}{\sqrt{\gamma ^2-\Delta ^2}-\gamma \sqrt{1-\bar{x}^2}} ~e^{-t \left[ \Delta \bar{x}+\gamma -\sqrt{\gamma^2-\Delta ^2}\sqrt{1-\bar{x}^2}\right]}, ~~~~~~~~~ \text{if}\ \gamma_1 < \gamma_2 \\
% \frac{1}{2 v s^{*} \sqrt{2 \pi t \phi^{''} (s^{*})}}\left(\bar{x}\sqrt{\frac{\gamma ^2-\Delta ^2}{1-\bar{x}^2}} +\Delta \right) % e^{t\left[-\Delta  \bar{x}+\phi (s^{*}) \right]},  \text{if}\ \gamma_1\leq \gamma_2 \text{\redw{verified}}
    \end{cases}
    \label{main_eq15}
\end{align} 
with $D_0 = \frac{v^2}{2 \gamma}$, $\bar{x}=\frac{|x|}{v t}$ and $\rho _{\pm} = \sqrt{\gamma - \sqrt{\gamma_1 \gamma_2}} \pm \bar{x}\sqrt{\frac{\sqrt{\gamma_1 \gamma_2}}{2}}$. 
%$~s^{*}=-\gamma +\sqrt{\frac{\gamma ^2-\Delta ^2}{1-\bar{x}^2}}$ and $\phi(s)=s-\bar{x} v \lambda (s)$
In Figure (\ref{asymptotic-alph-0}) we have compared these approximate results with the exact result in Eq. \eqref{main_eq14}.
\begin{figure}[t]
\includegraphics[scale=0.25]{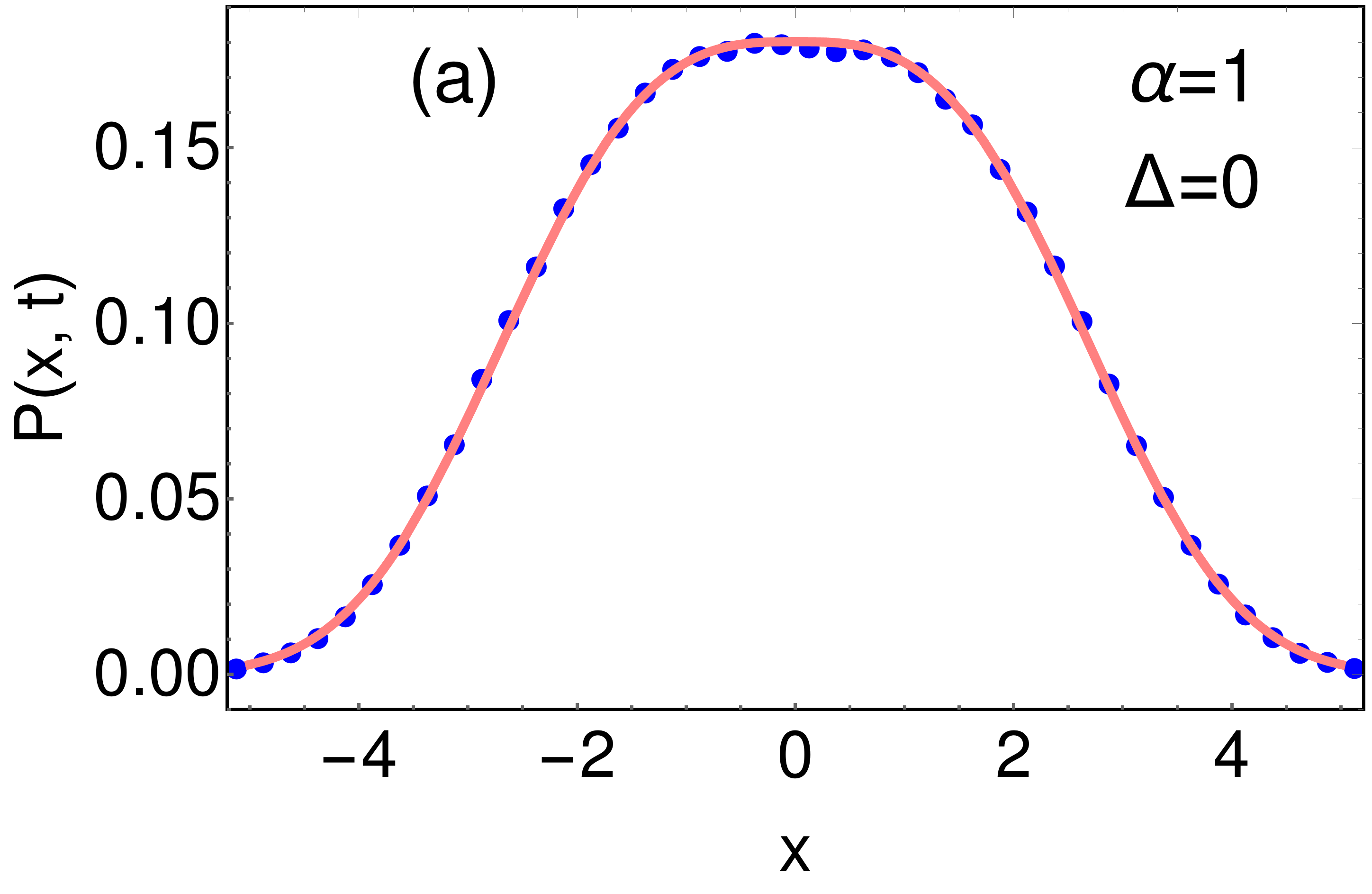}
\includegraphics[scale=0.27]{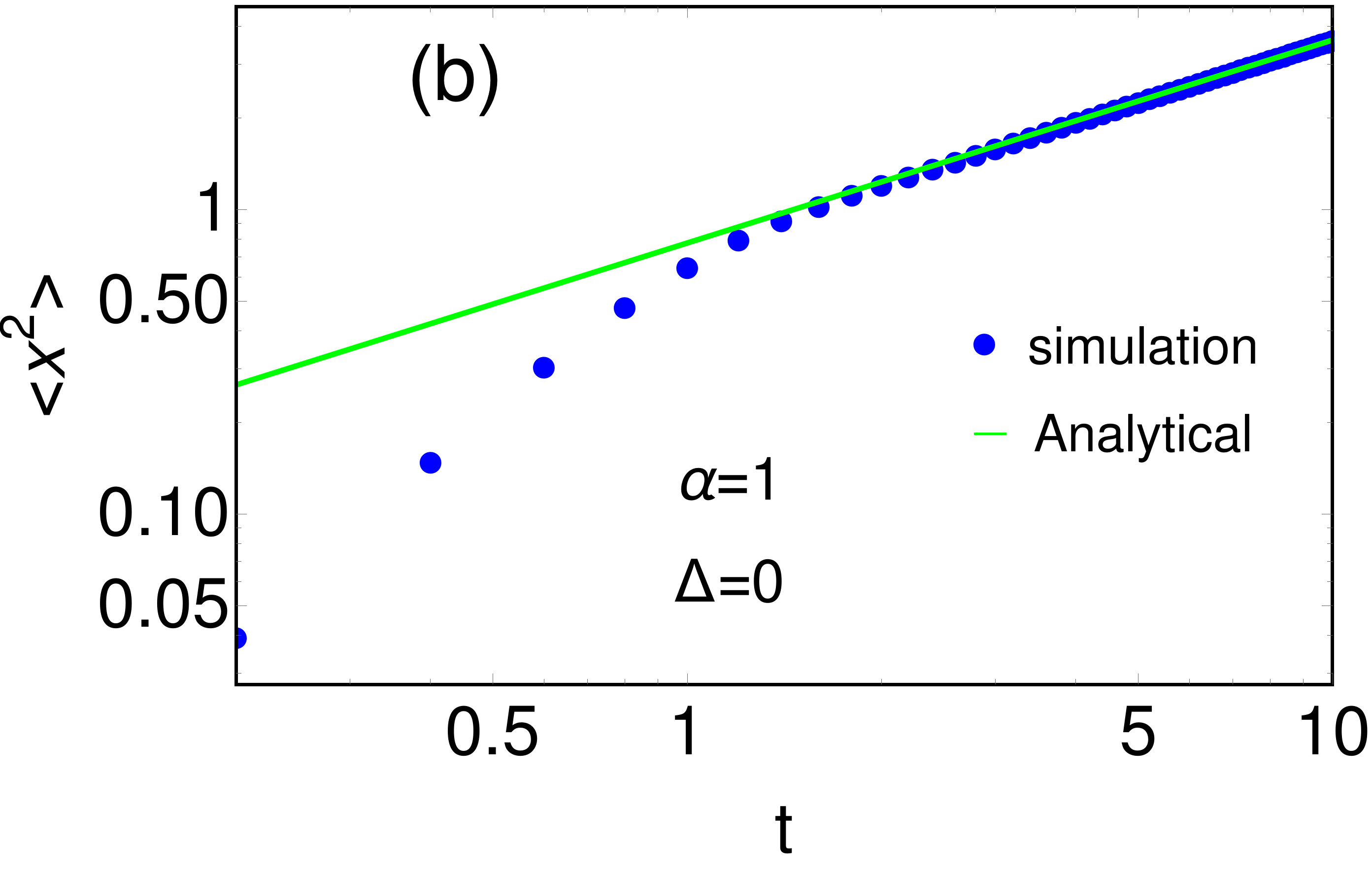}
\centering
\caption{Plots for the case $\alpha=1$ and $\Delta=0$. In figure (a) we plot $P(x,t)$ obtained from Eq.~\eqref{main_eqqq4} (solid line) and compare it with the numerical simulation of Eq.~\eqref{langevin} (filled circles) for $t=10$. In (b), we verify the scaling of $x \sim t^{\frac{1}{3}}$ by plotting $\langle x^2 \rangle$ against $t$. The Green solid line is the analytic result obtained by using expression for $P(x,t)$ in Eq.~\eqref{main_eqqq4} while blue filled circles are results of numerical simulation. For both plots, the chosen parameters are $v=1, \gamma_1=\gamma_2 = 1.5 \text{ and } l=1.$ }
\label{alph_1_Del_0}
\end{figure}

\subsection{Case II: $\alpha =1$}
\label{alpha=1}
We now focus on the second analytically solvable case $\alpha=1$ in which the equation \eqref{main-eq-G} becomes
\begin{align}
\partial_x^{2} G-\left[ \frac{\Delta}{v l}+\frac{2 \gamma s \mid x \mid}{v^2 l}+\frac{s^2}{v^2}+\frac{\Delta ^2 x^2}{v^2 l^2}\right]G=0.
\label{main_eqG}
\end{align}
We first look at the $\Delta=0$ case.
%We will proceed to solve this equation for $\Delta=0$ and $\Delta \neq 0$ cases separately. We will particularly use the notation and $G^{*}(x,s)$ and $\bar{P}^{*}(x,s)$ where $*$ is replaced by $0$ and $\Delta$ for $\Delta =0$ and $\Delta \neq 0$ respectively.

\subsubsection{$\Delta =0$}$ $\\
For this case Eq.~\eqref{main_eqG} becomes,
\begin{align}
\partial_x^{2} G(x,s)-\left[\frac{2 \gamma s \mid x \mid}{v^2 l}+\frac{s^2}{v^2} \right]G(x,s)=0
\label{G-alpha-1-Del-0}
\end{align}
We identify this equation as Airy differential equation whose general solutions are Airy functions. Satisfying appropriate boundary conditions, one finally gets $G(x,s)$ (see \ref{al-1-D-eq-0} for details) using which in  Eqs.\eqref{main_eq2} and \eqref{main_eq4} provides 
\begin{align}
\bar{P}(x,s)=-\frac{1}{2s} \frac{d}{d \mid x \mid} \left[\frac{\text{Ai} \left(c~ s^{\frac{1}{3}}\mid x \mid + d~ s^{\frac{4}{3}} \right)}{\text{Ai}\left(d ~s^{\frac{4}{3}} \right)} \right],
\label{main_eqqq2}
\end{align}
where $c=\left(\frac{2 \gamma }{v^2 l}\right)^{\frac{1}{3}}$, $d= \frac{c~l}{2 \gamma}$ and $\text{Ai}(x)$ is the Airy function of the first kind. Although it is possible to perform inverse Laplace transform for arbitrary $t$, it is however more interesting to look at the behaviour at large $t$, to obtain which one can neglect $O(s^{\frac{4}{3}})$ terms inside the argument of the Airy function in the numerator of Eq.~\eqref{main_eqqq2}. Using 
$\text{Ai}(z)=\frac{\sqrt{z}}{\sqrt{3} \pi} K_{\frac{1}{3}}\left(\frac{2}{3} z^{\frac{3}{2}} \right)$ we get
%We were not able to invert Eq.~\eqref{main_eqqq2} and get $P^{0}(x,t)$ for all finite $t$. However for large $t$ which correspond to small $s$, one can ignore terms $O(s^{\frac{4}{3}})$ and get
\begin{align}
\bar{P}(x,s)& \simeq -\frac{1}{2s} \frac{d}{d \mid x \mid} \left[\frac{\text{Ai} \left(c~ s^{\frac{1}{3}}\mid x \mid \right)}{\text{Ai}(0)} \right]
%, \nonumber \\&
=\frac{c^2 \mid x \mid}{2\sqrt{3} \pi \text{Ai}(0)s^{\frac{1}{3}}} K_{\frac{2}{3}} \left(\frac{2}{3} \left( c \mid x \mid\right)^{\frac{3}{2}}\sqrt{s} \right),
\label{main_eqqq21}
\end{align}
where $K_{\frac{2}{3}}\left(y \right)$ is Modified Bessel function. Performing the inverse Laplace transform we get
\begin{align}
P(x,t)=\frac{1}{4 \pi \text{Ai}(0) t^{\frac{1}{6}}} \sqrt{\frac{3 c}{ \mid x \mid}} e^{-\frac{\mid x \mid ^3 c^3}{18t}} W_{\frac{1}{6},\frac{1}{3}} \left( \frac{\mid x \mid ^3 c^3}{9t}\right),
\label{main_eqqq4}
\end{align}
where $W_{m, \nu}(x)$ is the Whittaker function. To arrive at the above result we have used the following result 
%\begin{align}
%\text{Inverse~Laplace~Transform}\left[2 \sqrt{g} ~s^{m -1} K_{2\nu} \left(2\sqrt{g s} \right)\right]=t^{\frac{1}{2}-m} e^{-\frac{g}{2 t}}~ W_{m-\frac{1}{2},~ \nu} \left(\frac{g}{t} \right),
%\label{main_eqqq3}
%\end{align}
\begin{align}
L^{-1}_{s\to t}\left[2 \sqrt{g} ~s^{m -1} K_{2\nu} \left(2\sqrt{g s} \right)\right]=t^{\frac{1}{2}-m} e^{-\frac{g}{2 t}}~ W_{m-\frac{1}{2},~ \nu} \left(\frac{g}{t} \right),
\label{main_eqqq3}
\end{align}
for $g>0$.
% and $W_{m, \nu}(x)$ is the Whittaker function. Using this, one can easily invert $\bar{P}^{0}(x,s)$ in Eq.~\eqref{main_eqqq2} to get the distribution in time domain as
In Figure \ref{alph_1_Del_0}(a), we compare our result for the distribution $P(x,t)$ in Eq.~\eqref{main_eqqq4} against numerical simulation where we notice excellent agreement. For very large $t$, one can  simplify the expression for $P(x,t)$ further by using the asymptotic of $W_{\frac{1}{6},\frac{1}{3}} (z) \approx z^{\frac{1}{6}} e^{-\frac{z}{2}}$ for $z \to 0$ and $\text{Ai}(0)=\frac{\Gamma \left(\frac{1}{3}\right)}{2 \pi 3^{\frac{1}{6}}}$. We get
\begin{align}
P(x,t) \simeq \frac{1}{t^\frac{1}{3}}f_0\left(\frac{x}{t^\frac{1}{3}}\right),~\text{where},~f_0(y)=\frac{3^{\frac{1}{3}} c}{2 \Gamma\left(\frac{1}{3} \right)} e^{-\frac{\mid y \mid ^3 c^{3}}{9}}.
\label{sc-sol-a-1-D-0}
\end{align}
Remember $c=\left(\frac{2 \gamma }{v^2 l}\right)^{\frac{1}{3}}$. This result implies that for $\alpha=1$ with $\Delta=0$, the position $x$ of the particle 
scales as  $x \sim t^{\frac{1}{3}}$ for large $t$ which is different from the $\alpha=0$ case for which $x \sim \sqrt{t}$. To numerically verify this, we plot variance $\langle x^2 \rangle$ against $t$ in Figure \ref{alph_1_Del_0}(b) and at large $t$ we indeed observe the scaling behaviour $\langle x^2 \rangle \sim t^{\frac{2}{3}}$. We will later see that the above scaling behaviour gets generalised for general $\alpha > 0$.

\begin{figure}[t]
\includegraphics[scale=0.31]{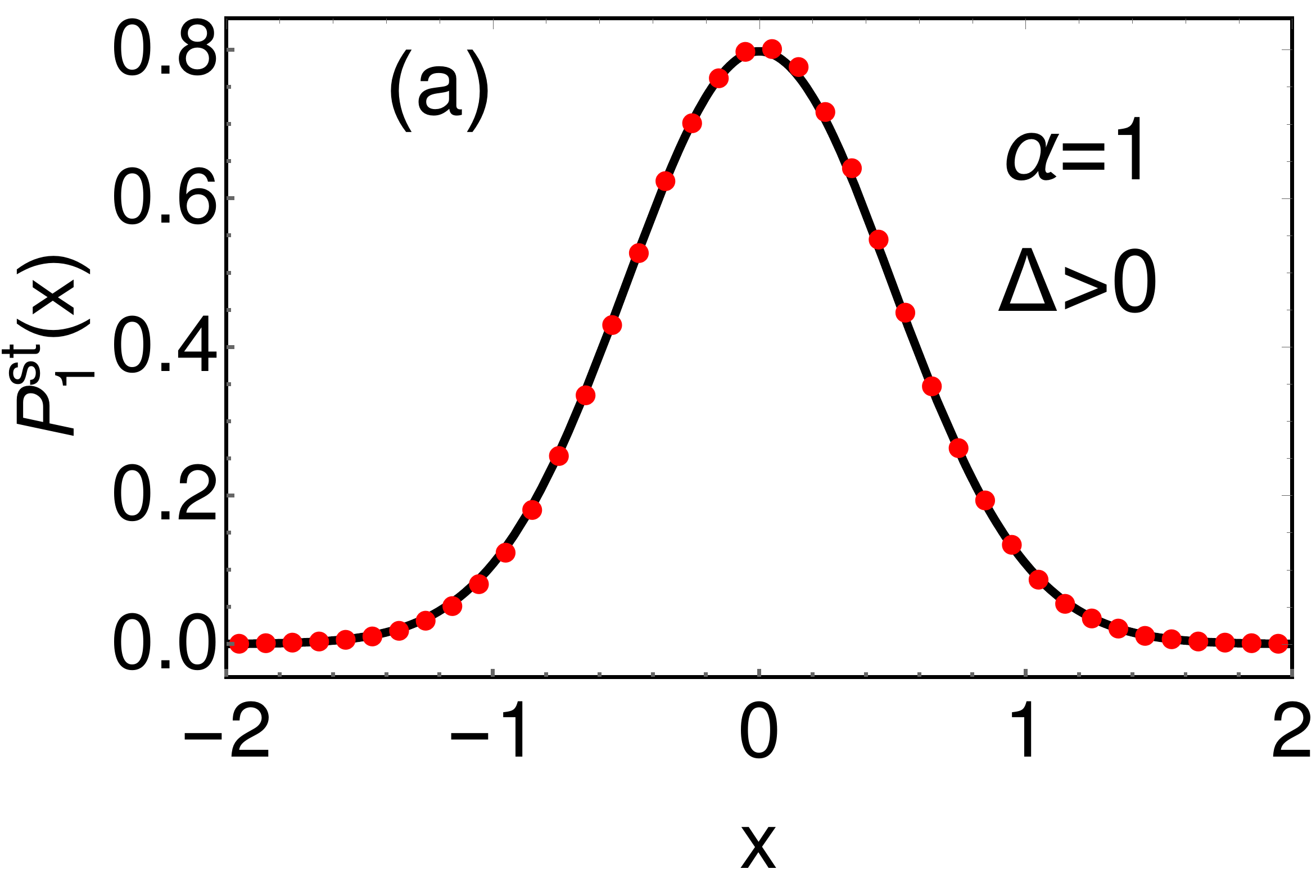}
\includegraphics[scale=0.27]{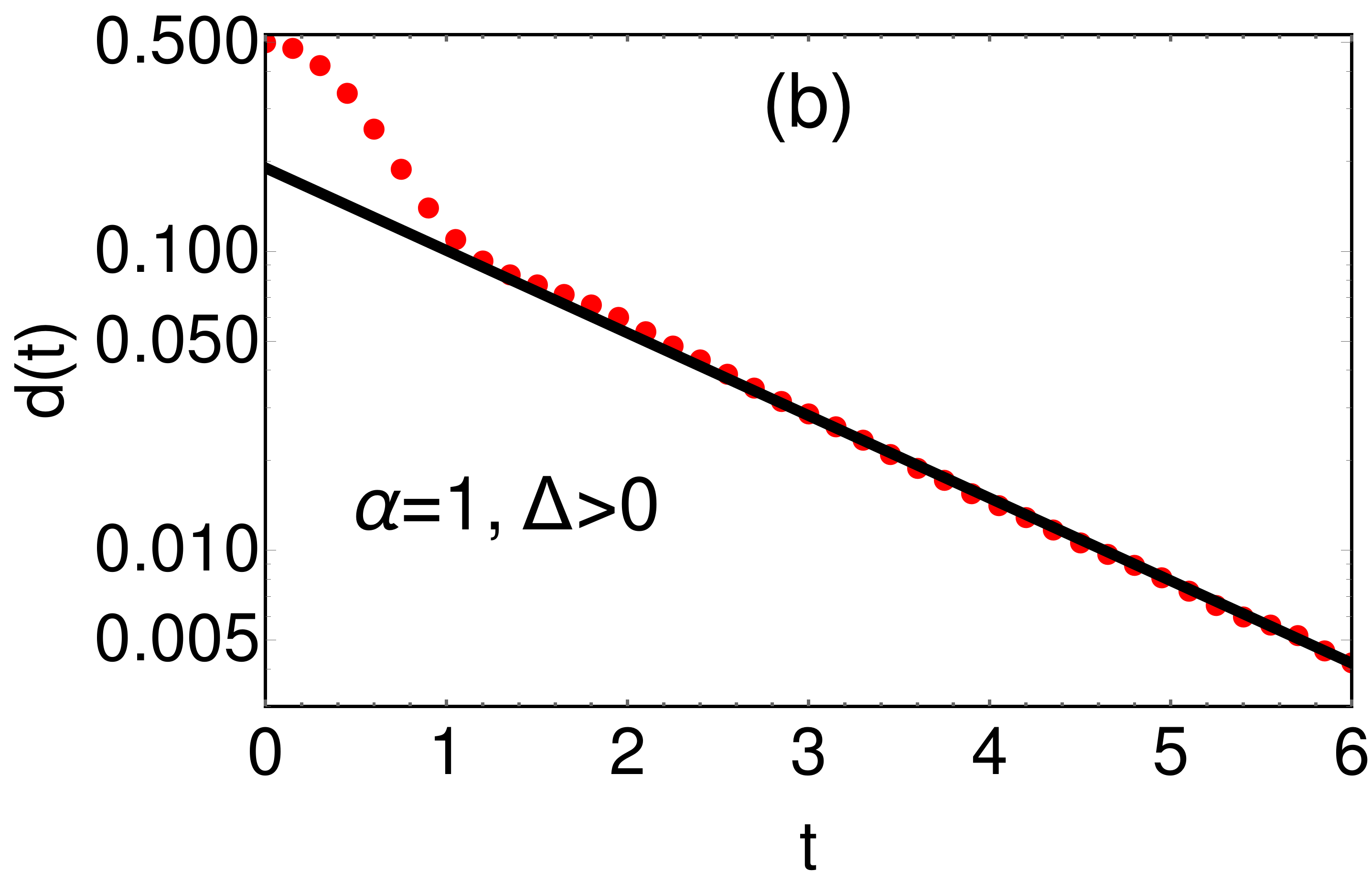}
\centering
\caption{(a) Comparison of the stationary distribution $P^{st}_{1}(x)$ in Eq.\eqref{Pss-a-1} for $\alpha =1$ (solid line) with simulation of the Langevin equation (shown by filled circles). The parameters chosen are $v=1,l=0.5, \gamma_1 = 3 \text{ and }\gamma_2 = 1$. The histogram is constructed for $10^6$ realisations at time $t=30$. (b) Here we plot $d(t)=\text{var}(\infty)-\text{var}(t)$ where the variance is $\text{var}(t)=\langle x(t)^2\rangle -\langle x(t) \rangle^2$. The simulation result (shown by filled circles) is plotted with $d(t)\sim e^{-\zeta t}$ (shown by thick line) with $\zeta=0.636$ is given by solution of Eq.~\eqref{D-zero} with largest real part. The parameters chosen are  $\gamma_1=3.5, \gamma_2=1.5, v=1 \text{ and } l=1$.}
\label{new_steadypic_alph_1}
\end{figure}

\subsubsection{$\Delta \neq 0$} $ $\\
Here we consider the $\Delta \neq 0$ case for $\alpha=1$, for which we solve Eq.~\eqref{main_eqG}. Note that the general solutions of this equation can be expressed in terms of parabolic cylinder functions $D_{\mu-1/2}(x)$. Choosing the integration constants appropriately to satisfy the boundary conditions, one obtains $G(x,s)$, using which in Eqs.\eqref{main_eq2} and \eqref{main_eq4} we get ( see \ref{al-1-D-neq-0} for details) 
\begin{align}
\bar{P}(x,s)=-\frac{1}{2s} \frac{d}{d\mid x \mid} \left[ e^{-\frac{\Delta }{2vl}x^{2}} \frac{D_{\beta s^2-\frac{1+\text{sgn}(\Delta)}{2}}\left( \sqrt{\frac{2 \mid \Delta \mid}{v l}}\left(\mid x \mid+\frac{\gamma s l}{\Delta ^2} \right)\right)}{D_{\beta s^2-\frac{1+\text{sgn}(\Delta)}{2}}\left( \sqrt{\frac{2 \mid \Delta \mid}{v l}}\frac{\gamma s l}{\Delta ^2} \right)}\right],
\label{main_eq17}
\end{align} 
where $\beta=\frac{ l(\gamma^2-\Delta^2)}{2 v \mid \Delta \mid^3}$.
%$\mu(s)=\frac{s^2 l(\gamma^2-\Delta^2)}{2 v \mid \Delta \mid^3}-\frac{\text{sgn}(\Delta)}{2}$. 
Note that for $\Delta>0$, we have $\mu(s)=\nu s^2$Just like the 
$\alpha=0$ case, in this case also we anticipate stationary state for $\Delta >0$ which can be determined from the limit $\lim_{s \to 0}\left[s \bar{P}(x,s)\right]$. Using $D_{-1}(z)=\sqrt{\frac{\pi}{2}} e^{\frac{z^2}{4}} \text{Erfc}(\frac{z}{\sqrt{2}})$, we get 
\begin{align}
P^{st}_{1}(x)=\sqrt{\frac{\gamma_1-\gamma_2}{2\pi v l}}e^{-\frac{\gamma_1-\gamma_2}{ 2v l} x^2},
%~\magentaw{l~dependence~ checked}
\label{Pss-a-1}
\end{align} 
for $\Delta >0$ where the subscript $1$ in $P^{st}_{1}(x)$ stands for $\alpha =1$. In Figure \ref{new_steadypic_alph_1}(a), we have plotted $P^{st}_1(x)$ and compared it with the direct numerical simulation of the microscopic equation \eqref{langevin}. We find excellent agreement between the two. To understand the 
relaxation to this stationary state one needs to take into account the contribution from the pole with second largest real part (largest pole is $s=0$ which gives the steady state) in the Laplace inversion procedure. The poles of $\bar{P}(x,s)$ in Eq.~\eqref{main_eq17}, come from the zeros of $D_{\beta s^2-1}\left( \sqrt{\frac{2 \mid \Delta \mid}{v l}}\frac{\gamma s l}{\Delta ^2} \right)$ which lie on the negative real axis. 
The pole $\zeta$ with largest real part other than $0$ will set the time scale $|\zeta|^{-1}$ for the exponential relaxation which can be determined by solving 
\begin{align}
D_{\beta \zeta^2-1}\left( \sqrt{\frac{2 \mid \Delta \mid}{v l}}\frac{\gamma \zeta l}{\Delta ^2} \right)=0,
\label{D-zero}
\end{align}
numerically for $\Delta >0$. To verify this result we compute $d(t)=\text{var}(\infty)-\text{var}(t)$ where $\text{var}(t)=\langle x^2 \rangle-\langle x \rangle ^2$ is the variance obtained from the numerical simulation, which should decay to zero as $\sim e^{-\zeta t}$. In Figure \ref{new_steadypic_alph_1}(b) we plot $d(t)$ as a function of $t$ and indeed observe the exponential decay with time scale $|\zeta|^{-1}$.

\begin{figure}[t]
\includegraphics[scale=0.35]{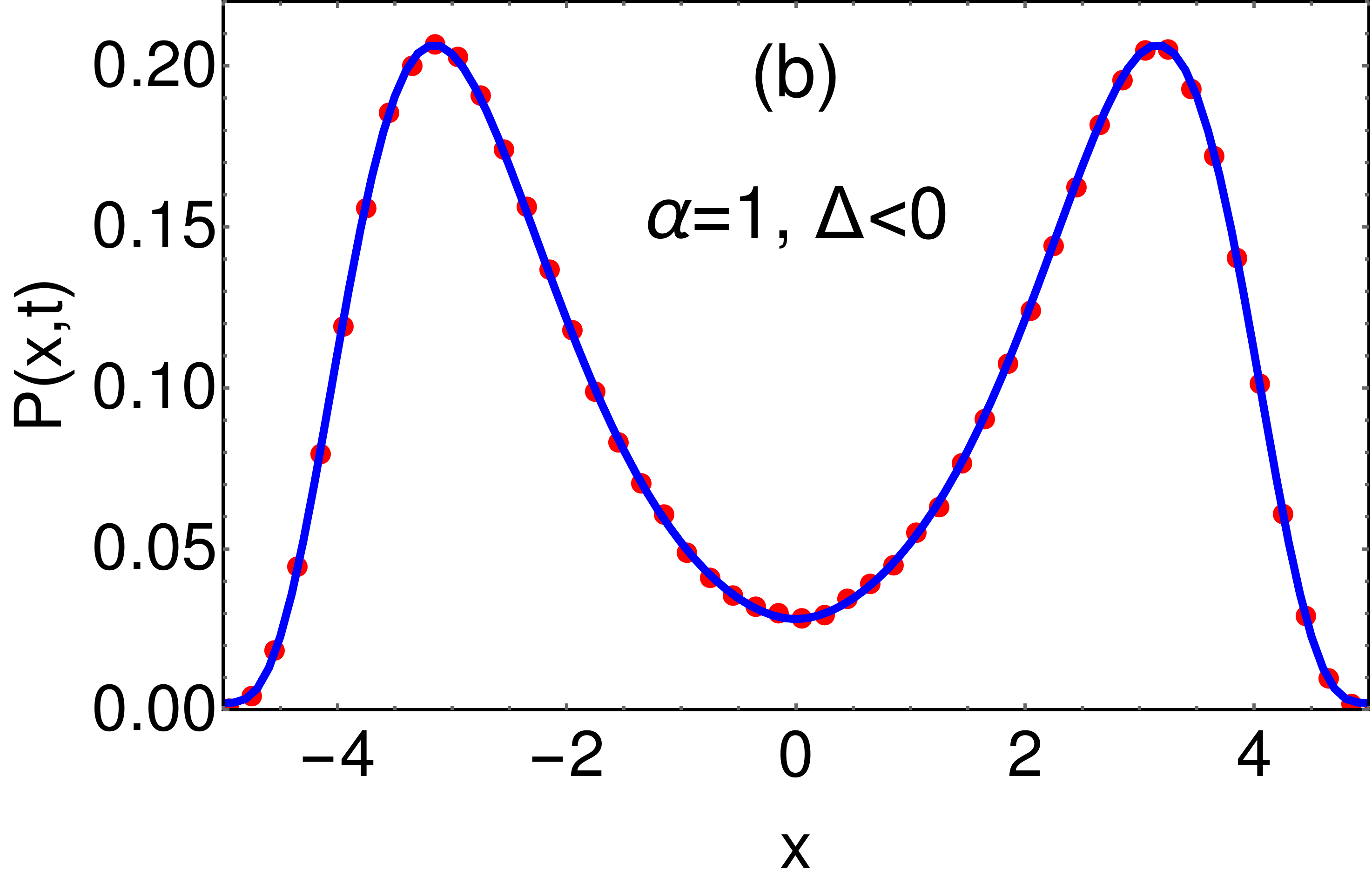}
\centering
\caption{Comparision of the probability distribution $P(x,t)$ for $\alpha=1$ and $\Delta<0$ with the numerical simulation. The theretical curve (solid line) is obtained by performing inverse Laplace transform in Eq.~\eqref{bar_P_alph-1-Del<0}. For this plot the parameters we have taken are $\gamma_1=1,~\gamma_2=3,~v=1~l=1$ and $t=5$.} 
\label{alpha-1-Delta<0}
\end{figure}

\begin{figure}[t]
\includegraphics[scale=0.3]{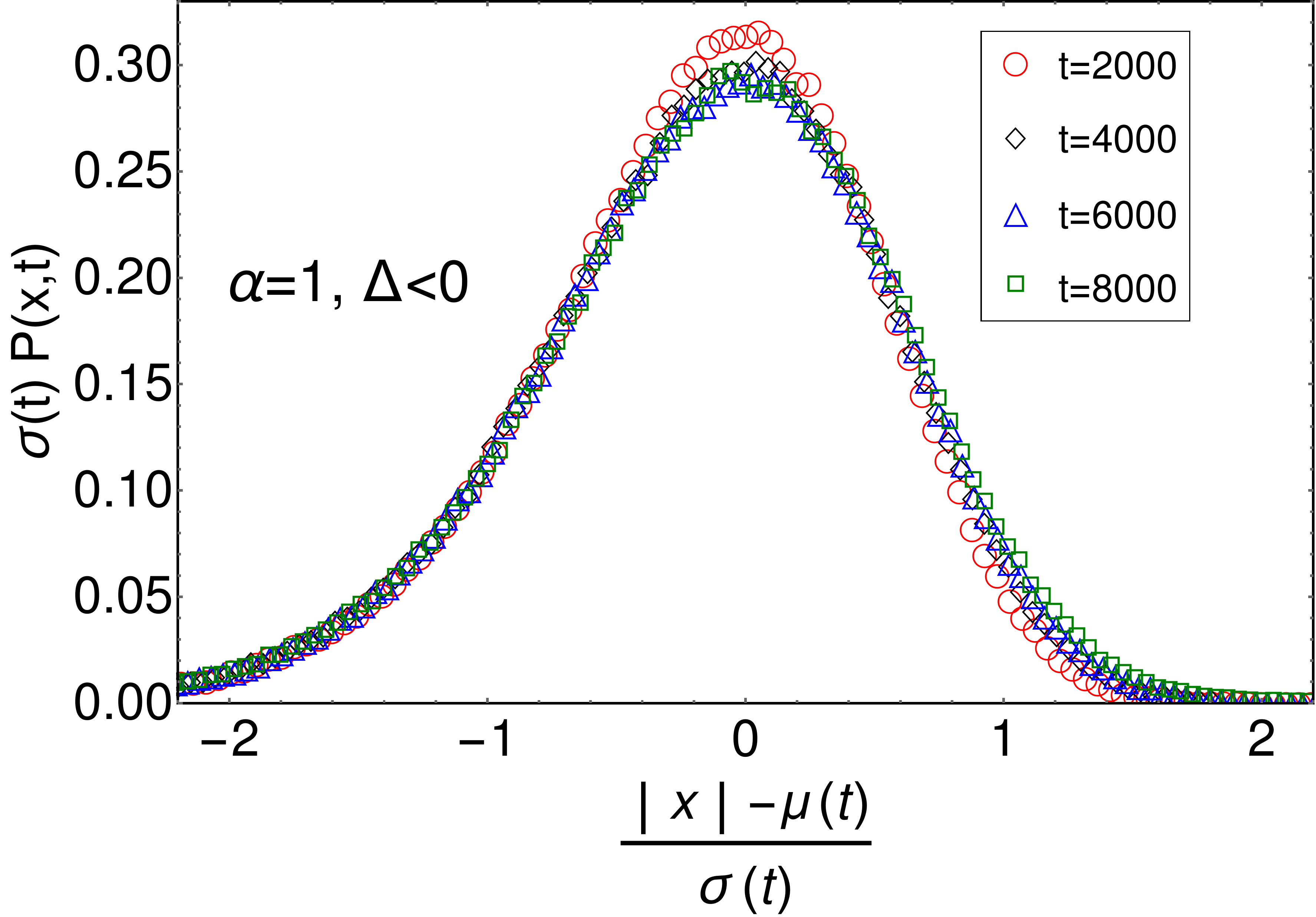}
\centering
\caption{Verification of the scaling behaviour of the distribution $P(x,t)=\frac{1}{2\sigma_1(t)}\mathcal{G}_1\left(\frac{|x|-\mu(t)}{\sigma_1(t)}\right)$ for $\alpha=1$ where $\mu(t)=\langle |x| \rangle$ and $\sigma_1^2(t)=\langle x^2 \rangle-\langle |x| \rangle^2$. Here we have shown the scaling behaviour only for positive $x$ as the distribution $P(x,t)$ is symmetric. Other parameters of the plot are $\gamma_1=1.5,~\gamma_2=1.6,~v=1$ and $l=1$.} 
\label{scaling-alpha-1}
\end{figure}

For $\Delta <0$, as we have argued earlier there is no stationary state. In this case, the solution $\bar{P}(x,s)$ in the Laplace space, given in Eq.~\eqref{main_eq17}, becomes
\begin{align}
\bar{P}(x,s)=-\frac{1}{2s} \frac{d}{d\mid x \mid} \left[ e^{-\frac{\Delta }{2vl}x^{2}} \frac{D_{\beta s^2}\left( \sqrt{\frac{2 \mid \Delta \mid}{v l}}\left(\mid x \mid+\frac{\gamma s l}{\Delta ^2} \right)\right)}{D_{\beta s^2}\left( \sqrt{\frac{2 \mid \Delta \mid}{v l}}\frac{\gamma s l}{\Delta ^2} \right)}\right],
\label{bar_P_alph-1-Del<0}
\end{align} 
where $\beta=\frac{ l(\gamma^2-\Delta^2)}{2 v \mid \Delta \mid^3}$.
Performing the inverse Laplace transform , we can obtain $P(x,t)$. For the parameters $\gamma_1=1,~\gamma_2=3,~v=1,~l=1$ and $t=5$ we perform the inverse Laplace transform numerically to get $P(x,t)$ at $t=5$ which we compare with simulation results in Fig.~\ref{alpha-1-Delta<0} and observe excellent agreement. The convergence of the numerical inversion procedure becomes poor with increasing $t$.  However, following a different approximate procedure, explained in the next section, we find that for large $t$, the distribution $P(x,t)$ has the following scaling form $P(x,t) \simeq \frac{1}{2\sigma_1(t)}\mathcal{G}\left(\frac{|x|-\mu(t)}{\sigma_1(t)}\right)$ with $\mu(t)=\langle |x| \rangle=\frac{v|\Delta|}{\gamma}t$ and $\sigma_1^2(t)=\langle x^2 \rangle-\langle |x| \rangle^2=(v l/|\Delta |)~\ln(t)$. In Fig.~\ref{scaling-alpha-1} we verify this scaling behaviour numerically. In the next section we show that $\mathcal{G}(u)$ is a mean zero and unit variance Gaussian.

\begin{figure}[t]
\includegraphics[scale=0.24]{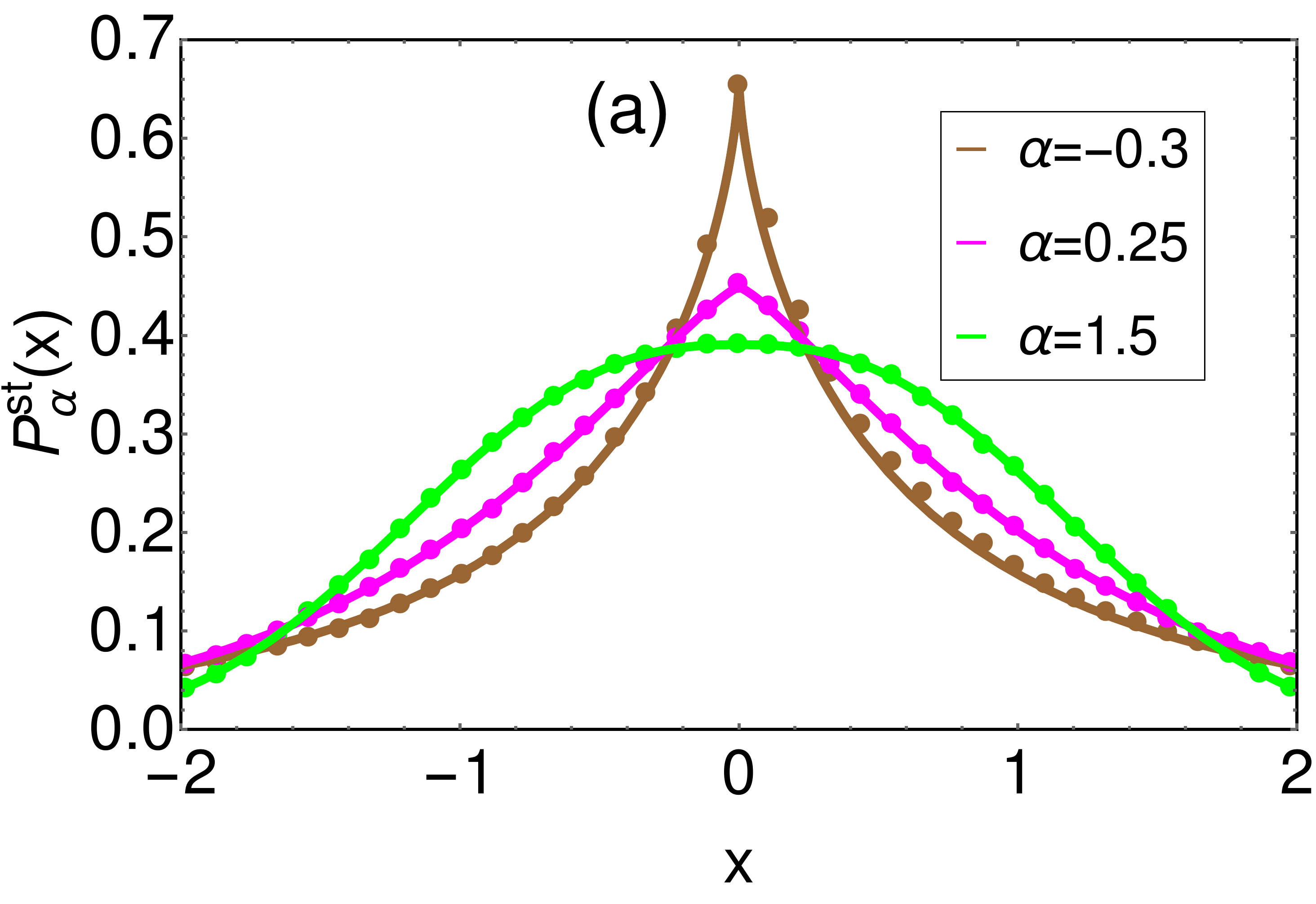}
\includegraphics[scale=0.28]{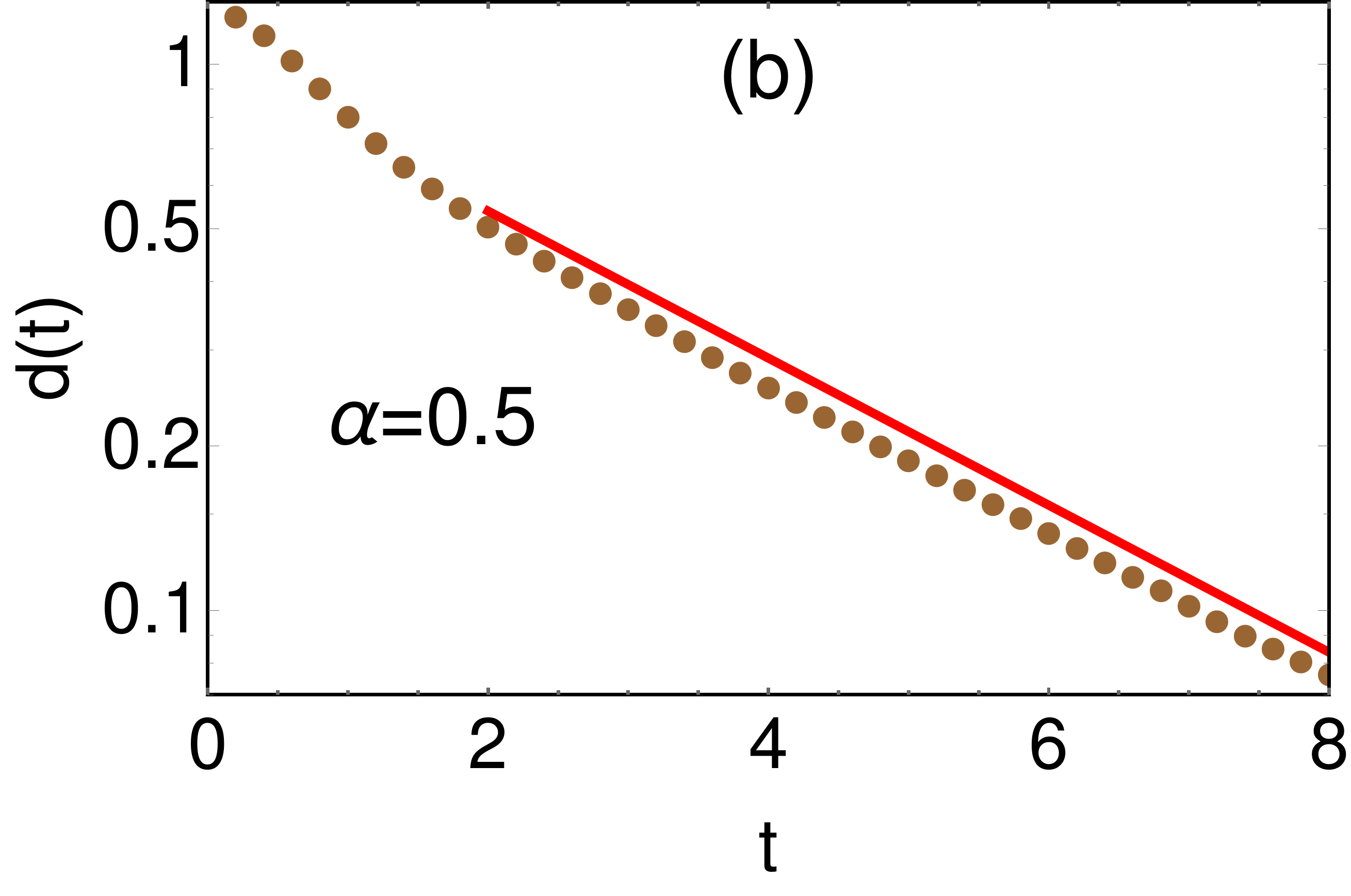}
\centering
\caption{(a)Comparision of the stationary state distribution $P_{\alpha}^{st}(x)$ obtained in Eq.~\eqref{stdy_sol} (solid lines) with the numerical simulation data (filled circles) for three values of $\alpha$. The histogram has been constructed using $10^6$ realisations at $t=15$.(b)Numerical simulation of $d(t)$ defined as $d(t)=\text{var}(\infty)-\text{var}(t)$ where $\text{var}(t)=\langle x(t)^2\rangle -\langle x(t) \rangle^2$. We find $d(t) \sim e^{-\zeta t}$(shown by red line) where we numerically find $\zeta=0.31$. For both plots, we have taken $\gamma_1=2,~\gamma_2=1~l=1~\text{and }v=1.$} 
\label{gen-alp-delgt0}
\end{figure}

%Coming to $P^{\Delta}(x,t)$, we were not able to invert the Laplace transforms in Eq.~\eqref{main_eq17}. However in simulations for $\Delta >0$ (as shown in Figure \ref{new_steadypic_alph_1}(b)), we observe that the particle relaxes to the stationary state exponentially as $e^{-\zeta t}$. The exponent $\zeta$ is given by the second largest pole of Eq.~\eqref{main_eq17} which comes from zeros of $D_{\mu(s)-\frac{1}{2}}\left( \sqrt{\frac{2 \mid \Delta \mid}{v l}}\frac{\gamma s l}{\Delta ^2} \right)$. Note that $s=0$ is the largest pole which will give rise to steady state contribution while all other poles will have negative real part else probability distribution will diverge with $t$. In Figure \ref{new_steadypic_alph_1}(b), we have plotted $d(t)=\text{var}(\infty)-\text{var}(t)$ where $\text{var}(t)=\langle x^2 \rangle-\langle x \rangle ^2$ is the variance obtained from the numerical simulation. As discussed above $d(t) \sim e^{-\zeta t}$ for large $t$ with $\zeta$ given largest solution of Eq.~\eqref{main_eqqqqqq1}. This is validated by numerical simulation to an excellent extent as shown in Figure \ref{new_steadypic_alph_1}(b). 

\subsection{Case III: General $\alpha$}
\label{gen-alpha}
We now look at the general $\alpha~(>0)$ case. For this case making concrete analytical progress from Eq.~\eqref{main_eqG} for any $\Delta $ is difficult. However, it is possible to obtain some results for the occupation probability distribution in asymptotically large times.  To proceed, in this case,  it seems convenient to start from the original master equations in \eqref{fokker} which, by defining $P(x,t)=P_+(x,t)+P_-(x,t)$ and $Q(x,t)=P_+(x,t)-P_-(x,t)$, can be rewritten, in terms of   $R_\pm(x)=\frac{R_1(x) \pm R_2(x)}{2}$, as 
\begin{align}
%\begin{split}
\partial_t P(x,t) &= -v~\partial_xQ(x,t),  \label{FP-P}\\
\partial_t Q(x,t) &= -R_+(x)Q(x,t)-R_-(x)P(x,t)-v~\partial_xP(x,t).
%\end{split}
\label{FP-Q}
\end{align}

\subsubsection{$\Delta >0:$}
We first present the $\Delta >0$ case for reasons that will be self-evident later.  In this case the particle reaches a stationary state, to obtain which we equate the time derivative on the left hand side of Eqs.~\eqref{FP-P} and \eqref{FP-Q} to zero and then solve for the $x$ dependence. We get the following expression for the stationary state distribution
\begin{align}
 P^{st}_{\alpha}(x) &=\frac{1}{2~ \Gamma\left(1+\frac{1}{\alpha+1}\right)}\left[\frac{(\gamma_1-\gamma_2)}{v~l^{\alpha}~(\alpha+1)}\right]^{\frac{1}{\alpha+1}}e^{-\frac{(\gamma_1-\gamma_2)}{v(\alpha+1)l^\alpha} \mid x \mid^{\alpha+1}}.
% ~\magentaw{l~dependence~checked}
\label{stdy_sol}
\end{align}
Note that for $\alpha =0$ and $\alpha =1$, this expression correctly reduces to the exponential and Gaussian distributions given in Eqs.~\eqref{Pss-a-0} and \eqref{Pss-a-1} respectively.
In Fig.~\ref{gen-alp-delgt0}a we numerically verify the above form of the steady state distribution $P^{st}_\alpha(x)$ for three choices of $\alpha$ different from $\alpha=0$ and $1$. 
%\greenw{At this point we would like to mention that the above stationary state distribution is valid for a wider range of $\alpha$ { i.e} for $\alpha >-1$. In \redw{Fig.~\ref{gen-alp-delgt0}a} we show a numerical verification of this fact in the case of $\alpha=-0.3$.} 
Approach to this steady state can in principle be understood by looking at the time dependent solutions of Eqs.~\eqref{FP-P} and \eqref{FP-Q} at large times, however finding such solutions is a difficult task  for which one has to solve the eigenvalue equation ~\eqref{main-eq-G}. Note that this eigenvalue equation looks similar to Schroedinger equation but it is actually different from it. In the $\alpha=0$ and $\alpha=1$ case we have seen that for $\Delta >0$ the approach to steady state is exponential. For general $\alpha \geq 0 $ also we expect exponential relaxation with a time scale determined from the structure of effective potential.  In Fig.~\ref{gen-alp-delgt0}b, we indeed observe that the relaxation is exponential where we plot the convergence of the $\text{var}(t)=\langle x(t)^2 \rangle$ to its value at $t \to \infty$.

\begin{figure}[t]
\includegraphics[scale=0.235]{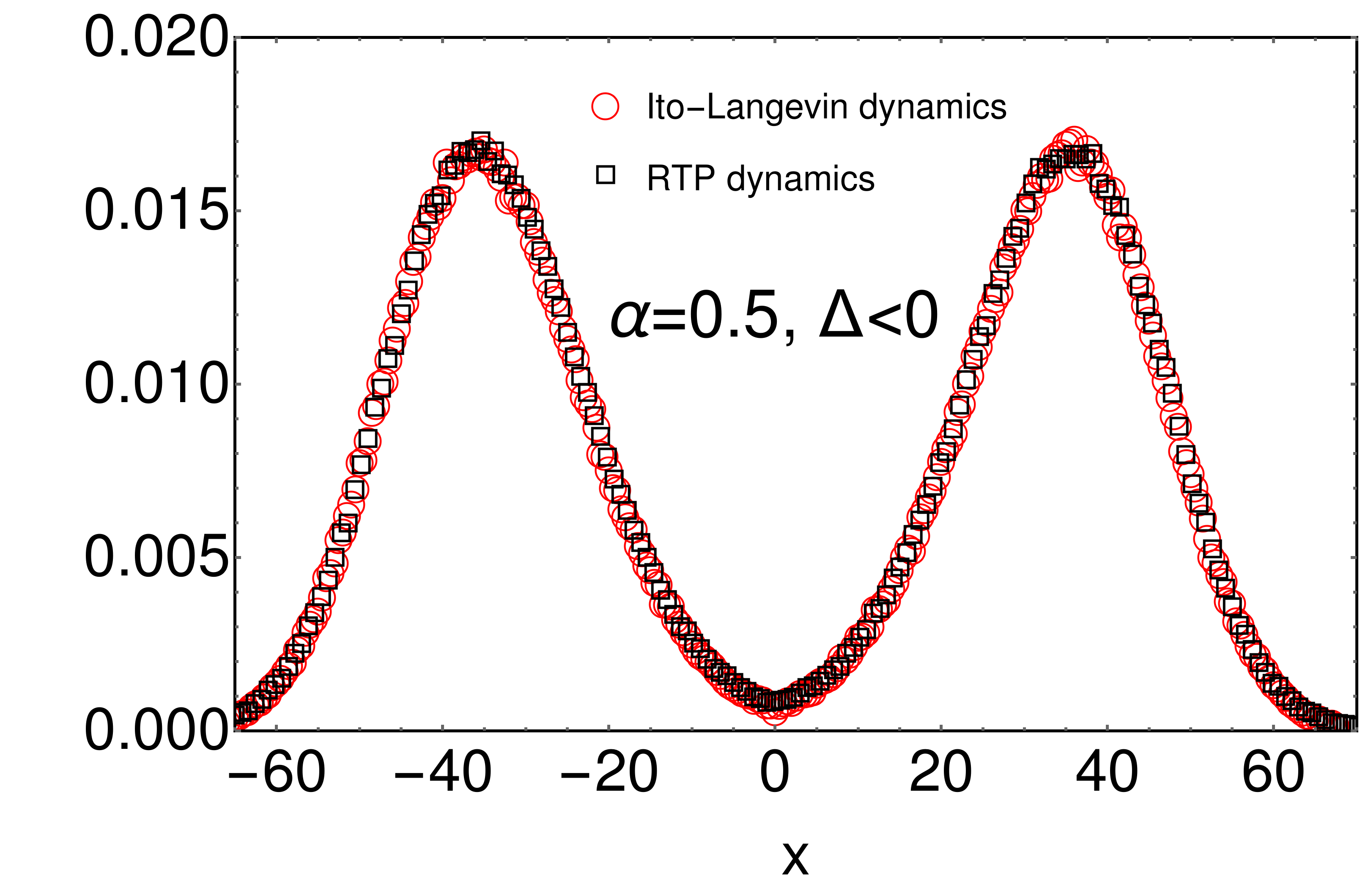}
\includegraphics[scale=0.26]{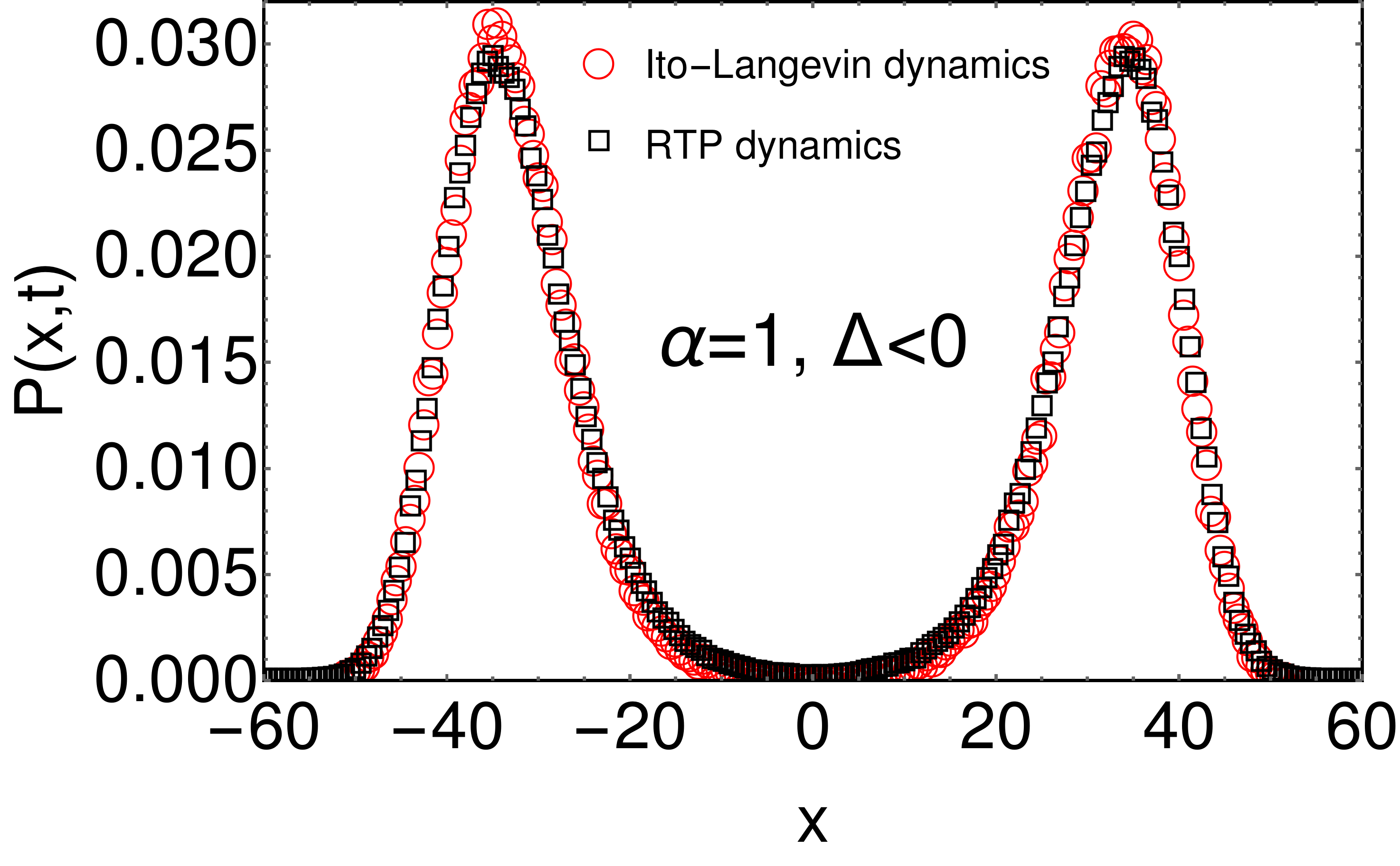}
\centering
\caption{
%(a) Scaling function $f_{\alpha}(z)$ in Eq.~\eqref{main_eq21} is plotted and compared with the results of simulation of Langevin equation \eqref{langevin} for three different times $t=50,60,70$. We have plotted for $\alpha =0.75$ and $\alpha=0.5$ (inset) with $\gamma_1=\gamma_2=1.5$. (b)
Comparison of the probability distribution obtained from simulation of the effective Langevin equation \eqref{eff-lang} (red circles) with the same from the original RTP dynamics \eqref{langevin} (black squares) for  $\alpha=0.5$ and $\alpha=1.0$. For both values of $\alpha$ the histograms are obtained at $t=1000$ with $\gamma_1=1.5$ and $\gamma_2=1.6$. Other common parameters are $v=1$ and $l=1$.}
\label{RTP-lang-comparison}
\end{figure}

\subsubsection{$\Delta \leq 0:$}
As seen in the previous two exactly solvable cases $\alpha=0$ and $\alpha=1$, for general $\alpha \geq 0$ also we expect that  the distribution $P(x,t)$ for general $\alpha$ also does not reach a stationary state for $\Delta \leq 0$. To proceed, we first note in Eqs.~\eqref{FP-P} and \eqref{FP-Q} that the equation for $P(x,t)$ is in the form of a continuity equation. On the other hand the equation for $Q(x,t)$ is not in this form, but has decay terms proportional to the rates $R_\pm(x)$ which are non-negative functions. As a result, at large times the difference distribution $Q(x,t)$ would not depend on time explicitly. Only time dependence would come from $P(x,t)$. Neglecting $\partial_tQ(x,t)$ for large $t$, we get $R_+(x)Q(x,t)=-R_-(x)P(x,t)-v\partial_xP(x,t)$, inserting which in Eq.~\eqref{FP-P} we get 
\begin{align}
\partial _t P(x,t)=\frac{v^2l^\alpha}{2 \gamma} \partial_x\left(\mid x \mid^{-\alpha} \partial_x P(x,t)\right)+\text{sgn}(x)\frac{ v \Delta}{\gamma} \partial_x P(x,t).
\label{eff-diff}
\end{align}
This equation can also be derived from Eq.~\eqref{eq-bar-P}. Performing inverse Laplace transform over the $s$ variable on both sides of this equation, we get 
\begin{align}
\partial_tP(x,t)=\partial_x \int_0^t dt' e^{-2\gamma\frac{|x|^\alpha}{l^\alpha}(t-t')}\left[ v^2 \partial_xP(x,t') 
+ \frac{2 v \Delta ~\text{sgn}(x)|x|^\alpha}{l^\alpha}P(x,t) \right].
\end{align}
For large $|x|$, the exponential term in the above equation can be approximated by $\sim \frac{l^\alpha}{2 \gamma |x|^\alpha}~\delta(t-t')$ and as a result the above equation reduces to Eq.~\eqref{eff-diff}. Note that this approximation does not work for $\alpha = 0$. It works only for $\alpha >0$. 
The equation \eqref{eff-diff} can be interpreted as the Fokker-Planck equation of a particle diffusing in an inhomogeneous environment of diffusion constant $\mathcal{D}(x)=\frac{v^2 l^{\alpha}}{\gamma \mid x \mid ^{\alpha}}$ and drift $\mathcal{V}(x)=-\text{sgn}(x) \left[ \frac{v\Delta}{\gamma}+\frac{\alpha v^2 l^{\alpha}}{2 \gamma \mid x \mid ^{\alpha+1}} \right]$. The corresponding Ito-Langevin equation is given by
\begin{align}
\frac{dx}{dt}=\mathcal{V}(x)+ \sqrt{\mathcal{D}(x)}~\eta (t),
\label{eff-lang}
\end{align}
where $\eta (t)$ is the Gaussian white noise with $\langle \eta(t) \rangle=0$ and $\langle \eta(t) \eta(t') \rangle =\delta (t-t')$. Comparison between the two dynamics is shown in Fig.~\ref{RTP-lang-comparison} where we have plotted $P(x,t)$ {\it vs.} $x$ obtained from the simulation of the effective Langevin equation \eqref{eff-lang} and the same from the original RTP dynamics \eqref{langevin} at $t=1000$ for two values of $\alpha=0.5$ (Fig.~\ref{RTP-lang-comparison}b) and $\alpha=1.0$ (Fig.~\ref{RTP-lang-comparison}a).

\begin{figure}[t]
\includegraphics[scale=0.24]{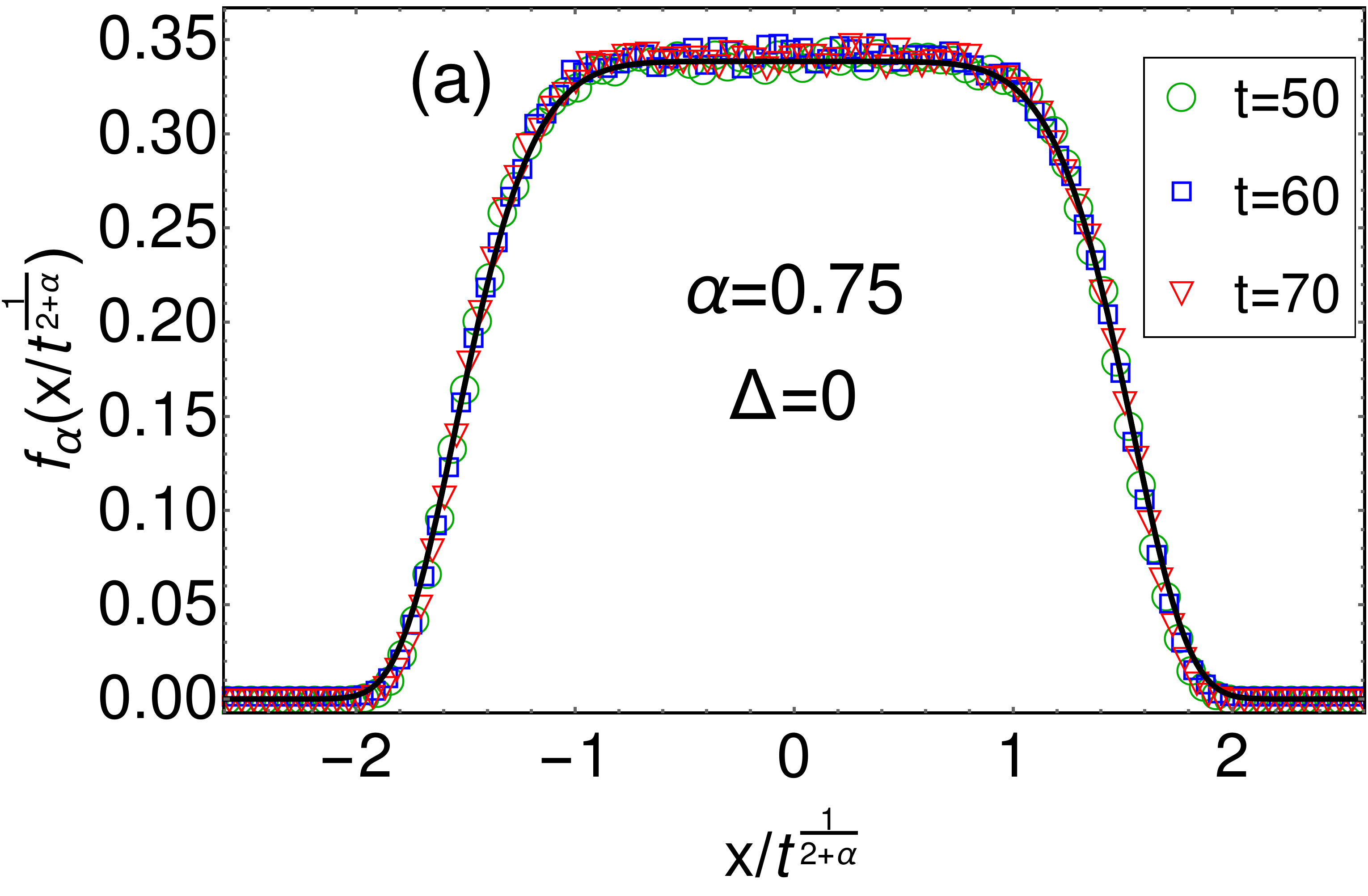}
\includegraphics[scale=0.26]{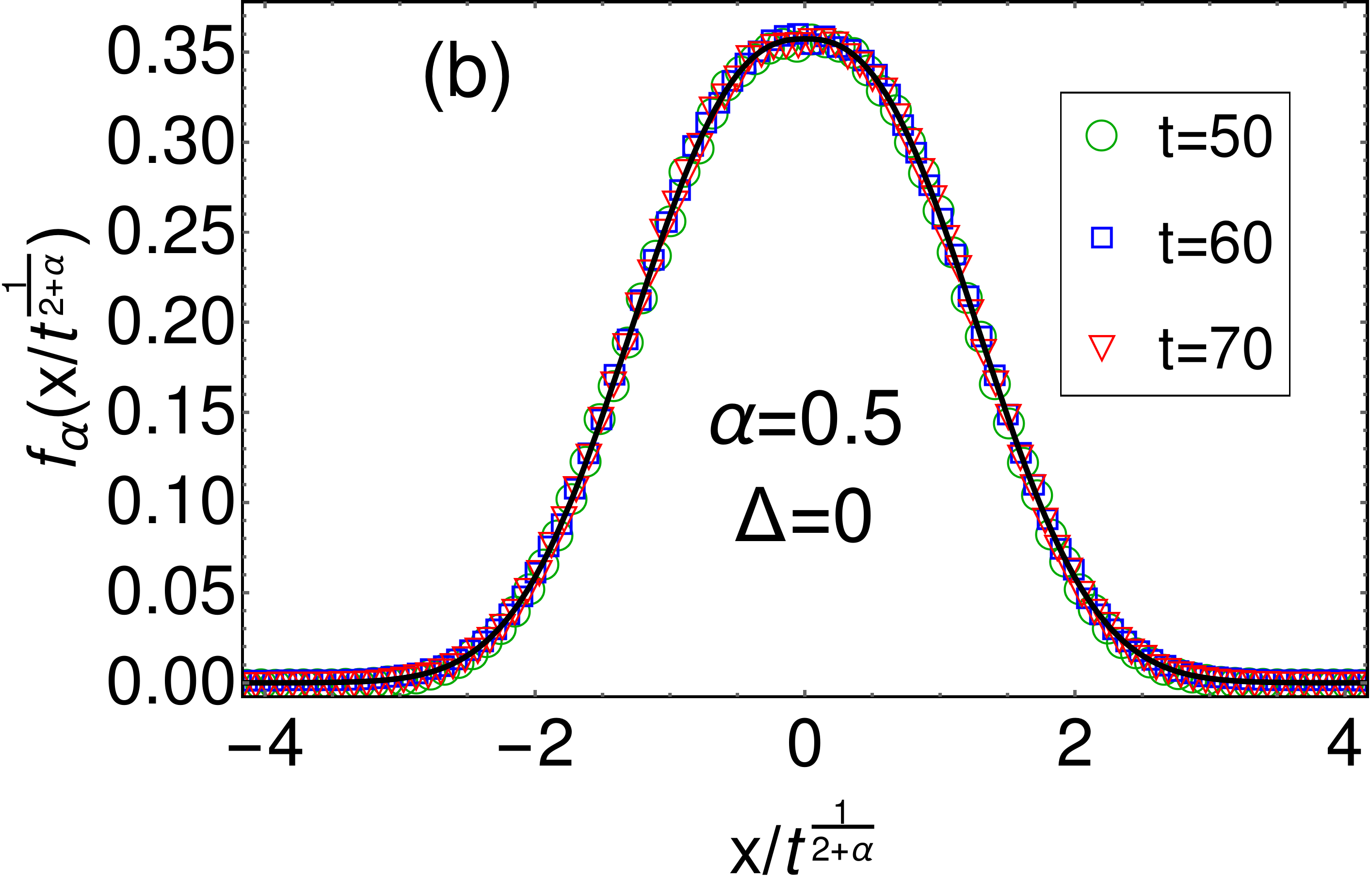}
\centering
    \caption{(a) Scaling function $f_{\alpha}(z)$ in Eq.~\eqref{main_eq21} is plotted ( solid black line) and compared with the results (symbols) obtained by simulating the RTP equation \eqref{langevin} at three different times $t=50,60,70$ for $\alpha =0.75$ (a) and $\alpha=0.5$ (b) with $\gamma_1=\gamma_2=1.5$.  Other common parameters are $v=1$ and $l=1$
}
\label{general_alph}
\end{figure}

~\\
\noindent
$\Delta=0$: For this the second term on the right hand side of Eq.~\eqref{eff-diff} is absent. It is easy to check that the solution of this equation, for large $t$, is given by \cite{Hentschel_1984}
\begin{align}
P(x,t) \simeq \frac{1}{t^{\frac{1}{2+\alpha}}} f_{\alpha}\left(\frac{\mid x \mid}{t^{\frac{1}{2+\alpha}}}\right),~~\text{with } f_{\alpha}(z)= \frac{(2+\alpha)^{\frac{\alpha}{2+\alpha}}}{2 \Gamma \left(\frac{1}{2+\alpha}\right) D_{\alpha}^{\frac{1}{2+\alpha}}}e^{-\frac{z^{2+\alpha}}{(2+\alpha)^2 D_{\alpha}}},
\label{main_eq21}
\end{align}
where $D_{\alpha}=\frac{v^2~l^{\alpha}}{2 \gamma}$.  Note that for $\alpha=1$ the scaling function $f_\alpha(z)$ correctly reduces to the scaling function in Eq.~\eqref{sc-sol-a-1-D-0}. The function $f_{\alpha}(z)$ is plotted in Fig.~\ref{general_alph} for two different values of $\alpha$ and compared with the numerical simulation for three different times. We observe excellent agreement between the theory and the simulation results.

%\cyanw{Note that this equation can be interpreted as the time independent Schroedinger equation (Is it true ???) of a quantum particle in ground state inside a potential 
%\begin{align}
%V(x)=\left[\frac{\Delta \alpha \mid x \mid ^{\alpha-1}}{v~ l^{\alpha}}+\frac{2 \gamma s\mid x \mid ^{\alpha}}{v^2~ l^{\alpha}}+\frac{\Delta ^2 \mid x \mid ^{2\alpha}}{v^2~ l^{2\alpha}} +\frac{s^2}{v^2}\right],~~\Delta>0, 
%\end{align}
%which is difficult to solve for arbitrary $\alpha$.} 
%However, it is possible to anticipate that the relaxation to the steady state for general $\alpha$ is also exponential with a time scale determined from the structure of $V(x)$ near its minima.
%\greenw{\it May be for $\alpha=0.5$ we can numerically find the relaxation. May be one can club the steady state plot and the relaxation together. \magentaw{done!}}\\

\begin{figure}[]
\includegraphics[scale=0.24]{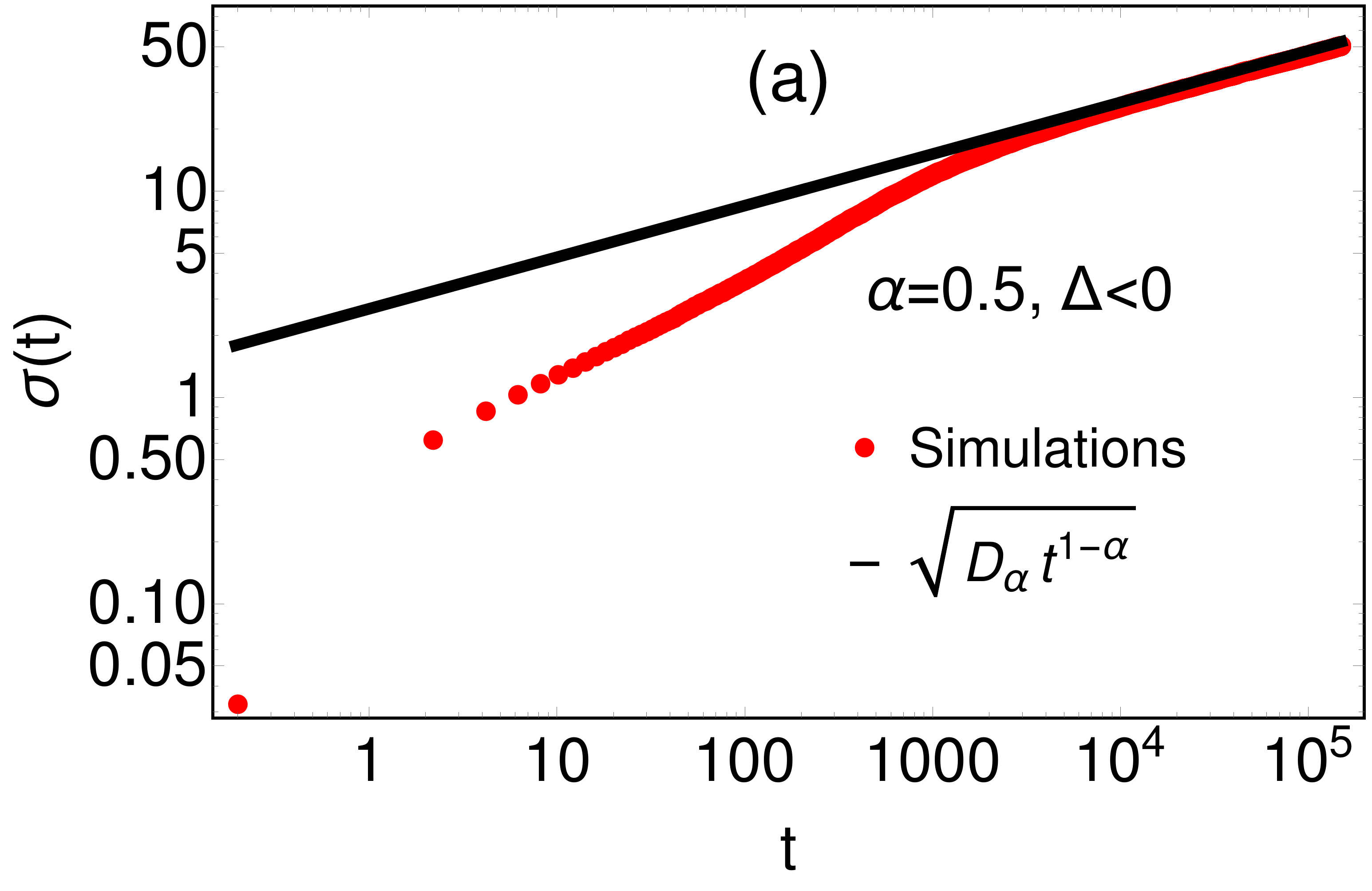}
\includegraphics[scale=0.26]{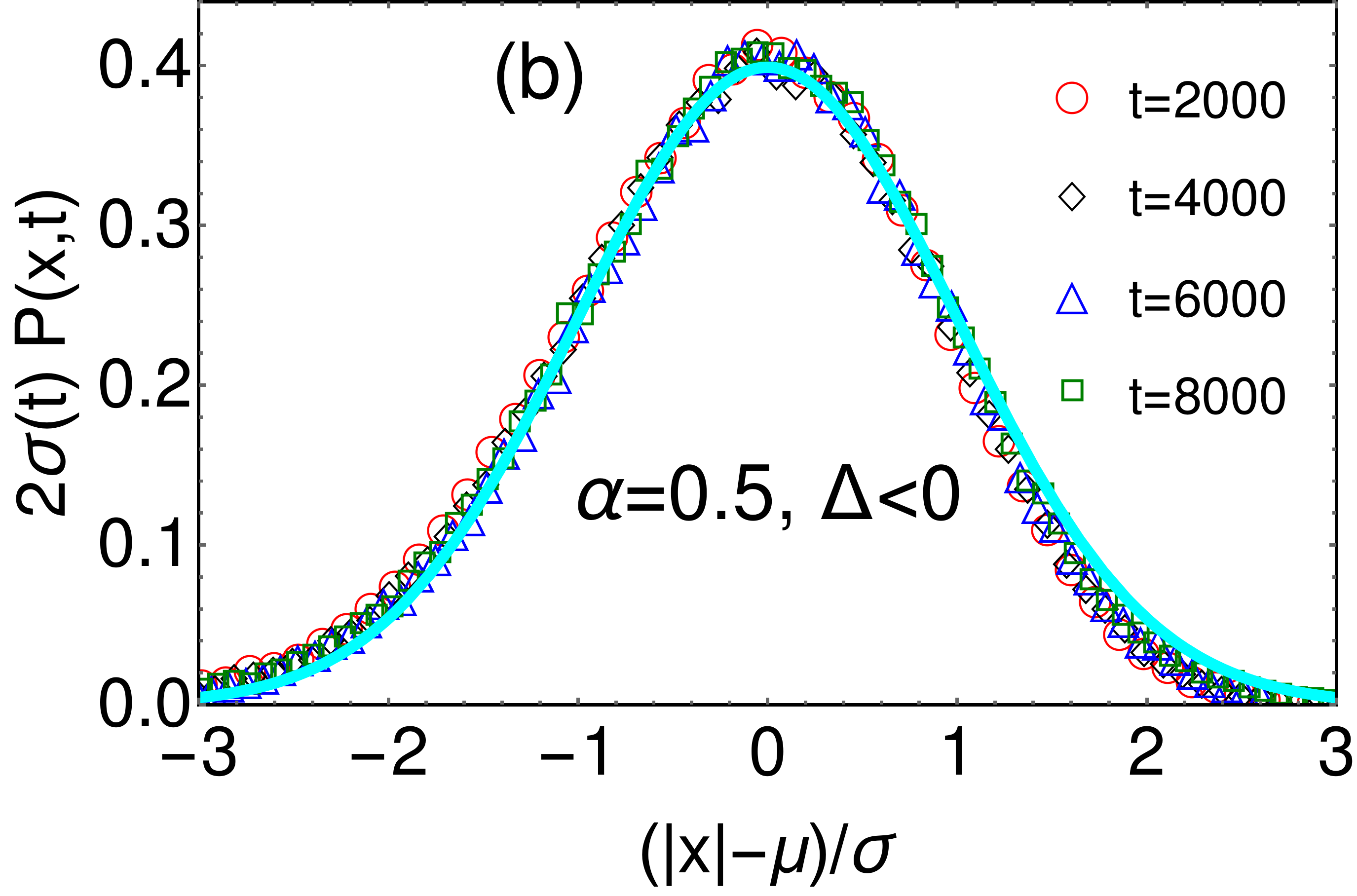}
\centering
\caption{(a) Numerical verification of $\sigma^2(t) \sim  \mathcal{D}_\alpha t^{1-\alpha}$ with $\mathcal{D}_\alpha=\frac{v^2\ell^\alpha}{\gamma(1-\alpha)}\left(\frac{\gamma}{v|\Delta|}\right)^\alpha$ for large $t$. (b) Numerical verification of the scaling behaviour in Eq.~\eqref{scaling-P_alpha} with $\mathcal{G}_\alpha(u)$ given by 
$\mathcal{G}_\alpha(u)=e^{-u^2/2}/\sqrt{2 \pi}$ (shown by solid line). For both plots, we have taken $\alpha=0.5, \gamma_1=1.5,~\gamma_2=1.6~l=1~\text{and } v=1.$} 
\label{gen-alp-dellt0}
\end{figure}

~\\
\noindent
$\Delta<0$: For this case the drift $\mathcal{V}(x)$ on the particle, as can be seen from Eq.~\eqref{eff-diff}, is away from the origin on both sides at large $x$. As a result the particle never reach a steady state as also has been argued earlier. We already have observed numerically in Fig.~\ref{RTP-lang-comparison} that at large times the dynamics of the RTP can be well described by the Langevin equation \eqref{eff-lang}. 
From this comparison, we also observe that the distribution $P(x,t)$ is symmetric with respect to the origin which implies that the mean position of the particle is zero, however the average value of the absolute position $\mu(t)=\langle |x| \rangle$ is not zero. The two symmetric peaks one the opposite sides of the origin are situated at $x=\pm \mu(t)$. For large $t$, this quantity increases linearly as $\langle |x| \rangle=\mu(t) \sim \frac{v|\Delta|}{\gamma}t$, which can be easily verified numerically. 
In particular,  it can be shown that, using the following variable transforms $|x|=\mu(t) +z$ and $\tau=t^{1-\alpha}$ the Eq.~\eqref{eff-diff} in the large $t$ limit becomes the following diffusion equation 
\begin{align}
\partial_\tau P(z,\tau)=\mathcal{D}_\alpha \partial_z^2 P(z,\tau), \label{P(z,tau)}
\end{align}
where $\mathcal{D}_\alpha=\frac{v^2\ell^\alpha}{\gamma(1-\alpha)}\left(\frac{\gamma}{v|\Delta|}\right)^\alpha$ for $0<\alpha <1$. This immediately implies that for large $t$, the distribution $P(x,t)$ has the following scaling form
\begin{align}
P(x,t) \simeq \frac{1}{2 \sigma_\alpha(t)}~\mathcal{G} \left(\frac{|x|-\mu(t)}{\sigma_\alpha(t)}\right)
\label{scaling-P_alpha}
\end{align}
where $\mathcal{G}(u)$ satisfies the differential equation $\partial_u^2 \mathcal{G}(u)+u \partial_u\mathcal{G}(u)+\mathcal{G}(u)=0$ and the variance $\sigma_\alpha^2(t) = {\langle x^2 \rangle -\langle |x| \rangle^2}$ is given by $ \sigma_\alpha^2(t)  \sim  \mathcal{D}_\alpha t^{1-\alpha}$. The solution of this equations is very simple and given by the zero mean and unit variance Gaussian, 
$\mathcal{G}(u)=e^{-u^2/2}/\sqrt{2 \pi}$. 
The same procedure can also be followed for $\alpha=1$ and one gets same scaling behaviour (as in Eq.~\eqref{scaling-P_alpha}) with same scaling function $\mathcal{G}(u)$ but not $\sigma_1^2(t) \sim (vl/|\Delta |)~\ln(t)$ for $\alpha=1$. The time dependence of the variance $\sigma^2_\alpha(t)$ is verified numerically in Fig.~\ref{gen-alp-dellt0}a for $\alpha < 1$. In the numerical simulation of the equation of motion \eqref{langevin} with the rates given in Eqs.~\eqref{rates} we have
chosen $\alpha=0.5$. It turns out that this value is optimal for the numerical verification. For given $\gamma$ and $\Delta$, the description given by the FP equation \eqref{eff-diff} starts becoming valid at (large) times  which increases with decreasing $\alpha$. Performing numerical simulation over such huge time duration turns out to be highly expensive. On the other hand, for larger $\alpha$, even though the effective inhomogeneous diffusion equation \eqref{eff-diff} starts becoming valid at time earlier than smaller alpha, but the rates (being $\propto |x|^\alpha$) also increases faster with time because for $\Delta <0$ the particles is effective drifted away from the origin. As a result, one requires a very small $dt$ in the numerical simulation in order to get good convergence, which in turn again makes the computation expensive. 

The scaling behavior in Eq.~\eqref{scaling-P_alpha} is demonstrated and verified in Fig.~\ref{gen-alp-dellt0}b numerically again for $\alpha=0.5$.  Note that this scaling behaviour is valid for $0< \alpha <1$. For $\alpha >1$, we numerically observe that the variance decreases with time. As a result at very large time we expect the distribution $P(x,t)$ to shrink to a sum of two delta functions at $x= \pm \mu(t)$ which one would naively guess when a similar procedure as done for $0<\alpha \leq 1$ is attempted  for $\alpha >1$ case.

%The effective Ito-Langevin equation does not only provide the large $t$ behaviour of the variance, it also provides the scaling bejaviour of the full solution of the FP equation \eqref{eff-diff} at large $t$. We find that, this solution at large $t$ has the following scaling form
%\begin{align}
%P(x,t) \simeq \frac{1}{2 \sigma_\alpha(t)}~\mathcal{G} \left(\frac{|x|-\mu(t)}{\sigma_\alpha(t)}\right)
%\label{scaling-P_alpha}
%\end{align}
%%\begin{align}
%%P(x,t) \simeq \frac{1}{t^{(1-\alpha)/2}}~\mathcal{G}_\alpha \left(\frac{|x|-v |\Delta| t/\gamma}{t^{(1-\alpha)/2}}\right)
%%\end{align}
%where $\mathcal{G}(u)$ satisfies the differential equation $\partial_u^2 \mathcal{G}(u)+u \partial_u\mathcal{G}(u)+\mathcal{G}(u)=0$. The solution of this equations is very simple and given by 
%$\mathcal{G}(u)=e^{-u^2/2}/\sqrt{2 \pi}$. % that the distribution $P(x,t)$ at moderately large time $t$, possesses the following scaling property
%%\begin{align}
%%P(x,t)\simeq \frac{1}{\sigma_t}~g\left ( \frac{|x|-\langle|x|\rangle}{\sigma_t}\right).
%%\end{align}

%\greenw{We have to add some comments on the simulation difficulties.}
%At  $t \to \infty$, we expect the $\sigma_t$ to go to zero. Consequently, the distribution becomes a sum of two delta functions situated at $x=\pm \mu(t)$.

%In Figure \ref{general_alph}(b), we simulate the effective Langevin equation in Eq.~\eqref{main_eq20} for $\Delta <0$ for $t=1000$ and compare it with the simulation of actual Langevin equation \eqref{langevin}. We see brilliant match.

\section{Survival probability}
\label{surv_prob}
In this section we study the motion of the RTP with space dependent rates in Eq.~\eqref{rates} in presence of an absorbing barrier. In many physical settings, how long does a particle survive from a given absorbing site is of primary interest \cite{Redner}. 
In particular, we will consider the absorbing site to be at the orgin and address the question of survival probability for the particle starting from some position $x_0>0$.  
%Also for many stochastic processes survival probabilty varies as $\sim t^{-\theta}$ for large times where $\theta$ is called the persistence exponent . The exponent $\theta$ has also been measured in many experiments \cite{satya}. 
Let $S_{\pm}(x_0,t)$ denote the survival probability for the RTP with initial position $x_0$ and initial velocity $\pm v$ in presence of an absorbing wall at $x=0$. 
%One can, in principle, compute the propagators in presence of an absorbing barrier and integrate over the final position to get the survival probability. 
We start by deriving the backward master equations satisfied by $S_{\pm}(x_0,t)$ and then solve them explicitly. Below we briefly discuss the derivation of the backward master equations.

The probability that RTP with initial velocity direction $+$ survives from the absorbing wall at $x=0$ till time $t+dt$ is $S_+(x_0,t+dt)$.  One can break the total time duration $t+dt$ into two parts (i) $[0, dt]$ and (ii) $[dt, t+dt]$. In the first interval of duration $dt$, the RTP can do two things: (a) without flipping its direction move to position $x_0+v dt$ with probability $(1-R_1(x_0) dt)$
 and, (b) flip its direction of motion with probability $R_1(x_0)dt$. After time $dt$, the RTP survives the remaining interval with probability $S_+(x_0+vdt,t)$, if event (a) occurs and with probability $S_-(x_0,t)$ if event (b) occurs. Adding all these probabilities with appropriate weights one gets
% (i)does not flip its direction with probability $(1-R_1(x_0) dt)$ and reaches $x_0+v dt$, starting from where the particle survives in the remaining interval $\left( S_+(x_0+v dt, t)\right)$ (ii)flips its direction with probability $R_1(x_0)dt$ and reaches $x_0-v dt$ and starting from where the particle survives in the remaining interval $\left( S_-(x_0-v dt, t)\right)$ . 
$S_+(x_0, t+dt)=(1-R_1(x_0) dt) S_+(x_0+ v dt, t)+ R_1(x_0) dt~ S_-(x_0,t)$. Similarly, if the initial velocity direction is negative, then one has $S_-(x_0, t+dt)=(1-R_2(x_0) dt) S_-(x_0- v dt, t)+ R_2(x_0) dt~ S_+(x_0+v dt,t)$. Performing Taylor series expansion in $dt$ and taking $dt \to 0$ limit, one gets the following backward master equations for $S_{\pm}(x_0, t)$
\begin{align}
\begin{split}
  \partial_t S_+(x_0,t) &= v \partial_{x_0} S_+(x_0,t) -R_1(x_0) S_+(x_0,t) + R_1(x_0) S_-(x_0,t), \\
  \partial_t S_-(x_0,t) &= -v \partial_{x_0} S_-(x_0,t) +R_2(x_0) S_+(x_0,t) - R_2(x_0) S_+(x_0,t).
 \end{split}
\label{surv_eq1}
\end{align}
To solve these equations, one needs to specify the initial condition as well as the boundary conditions. 
%The boundary conditions are $S_{\pm}(x_0 \to \infty,t)=1$ and $S_-(0,t)=0$. 
Note that if the particle starts from $x_0 \to \infty$ initially, then for all finite $t$ it survives regardless of its initial velocity direction. This gives rise $S_{\pm}(x_0 \to \infty,t)=1$. To understand the other boundary condition $S_-(0,t)=0$, the particle will be instantly absorbed if it starts at $x_0=0$ with $-$ velocity. However if the particle starts from $x_0=0$ with $+$ velocity, it will not get absorbed instantly and accordingly one gets $S_+(0,t)\neq0$. Hence for any $x_0 \neq0$ we have $S_{\pm}(x_0,0)=1$.
 
To solve Eqs.~\eqref{surv_eq1} we take Laplace transformation with respect to $t$ as $\bar{S}_{\pm}(x_0,s)=\int_{0}^{\infty}dt e^{-st} S_{\pm}(x_0,t)$ to get
\begin{align}
\begin{split}
&\left[-v \partial_{x_0} +R_1(x_0)+s \right] \bar{S}_+(x_0,s)=1+R_1(x_0) \bar{S}_-(x_0,s), \\
& \left[~v ~\partial_{x_0} +R_2(x_0)+s \right] \bar{S}_-(x_0,s)=1+R_2(x_0) \bar{S}_+(x_0,s),
\end{split}
\label{surv_lap}
\end{align} 
where we have used the the initial conditions $S_{\pm}(x_0,0)=1$, Eqs.~\eqref{surv_eq1}.  
Under this transformation the boundary conditions become $\bar{S}_{\pm}(x_0 \to \infty,s)=\frac{1}{s}$ and $\bar{S}_-(0,s)=0$. Note that the differential equations in Eqs.~\eqref{surv_lap} are inhomogeneous. To make them homogeneous we define $\bar{U}_\pm(x_0,s)$ such that 
\begin{align}
\bar{S}_{\pm}(x_0,s)=\frac{1}{s}+\bar{U}_{\pm}(x_0,s),
\label{surv_eq3}
\end{align}
which also simplifies the boundary conditions as $\bar{U}_{\pm}(x_0 \to \infty,s)=0$. The Eqs.~\eqref{surv_lap} now become
\begin{align}
&\left[-v \partial_{x_0} +R_1(x_0)+s \right] \bar{U}_+=R_1(x_0) \bar{U}_-, \nonumber \\
& \left[~v ~\partial_{x_0} +R_2(x_0)+s \right] \bar{U}_-=R_2(x_0) \bar{U}_+.
\label{surv_U_pm}
\end{align}
Further defining,
\begin{align}
&\bar{U}(x_0,s)=\bar{U}_+(x_0,s)+\bar{U}_-(x_0,s), \nonumber \\
&\bar{H}(x_0,s)=\bar{U}_+(x_0,s)-\bar{U}_-(x_0,s),
\label{surv_U-H-def}
\end{align}
we get
\begin{align}
&\partial_{x_0}^2\bar{H}- \frac{2 \Delta x_0^{\alpha}}{v l^{\alpha}} \partial_{x_0} \bar{H}-\left[\frac{2 \Delta \alpha x_0^{\alpha-1}}{v l^{\alpha}}+\frac{2 \gamma s x_0^{\alpha}}{v^2 l^{\alpha}}+\frac{s^2}{v^2} \right]\bar{H}=0, \label{surv_H-diff}\\
&~~~~~\text{and,}~~~~~~~~\bar{U}(x_0,s)=\frac{v}{s} \partial_{x_0}\bar{H}-\frac{2 \Delta x_0^{\alpha}}{s ~l^{\alpha}} \bar{H}. \label{surv_U-diff}
%\\&~~~\partial_{x_0}\bar{U}=\frac{2 \gamma |x_0|^\alpha}{v l^\alpha}\bar{H}+\frac{s}{v}\bar{H},
\end{align}
One can get rid of first order derivative in Eq.~\eqref{surv_H-diff} by making the transformation,
\begin{align}
\bar{H}(x_0,s)=e^{\frac{\Delta }{v(\alpha+1)l^{\alpha}}x_0^{\alpha+1}} F(x_0,s),
\label{surv_H}
\end{align}
using which in Eq.~\eqref{surv_H-diff}, one gets
\begin{align}
\partial_{x_0}^2 F(x_0,s)-\left[\frac{\Delta \alpha x_{0} ^{\alpha-1}}{v~ l^{\alpha}}+\frac{2 \gamma s x_{0}^{\alpha}}{v^2~ l^{\alpha}}+\frac{\Delta ^2 x_{0} ^{2\alpha}}{v^2~ l^{2\alpha}} +\frac{s^2}{v^2}\right] F(x_0,s)=0. \label{surv_F}
\end{align}
Note that this equation is identical to Eq.~\eqref{main-eq-G} for $G(x,s)$ obtained in the previous section except for the boundary conditions which are different for the two cases. In what follows, we will solve this equation for $\alpha=0$ and $\alpha=1$ separately and then address the case of general $\alpha$.

\subsection{Case I: $\alpha =0$}
\label{surv-alpha=0}
\begin{figure}[t]
\includegraphics[scale=0.25]{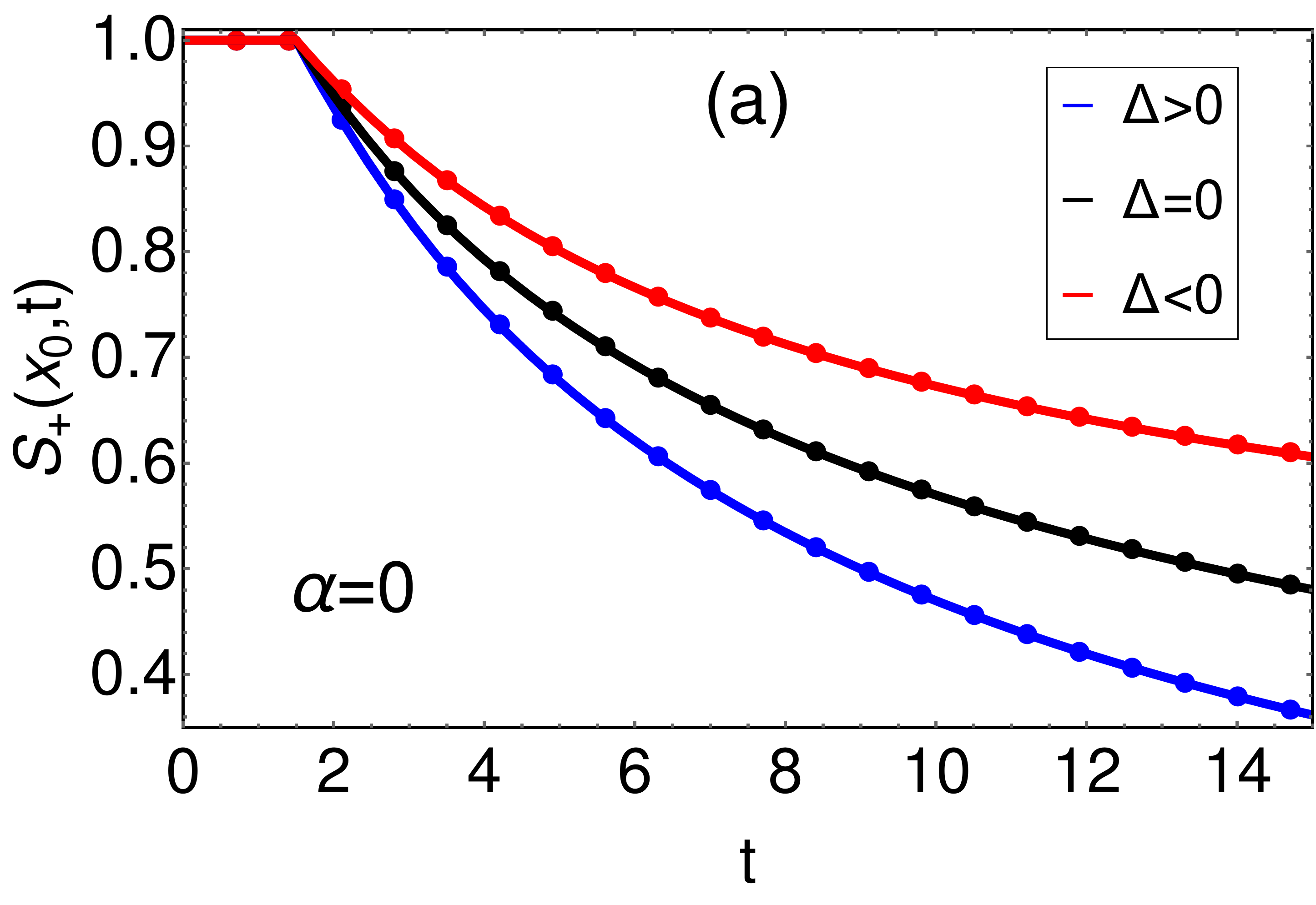}
\includegraphics[scale=0.24]{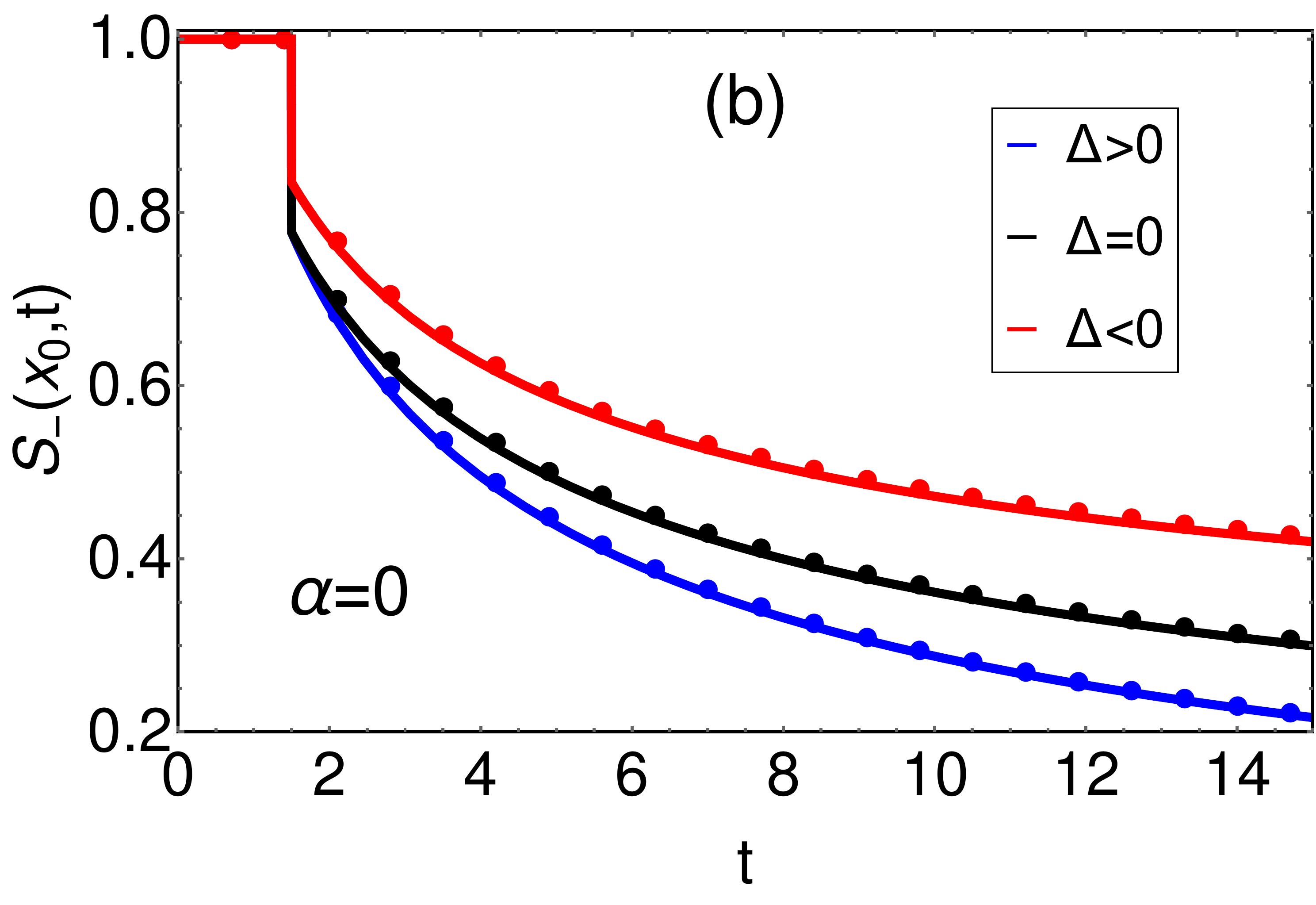}
\centering
\caption{Comparision of $S_{\pm}(x_0,t)$ given by Eqs. \eqref{surv_a-1} for $\alpha =0$ with the simulation results (filled circles) for various signatures of $\Delta$. In both (a) and (b), we have chosen (i)$\gamma_1=1.2 ~\gamma_2=1$ for blue, (ii)$\gamma_1= \gamma_2=1$ for black and (iii) $\gamma_1=1~ \gamma_2=1.2$ for red. Other parameters chosen are $x_0=1.5$,  $v=1$ and $l=1$.}
\label{new_surv_alph_0}
\end{figure}

We start with the simplest case of $\alpha =0$. For this case the rates $R_{1,2}(x)$ are actually $x$ independent.
%For $\alpha =0$, we have solved Eq.~\eqref{surv_H} in \ref{new_stdy_appen} (see $G(x,s)$ in \ref{new_stdy_appen}). 
Recall that $x_0$ is the initial position of the RTP which is greater than $0$. Hence noting that $F(x_0,s)$ is finite as $x_0 \to \infty$, one gets $F(x_0,s)\sim e^{-\lambda (s) x_0}$ where $\lambda(s)=\frac{1}{v} \sqrt{2 \gamma s+s^2+\Delta ^2}$ (see \ref{new_stdy_appen} for details). Inserting this in Eqs. \eqref{surv_U-diff} and \eqref{surv_H} and finally writing for $\bar{S}_{\pm}(x_0,s)$, the expressions read as,
\begin{align}
\bar{S}_{\pm}(x_0,s)=\frac{1}{s}+\frac{\mathcal{A}}{2s} e^{\left(\frac{\Delta}{v}- \lambda(s)\right)x_0}\left[ -\Delta-v \lambda(s)\pm s\right],
\label{surv_eq10}
\end{align}
where $\mathcal{A}$ is a constant independent of $x_0$. To evaluate $\mathcal{A}$, we use the boundary condtion $\bar{S}_-(0,s)=0$ which gives $\mathcal{A}(s)=\frac{2}{\Delta +s+v \lambda(s)}$. Finally inserting this in Eq. \eqref{surv_eq10} we get the following  the expressions for $\bar{S}_{\pm}(x_0,s)$ 
\begin{align}
&\bar{S}_-(x_0,s)=\frac{1}{s}-\frac{1}{s}e^{\left(\frac{\Delta}{v}-\lambda(s)\right)x_0}, \label{surv_eq11}\\
&\bar{S}_+(x_0,s)=\frac{1}{s}-\frac{s+\gamma-v\lambda(s)}{s \gamma_2}e^{\left(\frac{\Delta}{v}-\lambda(s)\right)x_0}. \label{surv_eq12}
\end{align}
Using the following results for inverse Laplace transformations,
\begin{align}
&L_{s\to t} \left[e^{- \lambda(s) x_0}\right]=-v e^{\frac{\Delta x_0}{v}} \frac{d}{dx_0} \left[ e^{-\gamma t}   I_0\left(\sqrt{\gamma_1 \gamma_2 \left(t^2-\frac{x_0^2}{v^2}\right)} \right)\Theta(vt-x_0)\right], \nonumber\\
&L_{s\to t} \left[(s+\gamma-v \lambda(s)) e^{- \lambda(s) x_0}\right]=\sqrt{\gamma_1 \gamma_2} \frac{e^{-\gamma t+\frac{\Delta x_0}{v}}}{t+\frac{x_0}{v}} \left[\frac{x_0 \sqrt{\gamma_1 \gamma_2}}{v} I_0\left(\sqrt{\gamma_1 \gamma_2 \left(t^2-\frac{x_0^2}{v^2}\right)} \right) \right. \nonumber\\
&~~~~~~~~~~~~~~~~~~~+\left. \sqrt{\frac{vt-x_0}{v t+x_0}}I_1\left(\sqrt{\gamma_1 \gamma_2 \left(t^2-\frac{x_0^2}{v^2}\right)} \right)\right] \Theta(vt-x_0)
\label{surv_eq13}
\end{align}
we get
\begin{align}
&S_-(x_0,t)=1+v e^{\frac{\Delta x_0}{v}} \frac{d}{dx_0} \int_0^{t} d \tau e^{-\gamma \tau}   I_0\left(\sqrt{\gamma_1 \gamma_2 \left(\tau^2-\frac{x_0^2}{v^2}\right)} \right) \Theta(v \tau-x_0), \nonumber \\
& S_+(x_0,t)=1-\sqrt{\frac{\gamma_1}{\gamma_2}} e^{\frac{\Delta x_0}{v}}\int_0^{t} d\tau\frac{e^{-\gamma \tau}}{\tau+\frac{x_0}{v}} \Theta(v\tau-x_0)\left[\frac{x_0 \sqrt{\gamma_1 \gamma_2}}{v} I_0\left(\sqrt{\gamma_1 \gamma_2 \left(\tau^2-\frac{x_0^2}{v^2}\right)} \right) \right. \nonumber\\
&~~~~~~~~~~~~~~~~~~~~~~~+\left. \sqrt{\frac{v\tau-x_0}{v \tau+x_0}}I_1\left(\sqrt{\gamma_1 \gamma_2 \left(\tau^2-\frac{x_0^2}{v^2}\right)} \right)\right] 
\label{surv_a-1}
\end{align}
For $\gamma_1=\gamma_2$ the above expressions match with the previously obtained results in \cite{Kanaya_2018, Singh2019}. In figure \ref{new_surv_alph_0} we have plotted our above theoretical results for $S_\pm(x_0,t)$  and compared with the numerical simulations. We observe excellent agreement between them. Notice that for both $S_{\pm}(x_0,t)$ remain $1$ till time $t_b=\frac{x_0}{v}$. This is because the RTP initially starting from $x_0$ will take at least time $t_b$ to reach the absorbing wall at $x=0$. Before $t_b$, the RTPs do not feel presence of the barrier. Once they reach the wall with velocity $-v$ at time $t_b^-$, a fraction of them will change the velocity from $-$ to $+$ and survive the wall, while others will get absorbed at time $t_b+$. This results in sudden drop in the population of RTPs as indicated by the sudden drop in $S_-(x_0,t)$.  However no sudden drop occurs in $S_+(x_0,t)$ because the particles do not reach the wall with $+$ velocity.

It is worth noting that the particle is drifted away from the origin for $\Delta<0$ which gives rise to non-zero $S_{\pm}(x_0,t)$ as $t \to \infty$. This can be, in principle, verified by putting $t \to \infty$ in the expressions of $S_{\pm}(x_0, t)$ given in Eqs.~\eqref{surv_a-1}. However it turns out more convenient to obtain this from the Laplace transforms given in Eqs. \eqref{surv_eq11} and \eqref{surv_eq12} by putting $s=0$ which corresponds to $t \to \infty$ limit. Hence for  $\Delta <0$ we get,  $\mathcal{S}_{\pm}(x_0)=S_{\pm}(x_0, t\to \infty)=\lim _{s \to0} \left[s\bar{S}_{\pm}(x_0,s)\right]$ given by 
\begin{align}
\begin{split}
\mathcal{S}_{+}(x_0)=&1-\frac{\gamma-\mid \Delta \mid}{\gamma+\mid \Delta \mid}e^{-\frac{2 \mid \Delta \mid}{v} x_0}, \\
\mathcal{S}_{-}(x_0)=&1-e^{-\frac{2 \mid \Delta \mid}{v} x_0}.
\end{split}
\label{surv_eq101}
\end{align}
One can similarly compute $\mathcal{S}_{\pm}(x_0)$ for $\Delta \geq 0$ which turns out to be $0$ as the particle will definitely reach origin after a  sufficient time interval. 

\begin{figure}[t]
\includegraphics[scale=0.235]{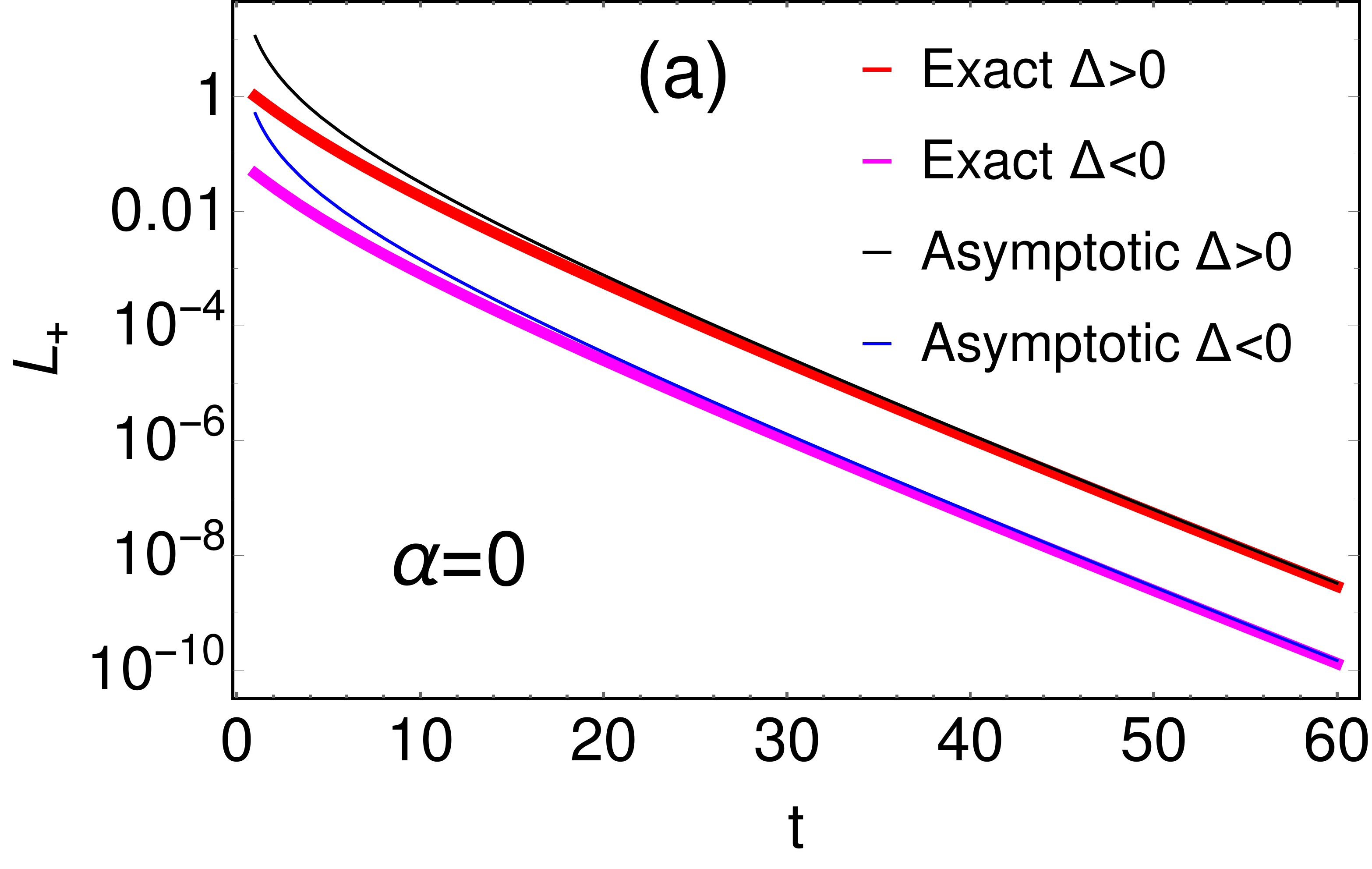}
\includegraphics[scale=0.23]{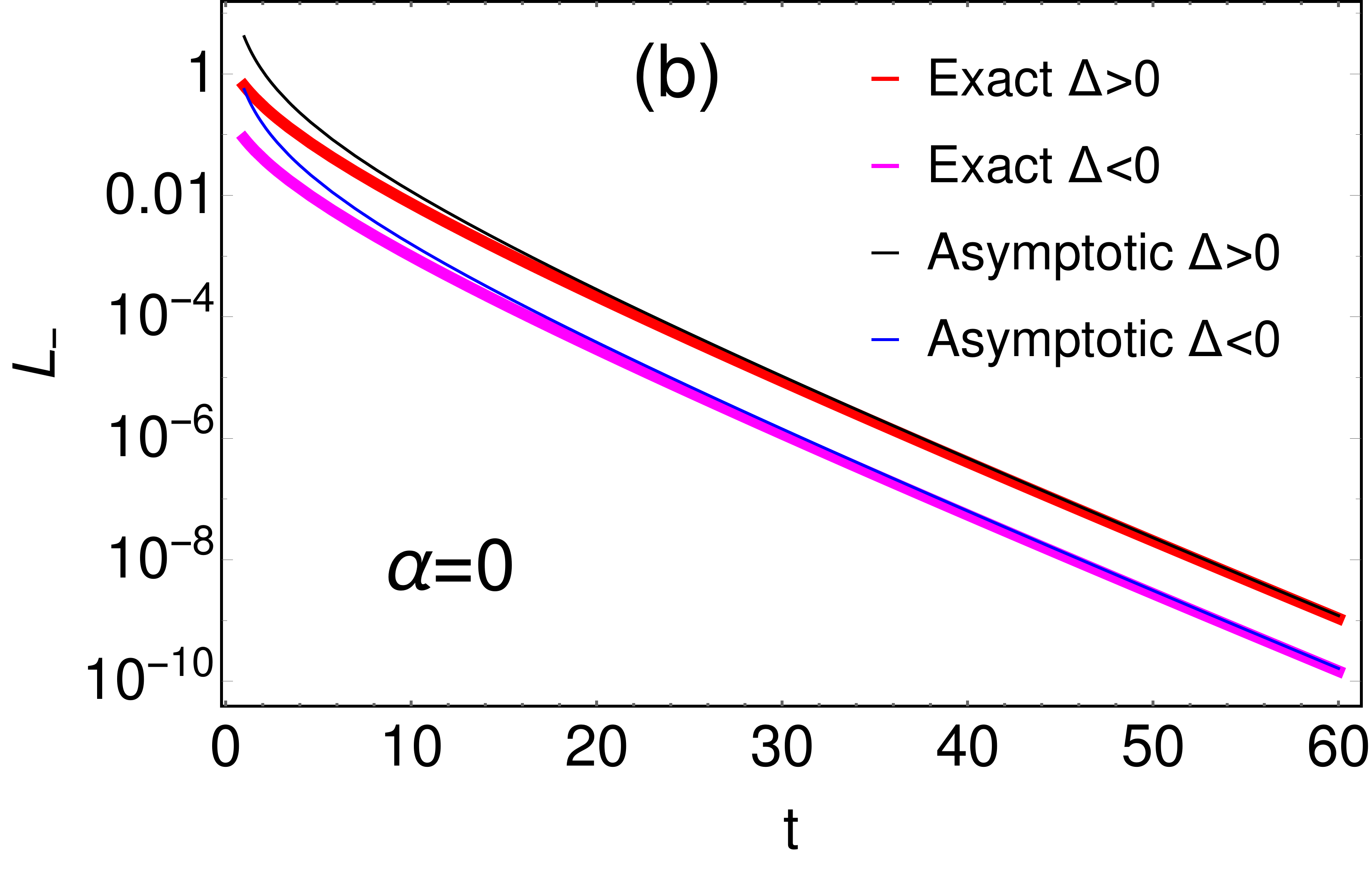}
\centering
\caption{Comparision of the asymptotic behaviour of $S_{\pm}(x_0,t)$ given by Eqs. \eqref{surv-a-001} and \eqref{surv-a-002} with the exact expression in Eq. \eqref{surv_a-1} for $\alpha =0$ and for both $\Delta>0$ and $\Delta <0$. We have plotted $L_{\pm}(x_0,t)=S_{\pm}(x_0,t)-S_{\pm}(x_0, t \to \infty)$. In both (a) and (b) we have chosen (i) $\gamma_1 =3,~\gamma_2=1$ for red and (ii) $\gamma_1 =1,~\gamma_2=3$ for magenta. Other common parameters are $v=1$ and $x_0=1$.} 
\label{asy_surv_alph_0}
\end{figure}

We now discuss the behaviour of $S_\pm(x_0,t)$ for large $t$ which would provide the relaxation  to this stationary value for $\Delta \geq 0$. For this, we take the large $t$ approximation in Eqs. \eqref{surv_a-1}. The detail of this calculation is given in \ref{appen-asy-surv-alph-0} and we present only the final results here. Defining $L_{\pm}(x_0,t)=S_{\pm}(x_0,t)-\mathcal{S}_{\pm}(x_0)$, where $\mathcal{S}_{\pm}(x_0)$ is $0$ for $\Delta \geq 0$ and given by Eqs.~\eqref{surv_eq101} for $\Delta <0$, we obtain
\begin{align}
 L_+(x_0,t) \approx
    \begin{cases}
    \sqrt{\frac{\gamma _1}{\gamma _2}} \left(\frac{x_0}{v}+\frac{1}{\sqrt{\gamma_1 \gamma_2}} \right) \frac{(\gamma_1 \gamma_2)^{\frac{1}{4}}e^{\frac{\Delta x_0}{v}}}{\sqrt{2 \pi t^3} \left(\gamma- \sqrt{\gamma_1 \gamma_2} \right)} e^{-t \left(\gamma- \sqrt{\gamma_1 \gamma_2} \right)}
   , & \text{if}\ \gamma_1 \neq \gamma_2 \\
     \frac{1}{\sqrt{\pi t}}  \frac{\sqrt{2 \gamma}}{v} \left( x_0+\frac{v}{\gamma}\right)
  , & \text{if}\ \gamma_1=\gamma_2 .
\end{cases}
    \label{surv-a-001}
\end{align} 
\begin{align}
 L_-(x_0,t) \approx
      \begin{cases}
    \frac{x_0}{v} \frac{(\gamma_1 \gamma_2)^{\frac{1}{4}}e^{\frac{\Delta x_0}{v}}}{\sqrt{2 \pi t^3} \left(\gamma- \sqrt{\gamma_1 \gamma_2} \right)} e^{-t \left(\gamma- \sqrt{\gamma_1 \gamma_2} \right)}
   , & \text{if}\ \gamma_1 \neq \gamma_2 \\
       \frac{1}{\sqrt{\pi t}}  \frac{x_0\sqrt{2 \gamma}}{v}
   , & \text{if}\ \gamma_1=\gamma_2.
    \end{cases}
    \label{surv-a-002}
\end{align}
Our results for $\Delta =0$ match with that in \cite{Kanaya_2018, Doussal2019}. Note that for $\Delta >0$, the survival probabilities decay exponentially whereas for $\Delta =0$ case it decays as a power law $\sim 1/\sqrt{t}$. In fact the time scale $\tau_{r}=\frac{1}{\gamma - \sqrt{\gamma _1 \gamma _2}}$ associated  to this exponential decay diverges in the $\Delta \to 0~(\gamma _1 \to \gamma_2)$ limit which is consistent with the power law behaviour for $\Delta =0$.  For $\Delta <0$ case, we observe that $S_{\pm}$ relaxes exponentially to their stationary values over the same time scale $\tau_{r}=\frac{2 \gamma}{\Delta^2}$. The divergence of $\tau _r$ as $\Delta \to 0$ implies that the stationary survival probabilities do not exist for $\Delta =0$ case. From the expressions, we see that for $x_0=0$ while $S_-$ is exactly $0$, $S_+$ still has a non-zero value. The particle can survive if it starts from origin with positive velocity. In Figure (\ref{asy_surv_alph_0}) we have compared these asymptotic behaviours with the exact results in Eqs. \eqref{surv_a-1}. 

%For $ \Delta = 0$, the above asymptotic behaviour at large $t$ match with the results of \cite{Kanaya_2018, Satya-2RTP_2019}.

\subsection{Case II: $\alpha=1$}
\label{surv-alpha=1}
We now turn to the $\alpha=1$ case. For this case the rates $R_{1,2}(x)$ of the orientation flipping  decays as $\sim |x|^{-1}$ which, as we will see, makes the large time behaviour for the survival probability different from the $\alpha=0$ case. This difference is most prominent in the $\Delta=0$ case which we consider next. In the subsequent sections we discuss the $\Delta \neq 0$ cases. 
%We denote survival probabilities as $S_{\pm}^{*}(x_0,t)$ where $*$ is $0$ and $\Delta$ for $\Delta=0$ and $\Delta \neq 0$ respetively.

\subsubsection{$\Delta=0$}
$~$\\
{We start with  Eq.~\eqref{surv_F} which for $\alpha=1$ and $\Delta=0$ takes the form
\begin{align}
\partial_{x_0} F(x_0,s)-\left(\frac{2 \gamma s x_0}{v^2 l} +\frac{s^2}{v^2}\right) F(x_0,s)=0
\label{surv_F-alph-1-D-0}
\end{align}
Since the general solutions of this equation are same to Eq.~\eqref{G-alpha-1-Del-0} we take the solutions from \ref{appen_alph_1} and write a general solution for $F(x_0,s)$ in terms of Airy functions and the integration constants $C_\pm$.
%\cyanw{and fix the integration constants by demanding $F(x_0,s)$ to be finite at $x_0 \to \infty$.} 
Inserting the $F(x_0,s)$ in Eq. \eqref{surv_H} and then in Eq.~ \eqref{surv_U-diff}, and fixing the integration constants through the boundary conditions, we finally get
\begin{align}
\bar{S}_{\pm}(x_0,s)=\frac{1}{s}-\frac{1}{s} \frac{v~c~\text{Ai}'\left(c x_0 s^{\frac{1}{3}}+d s^{\frac{4}{3}} \right)\pm s^{\frac{2}{3}} \text{Ai}\left(c x_0 s^{\frac{1}{3}}+d s^{\frac{4}{3}}\right)}{v c \text{Ai}'(0)- s^{\frac{2}{3}} \text{Ai}(0)}
\label{surv_alph_1_eq2}
\end{align}
with $c=\left(\frac{2 \gamma }{v^2 l}\right)^{\frac{1}{3}}$ and $d= \frac{c~l}{2 \gamma}$ where in the end we have used the definitions in Eqs.~\eqref{surv_U-H-def} . Performing inverse Laplace transform for arbitrary $s$ is hard and also not so illuminating. We focus on the large $t$ behaviour, to get which we neglect the $ds^{4/3}$ term in the argument of the Airy function and get the following simpler equation 
%To get the behaviour
%For small $s$ which corresponds to large $t$, one can simplify Eq.~\eqref{surv_alph_1_eq2}.
\begin{align}
\bar{S}_{\pm}(x_0,s) \simeq \frac{1}{s}-\frac{1}{s v c \mid \text{Ai}'(0) \mid} \left[v c \text{Ai}'(c x_0 s^{\frac{1}{3}}) \pm s^{\frac{2}{3}}\text{Ai}(c x_0 s^{\frac{1}{3}})-\frac{s^{\frac{2}{3}}\text{Ai}(0)}{\mid \text{Ai}'(0) \mid} \text{Ai}'(c x_0 s^{\frac{1}{3}})\right]
\label{surv_alph_1_eq3}
\end{align}
\begin{figure}[t]
\includegraphics[scale=0.195]{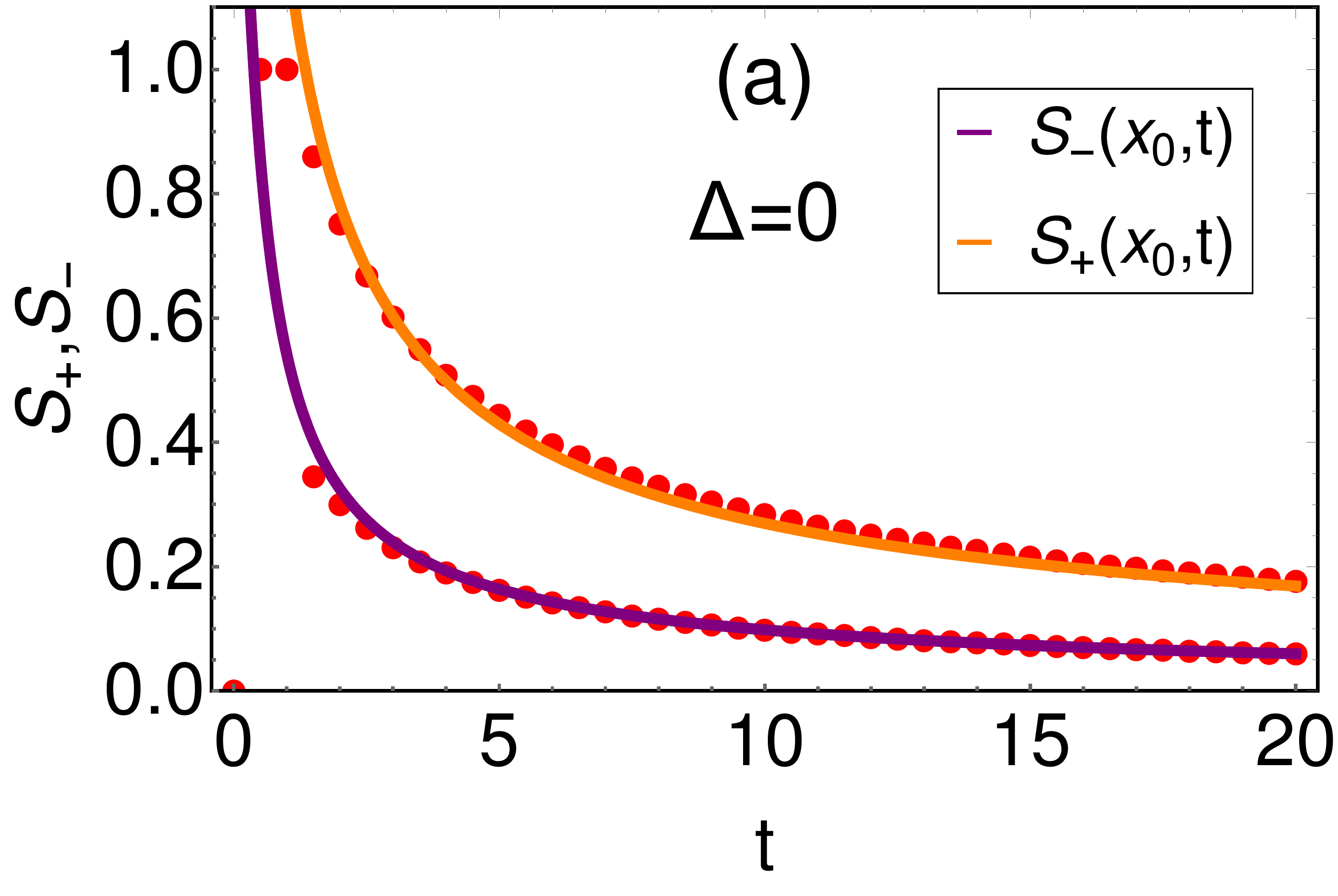}
\includegraphics[scale=0.202]{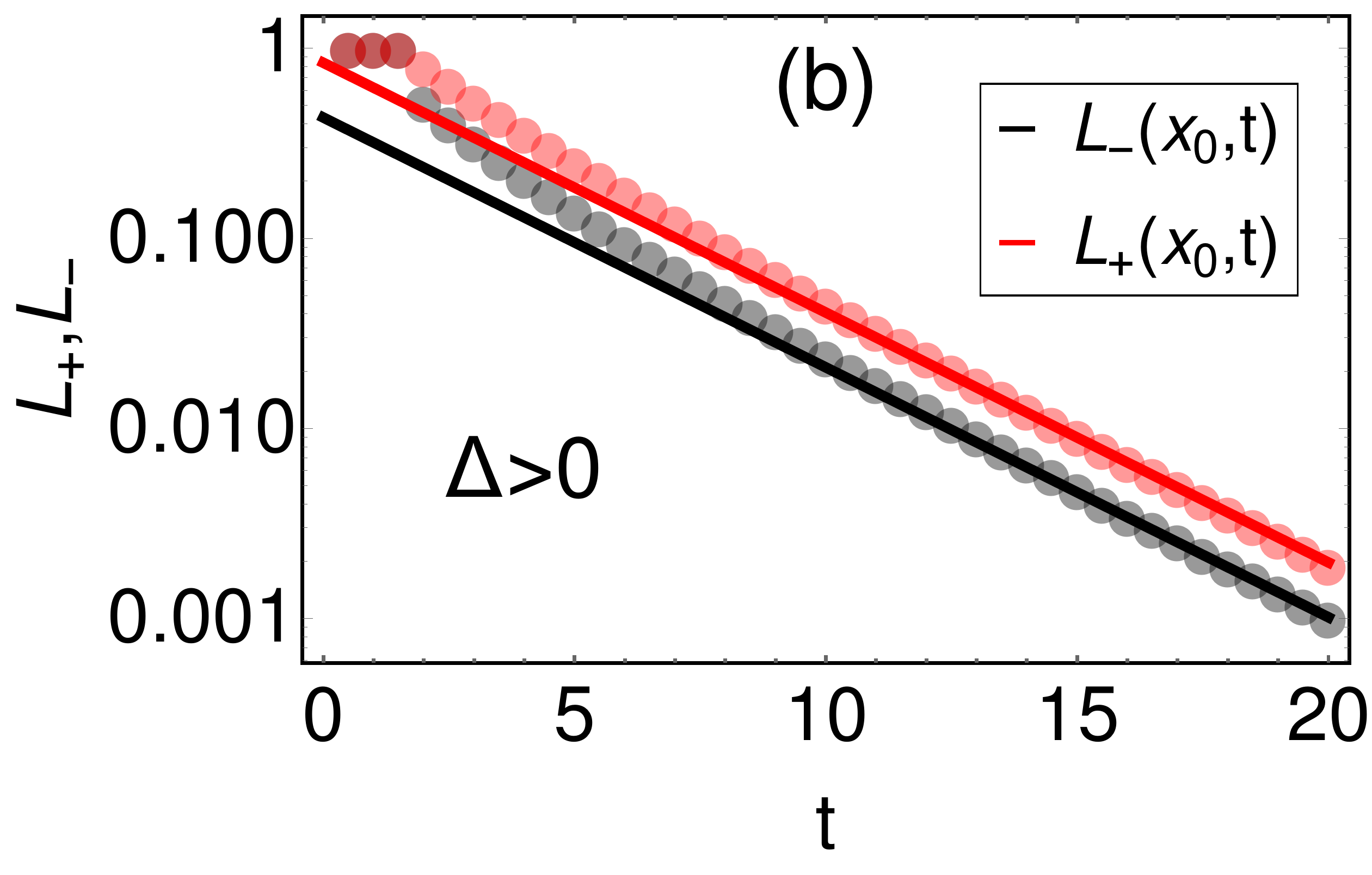}
\includegraphics[scale=0.202]{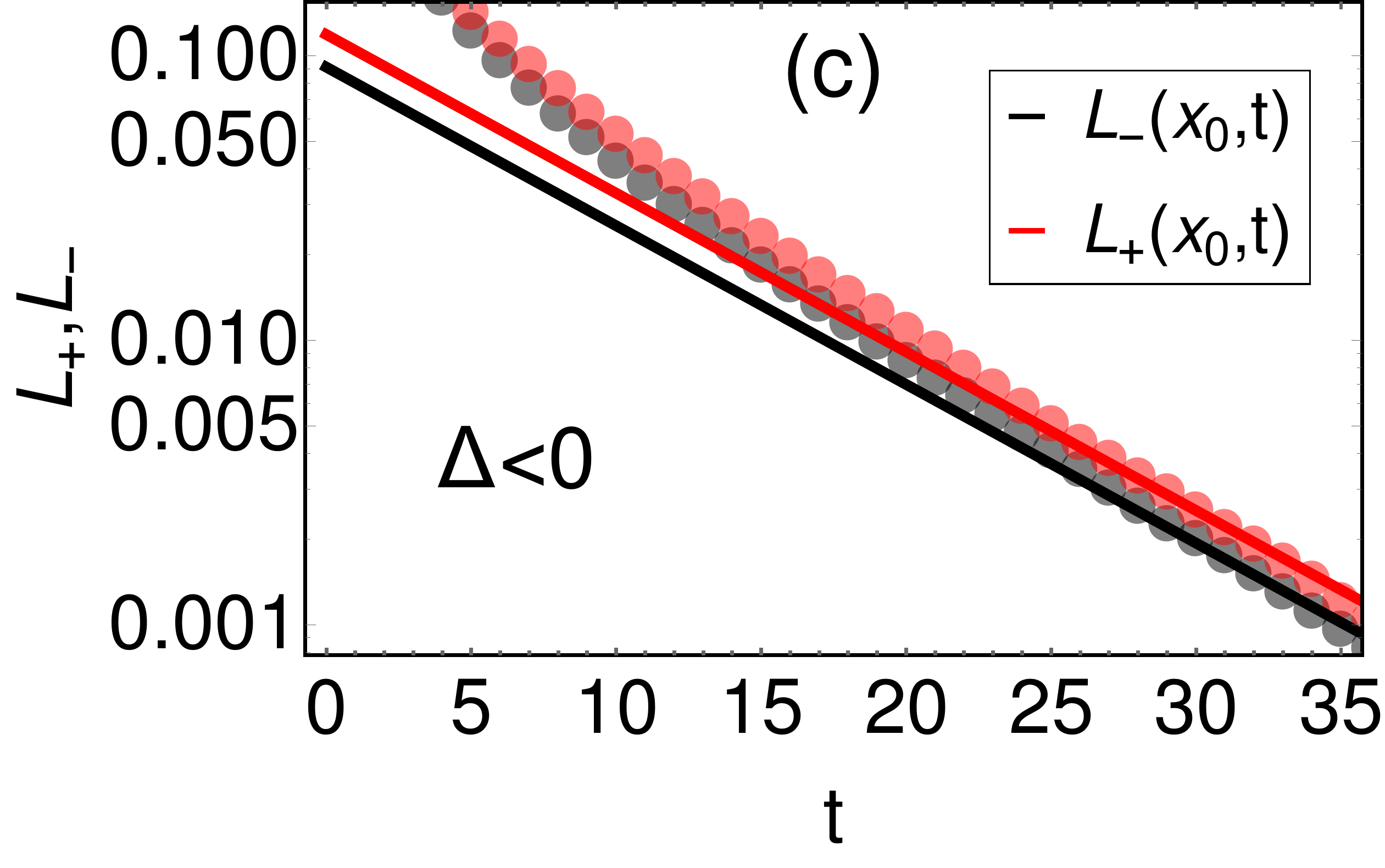}
\centering
\caption{Comparision of the analytic results of  $S_{\pm}(x_0,t)$ with the numerical simulation for $\alpha =1$ and  $\Delta=0$ (a), $\Delta >0$  (b) and $\Delta <0$ (c). In Figure (a), we have plotted Eq.~\eqref{surv_alph_eqq1} for $\gamma_1=\gamma_2=1$ and $x_0=1$ . For Figures (b) and (c), we define $L_{\pm}(x_0,t)=S_{\pm}(x_0,t)-\mathcal{S}_{\pm}(x_0)$ with $\mathcal{S}_{\pm}(x_0)$ equal to $0$ for $\Delta >0$ and given by Eqs.~\eqref{surv_eqqq6} for $\Delta <0$. The analytic expressions for $L_{\pm}(x_0,t)$, given by Eqs. \eqref{surv_eqqq7} for both $\Delta <0$ and $\Delta>0$, are plotted (solid line) and compared with the simulation data (filled circles). For Figure (b), we have taken $x_0=1.5,\gamma_1=2 \text{ and }\gamma_2=1$ while for Figure (c), we have taken $x_0=1.5,\gamma_1=1 \text{ and }\gamma_2=1.5$. For all three figures, other common parameters are $v=1$ and $l=1$.   }
\label{new_surv_alph_1}
\end{figure}
\noindent
Now once again using $\text{Ai}(z)=\frac{\sqrt{z}}{\sqrt{3} \pi} K_{\frac{1}{3}}\left(\frac{2}{3} z^{\frac{3}{2}} \right)$ and using Eq.~\eqref{main_eqqq3} we perform the inverse of the Laplace transform to get
\begin{align}
S_{\pm}(x_0,t)&\simeq1+\frac{e^{-\frac{x_0 ^3 c^3}{18t}}}{2 \pi \sqrt{3g} v c \mid \text{Ai}'(0) \mid} \left[\frac{c x_0 \text{Ai}(0)}{\mid \text{Ai}'(0) \mid \sqrt{t}}W_{\frac{1}{2},\frac{1}{3}}\left(\frac{x_0 ^3 c^3}{9t}\right)- v c^2 x_0 t^{\frac{1}{6}} W_{-\frac{1}{6}, \frac{1}{3}} \left( \frac{x_0 ^3 c^3}{9t}\right) \right. \nonumber\\
& ~~~~~~~~~~~~~~~~~~~\pm \frac{\sqrt{c x_0}}{t^{\frac{1}{3}}} W_{\frac{1}{3}, \frac{1}{6}} \left(\frac{x_0 ^3 c^3}{9t}\right).
\label{surv_alph_eqq1} 
\end{align}
In Figure \ref{new_surv_alph_1}(a), we have plotted our result of $S_{\pm}(x_0,t)$ alongwith with the results from numerical simulations where we observe perfect match at large  $t$. The mismatch at smaller $t$ is self-explanatory. Although the expression in Eq.~\eqref{surv_alph_eqq1} is nice but still less illuminating. It is more instructive to find the large $t$ asymptotic of $S_{\pm}(x_0,t)$. For this, we use the following representation of the Whittaker function $W_{k,m}(z)=e^{-\frac{z}{2}} z^{m+\frac{1}{2}} U(\frac{1}{2}+m-k, 1+2m,z)$, where $\mathcal{U}(a,b,z)$ is the confluent hypergeometric function of second kind which, for $z\to 0$ behaves as $\mathcal{U}(a,b,z)\approx\frac{\Gamma(b-1)}{\Gamma(a)}z^{1-b}+\frac{\Gamma(1-b)}{\Gamma(a-b+1)}$ with $\Gamma(a)$ being the Gamma function. Using this asymptotic form in Eq.~\eqref{surv_alph_eqq1} we get
\begin{align}
&S_{+}(x_0,t)\approx \frac{1}{2 \pi 3^{\frac{5}{6}}t^{\frac{2}{3}}} \left(-x_0^2 c^2 \Gamma\left(-\frac{2}{3} \right) + \frac{6\Gamma\left(\frac{1}{3} \right)}{v c}\right), \nonumber \\
&S_{-}(x_0,t)\approx -\frac{1}{2 \pi 3^{\frac{5}{6}}t^{\frac{2}{3}}} x_0^2 c^2 \Gamma\left(-\frac{2}{3} \right)
\label{surv_alph_eqq2}
\end{align}
suggesting a power law decay with persistent exponent $\theta = 2/3$. Note that this exponent is different from the exponent $\theta=1/2$ in the $\alpha=0$ case (see Eqs \eqref{surv-a-001} and \eqref{surv-a-002}). Another interesting feature to note is while $S_-(x_0,t)$ vanishes if $x_0=0$, $S_+(x_0,t)$ still has a non-zero value (which decays with $t$) as we have  seen  in the $\alpha =0$ case.
%, this was obtained in \cite{Kanaya_2018} whcih we had shown is also a feature of $\alpha =1$ case.  }

%\begin{figure}
%\includegraphics[scale=0.24]{survival_alph_1_del_gt_0.pdf}
%\includegraphics[scale=0.24]{survival_alph_1_del_lt_0.pdf}
%\centering
%\caption{Here we have plot $L_{\pm}(x_0,t)=S_{\pm}(x_0,t)-S_{\pm}(x_0,t)$ and compare it with the simulations. For $\Delta >0$, %%$S_{\pm}(x_0, t \to \infty)=0$ which means $L_{\pm}(x_0,t)=S_{\pm}(x_0,t)$ which is given in Eq.~\eqref{surv_eqqq10}. For $\Delta% <0$, we have non-zero $S_{\pm}(x_0, t \to \infty)$ and corresponding $L_{\pm}(x_0,t)$ is given at Eq.~\eqref{surv_eqqq7}. For %Fig(a)$\gamma_1=2,\gamma_2=1$ while for Fig(b)$\gamma_1=1,\gamma_2=1.5$. Other parameters chosen fro both figures are $x=1, l=1$ %and $x)=1.5$.}
%\label{new_surv_alph_1_delneq0}
%\end{figure}

%\begin{figure}
%\includegraphics[scale=0.24]{general_alph_del_0_surv.pdf}
%\includegraphics[scale=0.24]{survival_alph_1_del_lt_0.pdf}
%\centering
%\caption{Here we have plot $S(x_0,t)$ given in Eq.~\eqref{surv_genalph_eq4} and verify with numerical simulation. The parameters chosen are $x_0=10, v=1, l=1$ and $\gamma_1=\gamma_2=1$.}
%\label{new_surv_alph_1_delneq0}
%\end{figure}

\subsubsection{$\Delta \neq 0$}
$~$\\
We now consider $\Delta \neq 0$ case for which the Eq.~\eqref{surv_F} reduces to
\begin{align}
\partial_{x_0}^{2} F-\left[ \frac{\Delta}{v l}+\frac{2 \gamma s x_0}{v^2 l}+\frac{s^2}{v^2}+\frac{\Delta ^2 x_0^2}{v^2 l^2}\right]F=0.
\label{surv_eqqq3}
\end{align}
We note that this equation is identical to Eq.~\eqref{main_eqG} although the boundary conditions of the two equations are different. However, the general solutions of the two equations are same and can be expressed in terms of the parabolic cylinder functions $D_\mu(y)$  as shown in \ref{appen_alph_1}. Inserting this general solution in Eqs. \eqref{surv_U-diff} and \eqref{surv_H} and using the boundary conditions $\bar{S}_-(x_0 \to 0, s)=0$ we, after performing some simplifications, get 
\begin{align}
\bar{S}_{\pm}(x_0,s)=\frac{1}{s}-\frac{1}{s} e^{\frac{\Delta x_0^2}{2 v l}} \frac{\mathcal{ N}_{\pm} (x_0, s)}{\mathcal{N}_-(0,s)},
\label{surv_eqqq4}
\end{align}
where
\begin{align} 
\mathcal{N}_{\pm}(x_0,s)=&\sqrt{\frac{2 v \mid \Delta \mid}{l}}\left\{\beta s^2 \Theta(-\Delta)+\Theta(\Delta)\right\} D_{\beta s^2-\Theta(-\Delta)}\left(\sqrt{\frac{2 \mid \Delta \mid}{v l}}\left(x_0+\frac{\gamma s l}{\Delta ^2} \right)\right) \nonumber \\ 
&~~~-s \frac{\gamma \pm  \Delta }{\mid \Delta \mid}D_{\beta s^2-\Theta(\Delta)}\left(\sqrt{\frac{2 \mid \Delta \mid}{v l}}\left(x_0+\frac{\gamma s l}{\Delta ^2} \right)\right),
\label{surv_eqqq5}
\end{align}
with $\beta =\frac{ l (\gamma^2-\Delta ^2)}{2 v \mid \Delta \mid ^3}$ and $\Theta(x)$ is the Heaviside theta function. To get the survival probabilities in the time domain one needs to perform inverse Laplace transform. As we are interested in the behaviour at large $t$, we look at behaviour of  $ \bar{S}_{\pm} (x_0,s)$ for small $s$. In particular, the $\lim_{s \to 0}\left[s \bar{S}_{\pm} \right]$ gives the survival probability as $t \to \infty$. For $\Delta <0$ using $D_0(z)=e^{-\frac{z^2}{4}}$, we get
\begin{align}
\mathcal{S}_+(x_0)=& 1-\frac{\gamma -\mid \Delta \mid}{\gamma +\mid \Delta \mid} e^{-\frac{\mid \Delta \mid x_0^2}{v l}}, \nonumber\\
\mathcal{S}_-(x_0)=& 1- e^{-\frac{\mid \Delta \mid x_0^2}{v l}}.
\label{surv_eqqq6} 
\end{align}
where $\mathcal{S}_{\pm}(x_0)=S_{\pm}(x_0, t \to \infty)$. Similarly one can compute $\lim_{s \to 0}\left[s \bar{S}_{\pm} \right]$ for $\Delta > 0$ which turns out to be $0$ as the particle will definitely hit the absorbing wall at $x=0$ given sufficient time. 
%Coming to the inversion of the Laplace transforms in Eq\eqref{surv_eqqq4}, we were not able to invert them for all $t$. 
To get the approach to the stationary value $\mathcal{S}_{\pm}(x_0)$ for $\Delta<0$ and the decay  to $0$ for $\Delta>0$, at large $t$, we study the zeroes of $\mathcal{N}_{-}(0,s)$. Note from Eq.~\eqref{surv_eqqq4} that there is a  simple pole at $s=0$ which provides $\mathcal{S}_{\pm}(x_0)$ for $\Delta <0$ and $0$ for $\Delta >0$. Subtracting this part we define  $L_{\pm}(x_0,t)=S_{\pm}(x_0,t)-S_{\pm}(x_0,t \to \infty)$ which can be obtained from the poles of  $\bar{S}_{\pm}(x_0,s)$ other than $s=0$ on the negative $s$ axis. these poles come from the solution $\mathcal{N}_-(0,s)=0$. For large $t$, the solution of $\mathcal{N}_-(0,s)=0$ with largest real part (say $s^*$) sets the time scale for decay of $L_{\pm}(x_0,t)$. We get, 
\begin{align}
L_{\pm}(x_0,t)= -e^{\frac{\Delta x_0^2}{2 v l}}\frac{e^{s^{*} t}}{s^*} \frac{\mathcal{N}_{\pm}(x_0, s^*)}{\mathcal{N}_-^{'}(0, s^*)},
\label{surv_eqqq7}
\end{align}
where $s^*$ is the largest root of $\mathcal{N}_-(0,s^*)=0$. In Figure \ref{new_surv_alph_1}(b) and  \ref{new_surv_alph_1}(c), we have verified our analytic results with the numerical simulation and we observe excellent match.

\begin{figure}[t]
\includegraphics[scale=0.18]{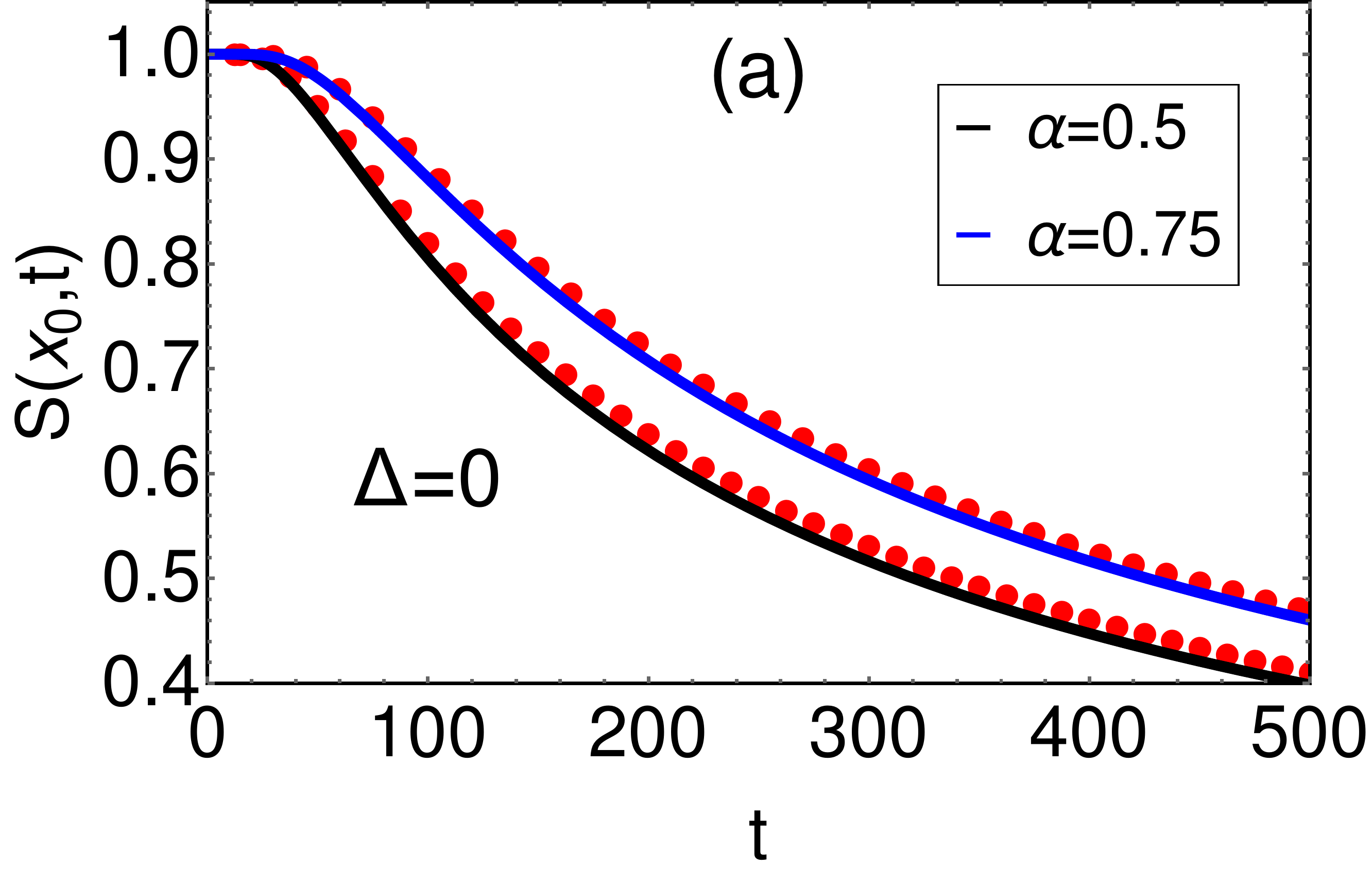}
\includegraphics[scale=0.21]{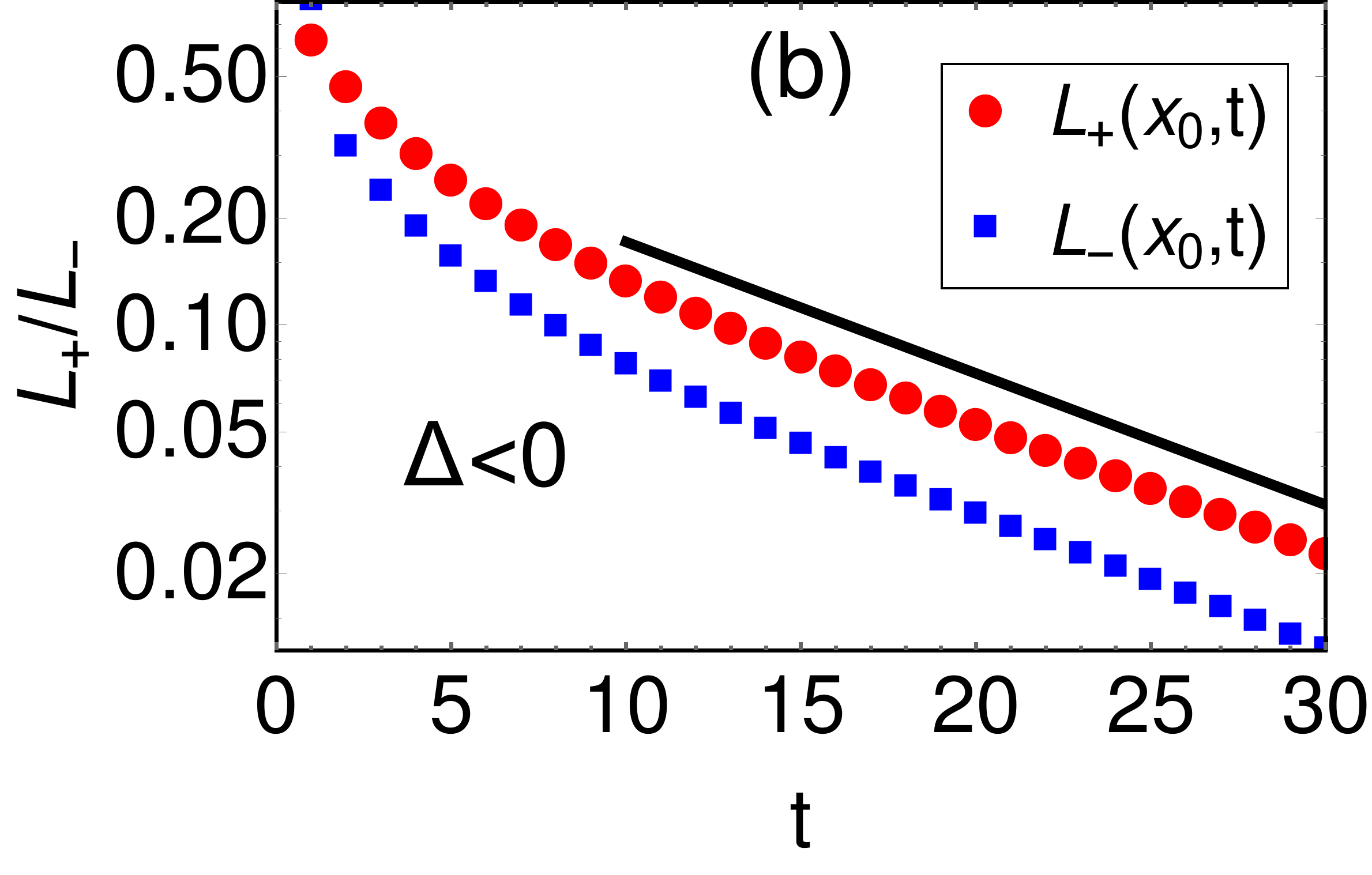}
\includegraphics[scale=0.175]{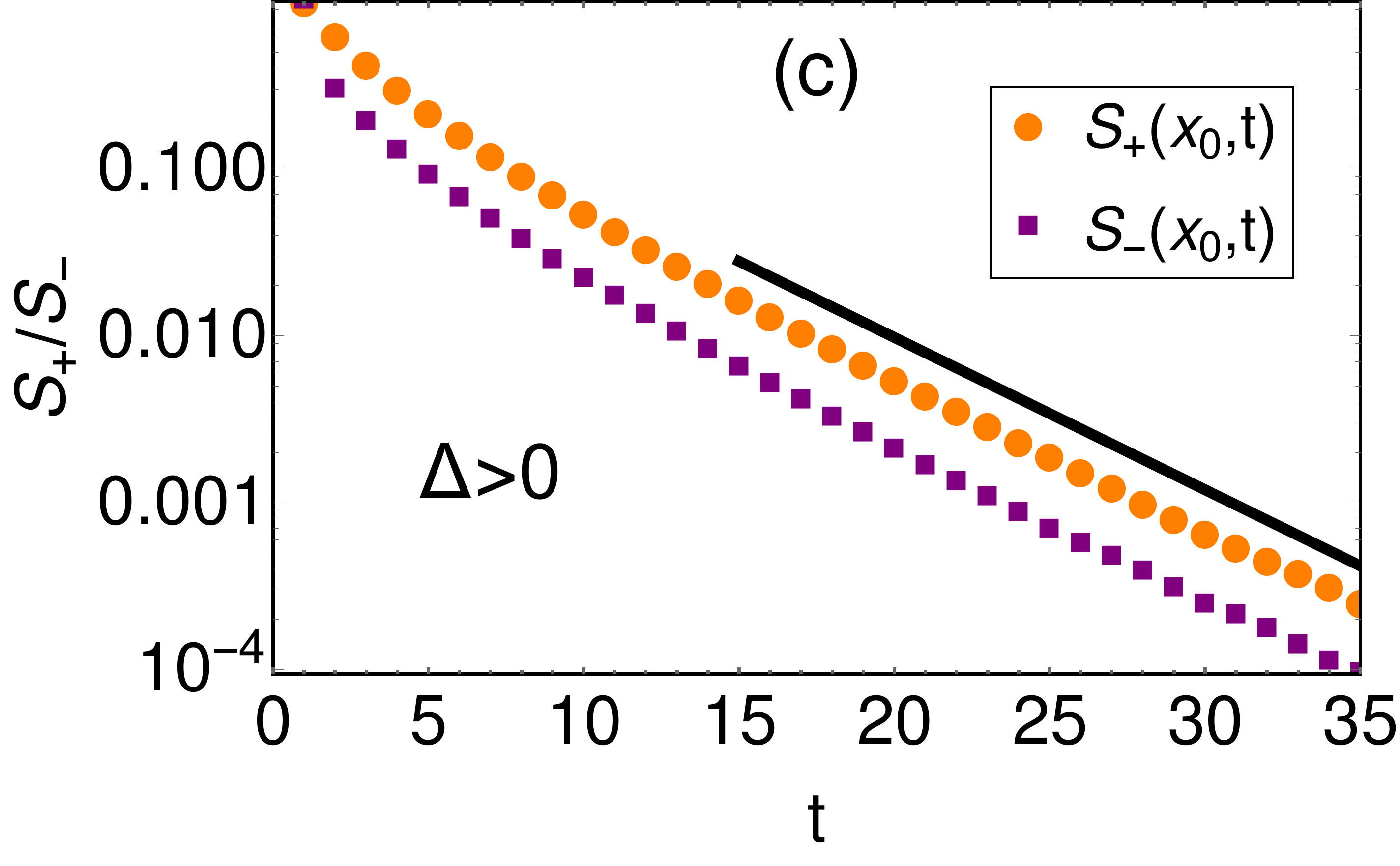}
\centering
\caption{(a) Comparision of the analytic results of  $S{\pm}(x_0,t)$ in Eq.~\eqref{surv_genalph_eq5} with the numerical simulation for two different $\alpha$ and $\Delta =0$. We have chosen $\gamma_1=\gamma_2=1,~v=1,~l=1$ and $x_0=10$. (b) Simulation results of $L_{\pm}(x_0,t)$ for $\alpha =0.5$ and $\Delta <0$. We observe exponential decay of the form $e^{-\zeta t}$ with $\zeta=0.085$. Parameters chosen for this plot are $\gamma_1=1.2,~\gamma_2=1.5,~v=1,~x_0=1,~\text{and }l=1.$ (c) Simulation results for $S_{\pm}(x_0,t)$ for $\Delta >0$. Here also we see exponential decay of the form $e^{-\zeta t}$ with $\zeta=0.2097$. We have chosen $\gamma_1=2,~\gamma_2=1,~v=1,~x_0=1,~\text{and }l=1.$}
\label{surv_gen_alph}
\end{figure}

\subsection{General $\alpha$}
\label{surv-gen-alpha}
For this case it seems convenient to solve Eqs.~\eqref{surv_U_pm} directly. Making analytical progress for arbitrary $t$ seems difficult. We instead look at the large time limit which necessarily requires $x_0$ to be large so that the particle survives for long time. In this limit, the difference $H(x_0,t)$ of survival probabilities $U_\pm(x_0,t)$ of the particle starting with $\pm$ velocities would decay fast (exponentially) which allows one to one to neglect the difference $H(x_0,t)$. Making the approximation, as we show in \ref{appen_efff_surv}, the equation for $\bar{U}(x_0,s)$ becomes,
\begin{align}
s \bar{U}(x_0,s)=\frac{v^2 l^{\alpha}}{2 \gamma} \partial_{x_0} \left(\frac{1}{x_0 ^{\alpha}} \partial_{x_0} \bar{U}\right)-\frac{ v \Delta}{\gamma} \partial_{x_0} \bar{U}.
\label{surv_genalph_eq1}
\end{align}
where $\bar{U}(x_0,s)=\bar{U}_+(x_0,s)+\bar{U}_-(x_0,s)$. To solve this equation, we need to specify the boundary conditions in terms of $\bar{U}(x_0,s)$. The boundary condition at $x_0=0$ discussed previously for $S_{\pm}(x_0,t)$ which can be translated in terms of $\bar{U}_{\pm}(x_0,s)$ as,
\begin{align}
\bar{U}(0,s)=-\frac{2}{s}.
\label{bcs1}
\end{align}
Note that we have neglected $\bar{H}(0,s)$ term in Eq.~\eqref{bcs1} as it decays faster than $\bar{U}(x_0,s)$. We now solve Eq.~\eqref{surv_genalph_eq1} separately for $\Delta =0$ and $\Delta \neq 0$ cases.

\subsubsection{$\Delta =0 :$}
For $\Delta =0$, Eq.~\eqref{surv_genalph_eq1} reduces to 
 \begin{align}
s \bar{U}(x_0,s)= D_{\alpha} \partial_{x_0} \left(\frac{1}{x_0 ^{\alpha}} \partial_{x_0} \bar{H}\right),~~~~~~\text{with }D_{\alpha}= \frac{v^2 l^{\alpha}}{2 \gamma}
\label{surv_genalph_eq2}
\end{align} 
This equation is solved in \ref{appen_efff_surv_sol} and we here write the solution
\begin{align}
\bar{U}(x_0,s)=-\frac{4x^{1+\alpha}}{\Gamma\left( \frac{1+\alpha}{2+\alpha}\right)}\frac{s^{-\frac{\alpha+3}{2(2+\alpha)}}}{((2+\alpha)\sqrt{D_{\alpha}})^{\frac{1+\alpha}{2+\alpha}}} K_{\frac{1+\alpha}{2+\alpha}}\left( 2 \sqrt{g_{\alpha}s}\right),
\label{surv_genalph_eq3}
\end{align}
where $K_{\nu}(z)$ is the modified Bessel function of second kind and $g_{\alpha}=\frac{x^{2+\alpha}}{D_{\alpha}(2+\alpha)^2}$. To find the probability in time domain, one has to perform the inversion of the Laplace transforms. Looking at the expression of $\bar{U}(x_0,s)$, we use Eq.~\eqref{main_eqqq3} to invert the Laplace transform.
\begin{align}
S(x_0,t)&=\frac{S_+(x_0,t)+S_-(x_0,t)}{2}, \nonumber\\
&=1-\frac{1}{2} L_{s \to t}^{-1} \left[ \bar{U}(x_0,s)\right],\nonumber \\
&=1-\frac{e^{-\frac{g_{\alpha}}{2t}}}{\Gamma\left( \frac{1+\alpha}{2+\alpha}\right)}\frac{x^{1+\alpha}~~ t^{\frac{1}{2(2+\alpha)}} }{\{(2+\alpha)\sqrt{D_{\alpha}}\}^{\frac{1+\alpha}{2+\alpha}}\sqrt{g_{\alpha}}} W_{-\frac{1}{2(2+\alpha)},\frac{1+\alpha}{2(2+\alpha)}}\left(\frac{g_{\alpha}}{t} \right)
\label{surv_genalph_eq4}
\end{align}
To find asymptotics, we use  the following representation of the Whittaker function $W_{m,k}(z)=e^{-\frac{z}{2}} z^{m+\frac{1}{2}} \mathcal{U}(\frac{1}{2}+m-k, 1+2m,z)$ in terms of the confluent hypergeometric function $\mathcal{U}(a,b,z)$ of second kind whose asymptotic behaviour as $z\to 0$ is $\mathcal{U}(a,b,z)\approx\frac{\Gamma(b-1)}{\Gamma(a)}z^{1-b}+\frac{\Gamma(1-b)}{\Gamma(a-b+1)}$ which gives,
\begin{align}
S(x_0,t)=\frac{\mid \Gamma\left(-\frac{1+\alpha}{2+\alpha}\right)\mid}{\Gamma\left(\frac{1+\alpha}{2+\alpha}\right) \Gamma\left(\frac{1}{2+\alpha}\right)} \frac{x_0^{1+\alpha}}{(2+\alpha)^{\frac{2(1+\alpha)}{2+\alpha}}} \frac{1}{(D_{\alpha} t)^{\frac{1+\alpha}{2+\alpha}}}
\label{surv_genalph_eq5}
\end{align}
In Figure \ref{surv_gen_alph}(a), we have plotted our analytic result in Eq.~\eqref{surv_genalph_eq4} and compared with the numerical simulation. We see excellent match between them for large $x_0$. For small $x_0$, our result in Eq.~\eqref{surv_genalph_eq4} does not match with simulation results as this expression is not valid although the power law decay $\left(1/t^{\frac{1+\alpha}{2+\alpha}} \right)$ is correctly predicted. This mismatch arises from the approximation $\bar{H}(x_0,s) \approx 0$ for large $t$, which is true only for large $x_0$.

\subsubsection{$\Delta < 0:$}
When $\Delta <0$, the particle is drifted away from the origin with higher probability implying a non-zero survival probability for the particle even for infinite $t$. To find this probability we solve the original Eqs.~\eqref{surv_eq1} directly for the stationary value of the survival probability by putting $\partial_tS_\pm(x_0,t)=0$. We present here the final expression of $S_{\pm}(x_0, t \to \infty)$ and relegate the details of derivation to \ref{surv_large_t}. Defining $\mathcal{S}_{\pm}(x_0)=S_{\pm}(x_0, t \to \infty)$, we get 
\begin{align}
\mathcal{S}_{+}(x_0)=& 1 - \frac{\gamma - \mid \Delta \mid}{\gamma + \mid \Delta \mid}~ e^{-\frac{2 \mid \Delta \mid}{v l^{\alpha} (\alpha+1)}x_0^{\alpha+1}}, \label{surv_genalph_dell0_eq1}\\
\mathcal{S}_{-}(x_0)=&1 - e^{-\frac{2 \mid \Delta \mid}{v l^{\alpha} (\alpha+1)}x_0^{\alpha+1}}. \label{surv_genalph_dell0_eq2}
\end{align}
To study the approach to the steady state, we numerically find $S_\pm(x_0,t)$  and plot $L_\pm(x_0,t)=S_\pm(x_0,t)-\mathcal{S}(x_0)$ as functions of $t$ in Figs.~\ref{surv_gen_alph} (b). From these plots we see that the approach to the stationary values of both $S_\pm(x_0,t)$ is exponential with same relaxation time. 

\subsubsection{$\Delta >0:$}
When $\Delta >0$, the particle is drifted towards the origin. Unlike the previous case, the particle will now definitely hit the origin. In  Figs.~\ref{surv_gen_alph} (c), we numerically find that the survival probability decays exponentially to zero.

\begin{figure}[t]
\includegraphics[scale=0.29]{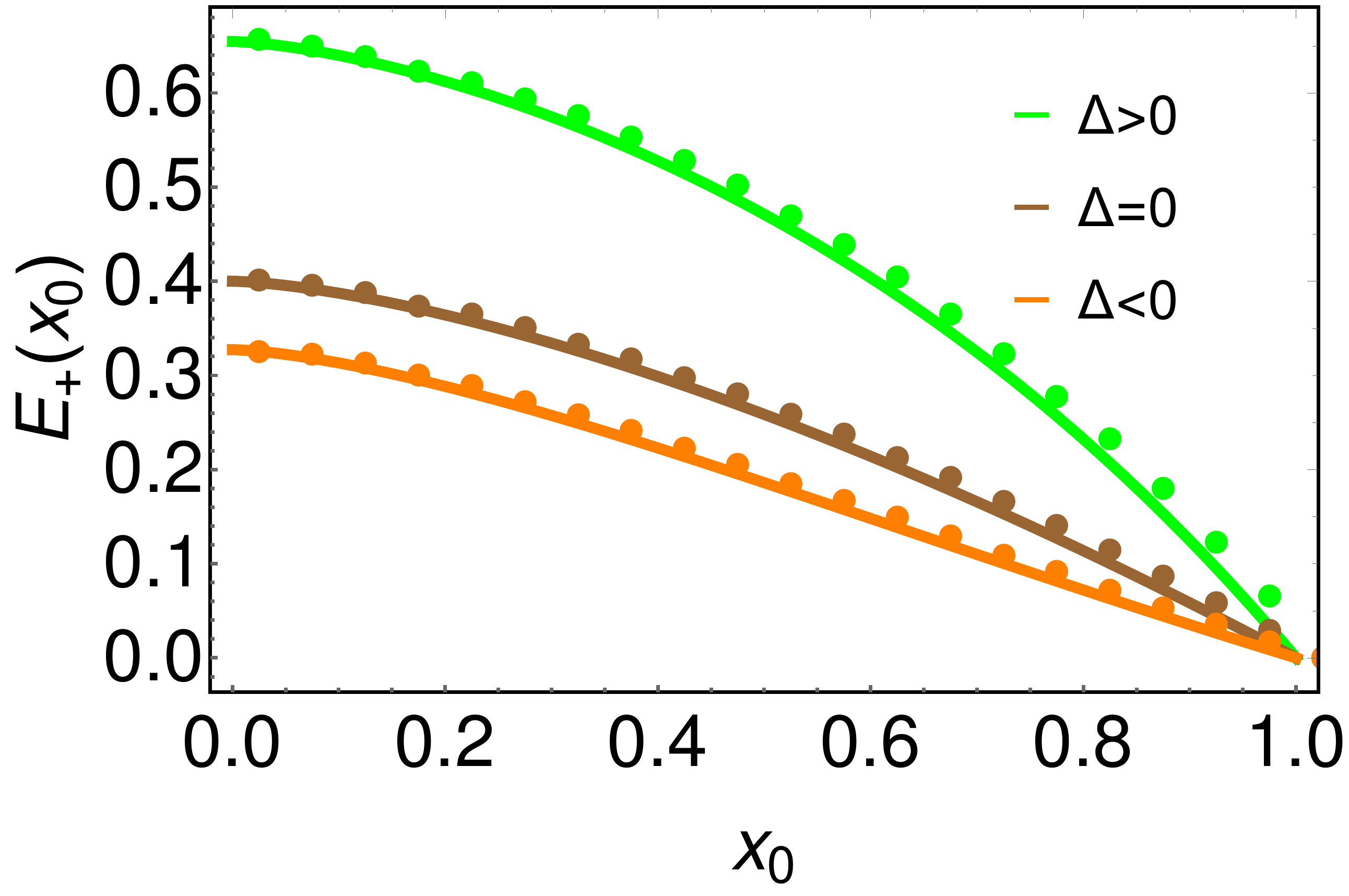}
\includegraphics[scale=0.3]{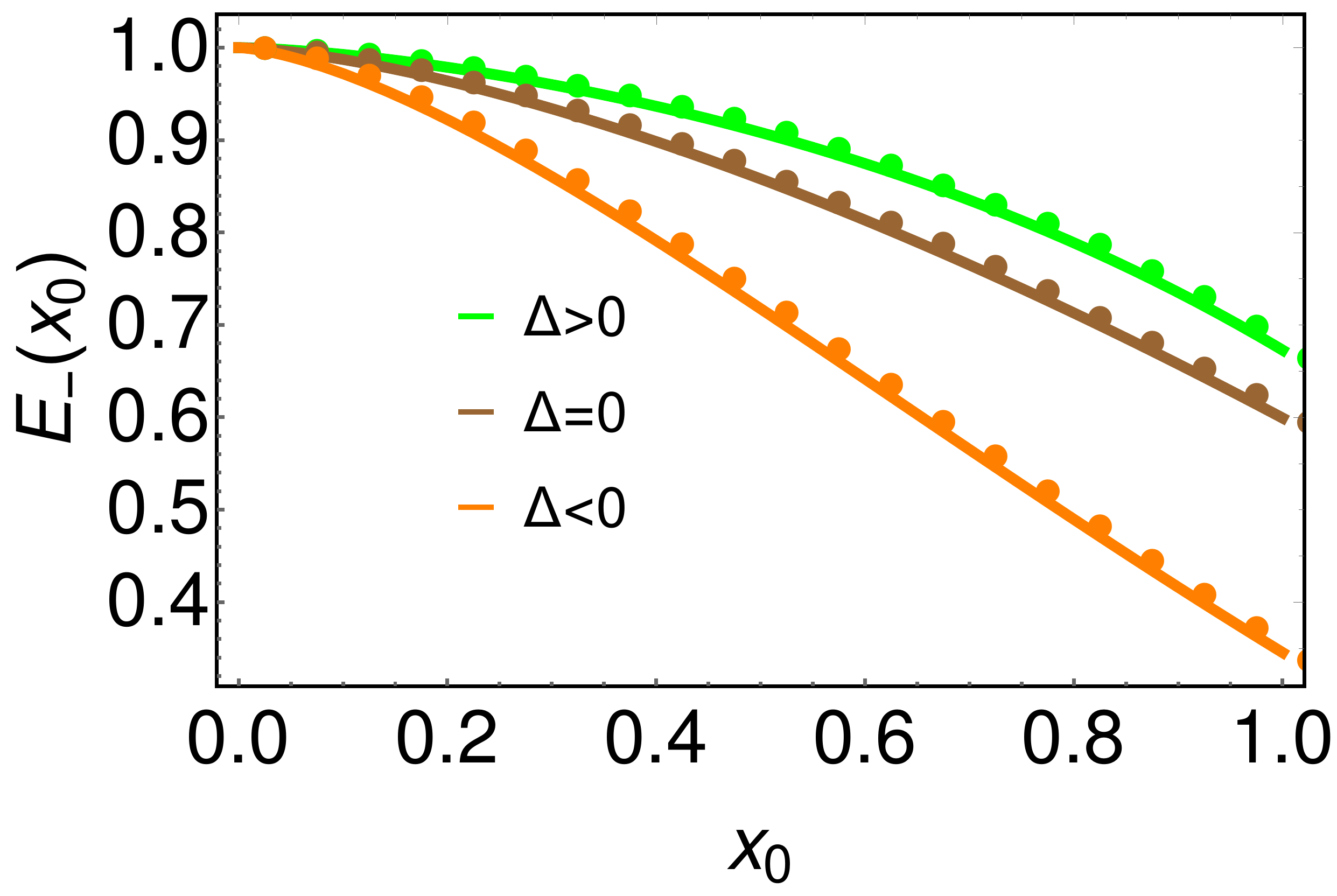}
\centering
\caption{Comparison of the exit probability $E_{\pm}(x_0)$ given in Eqs. \eqref{Exitp_eq13} and \eqref{Exitn_eq14} with the numerical simulation of the same (filled circles) for $\alpha=0.5$. The three curves correspond to (i) Green: $\gamma_1=2, ~\gamma_2=1$, (ii)Brown: $\gamma_1=1, ~\gamma_2=1$ and (iii)Orange: $\gamma_1=1,~ \gamma_2=2$. For both the figures, we have chosen $v=1\text{ and }l=1~$} 
\label{exit_pic1}
\end{figure}
\section{Exit probability of RTP from a finite interval for general $\alpha$}
\label{exit_prob_label}
In the previous sections, we have considered RTP in an infinite or semi-infinite line. This section deals with RTP in a finite interval $[0,M]$.  The question that is addressed in this section is: what is the probability that the RTP will exit from the side $x=0$ (or equivalently $x=M$) for general $\alpha$. Let $E_{\pm}(x_0)$ denote the exit probability of the particle from side $x=0$ starting from $x_0$ with velocity $\pm v$.  Following \cite{Redner}, one can write a coupled backward equations for $E_{\pm}(x_0)$ and solve them explicitly. Below we discuss the derivation of these equations.
%, $E_{\pm}(x_0)= \frac{1}{N} \sum_{p}P_p(x_0,\pm)$ where $P_p(x_0, \pm)$ is the paths that reach $x=0$ without touching $x=M$ starting from $x_0$ with veocity $\pm v$ and $N$ is the total number of paths. One can follow the particle for small time $\Delta t = \frac{\Delta x}{v}$ and in the remainder it reaches $x=0$ without touching $x=M$. Let us now focus on $E_+(x_0)=\frac{1}{N} \sum_{p}P_p(x_0,+)$.
 
Consider that the RTP starts at $x_0$ with $+v$. In the small time $dt$, RTP can (i) flip its velocity with probability $ R_{1}(x_0) \frac{dx}{v}$ and move to $x_0-dx$ or (ii) continue to move with $+v$ velocity with probability $ \left[1-R_{1}(x_0) \frac{dx}{v}\right]$ and reach $x_0+dx$. Starting from this new position, the particle then exits from $x=0$ without touching $x=M$. One can then write for $E_{\pm}(x_0)$,
\begin{align}
E_{+}(x_0)=\left[1-R_{1}(x_0) \frac{dx}{v}\right] E_+(x_0+d x)+ R_{1}(x_0) \frac{d x}{v} E_{-}(x_0-d x), \nonumber \\
E_{-}(x_0)=\left[1-R_{2}(x_0) \frac{dx}{v}\right] E_-(x_0-d x)+ R_{2}(x_0) \frac{d x}{v} E_{+}(x_0+d x).
\label{Exit_eq11}
\end{align}
Performing the Taylor's series expansion in $d x$ and then taking $\Delta x \to 0$ limit, one gets the backward equations for $E_{\pm}(x_0)$ which read as,
\begin{align}
v\partial_{x_0} E_+-R_1(x_0) E_++R_1(x_0) E_-=0, \nonumber\\
-v\partial_{x_0} E_-+R_2(x_0) E_+-R_2(x_0) E_-=0.
\label{Exit_eq12} 
\end{align}
Note that the rates $R_1(x_0)$ and $R_2(x_0)$ are defined in Eq.~\eqref{rates}. Before solving these equations, we need to specify the boundary conditions. The boundary conditions are $E_+(x_0=M)=0$ and $E_-(x_0=0)=1$. The first boundary condition comes from the fact that if the particle starts at $x_0=M$ with positive velocity, it will exit from $x=M$ wall in the next time-step. Likewise the second boundary condition appears because if the particle starts from $x_0=0$ with $-v$, it will, in the next time-step exit from $x=0$. Given these boundary conditions, one can solve these coupled differential equations in Eq.~\eqref{Exit_eq12} for general $\alpha$. After a  straightforward but tedious calculation, we find the following  final expressions for the exit probabilities
\begin{align}
&E_+(x_0)=\frac{e^{\bar{\Delta} M^{1+\alpha}}-e^{\bar{\Delta} x_0^{1+\alpha}}}{e^{\bar{\Delta} M^{1+\alpha}}-\frac{\gamma_2}{\gamma_1}}, \label{Exitp_eq13}\\
&E_-(x_0)=\frac{e^{\bar{\Delta} M^{1+\alpha}}-\frac{\gamma_2}{\gamma_1}e^{\bar{\Delta} x_0^{1+\alpha}}}{e^{\bar{\Delta} M^{1+\alpha}}-\frac{\gamma_2}{\gamma_1}}, \label{Exitn_eq14}
\end{align}
where $\bar{\Delta}=\frac{2 \Delta}{v(1+\alpha)l^{\alpha}}$. One can, in principle compute $E_{\pm}(x_0)$ also by integrating the the current $j(0,t)$ through side $x=0$ over all $t$. Although these two approaches yield the same result, the backward equation written in Eq.~\eqref{Exit_eq12} is more illustrative and instructive specially for general $\alpha$ where the computation of $j(x,t)$ with absorbing barriers at $x=0$ and $x=M$ is still a theoretical challenge. We also remark that taking $\Delta \to 0$ limit in Eqs. \eqref{Exitp_eq13} and \eqref{Exitn_eq14} correctly gives the results of \cite{Kanaya_2018} for $\alpha=0$. In Figure (\ref{exit_pic1}), we have plotted our results in Eq.~\eqref{Exitp_eq13} and \eqref{Exitn_eq14} with the numerical simulation of the same. The match between the two is excellent. In Figure (\ref{exit_pic1}), we notice that for a given $x_0$, $E_{\pm}$ is least for $\Delta <0$ and largest for $\Delta>0$ . For $\Delta >0 \left( \gamma_1>\gamma_2\right)$, the particle experiences an effective drift towards the origin which enhances the chance for particle to escape the origin. Similarly for $\Delta <0 \left( \gamma_1<\gamma_2\right)$ the particle is drifted away from the origin. 

Another interesting point to remark is that in the limit $M \to \infty$, $E_{\pm}(x_0)$ is equal to $1-\mathcal{S}_{\pm}(x_0)$. One can easily verify that Eqs. \eqref{Exitp_eq13} and \eqref{Exitn_eq14} in this limit indeed reduce to $1$ for $\Delta \geq 0$ and $1-\mathcal{S}_{\pm}(x_0)$ for $\Delta <0$ where $\mathcal{S}_{\pm}(x_0)$ is given by Eqs. \eqref{surv_genalph_dell0_eq1} and \eqref{surv_genalph_dell0_eq2} respectively.

\section{Conclusions}
\label{conc}
To summarise, we have studied the motion of a run and tumble particle in one dimensional inhomogeneous media. The inhomogeneity was introduced by considering the position and direction dependent rates of flipping given in Eqs. \eqref{rates}. For $\gamma_1> \gamma_2$, we have found that the particle reaches a stationary state in one dimension even in absence of any external confining potential. We have obtained an exact expression of the probability distribution given in Eq.~\eqref{stdy_sol}, which characterises this non-equilibrium stationary state.  
%Although, we have derived this result for $\alpha>0$, it can be verified that this result in Eq.~\eqref{} is also valid for $-1 < \alpha < 0$. This has been numerically verified in Figure \ref{gen-alp-delgt0}(a).  
The approach to this steady state is exponential for all $\alpha>0$. While for $\alpha =0$ and $\alpha=1$ we have 
been able to compute the full distribution $P(x,t)$ which indeed shows exponential relaxation, for general $\alpha>0$ performing 
exact calculation turned out to be difficult. We have provided heuristic argument for the exponential relaxation for general $\alpha>0$ with strong numerical evidence.

\begin{table}[h]
\centering
\begin{tabular}{|c|c|c|c|}
\hline

~&$\Delta >0$ & $\Delta =0$ & $\Delta <0$
\\ [0.5ex]
\hline 
& Stationary state  
 &No stationary  & No stationary  \\
%[-1ex]\raisebox{1.5ex}{} 
 & $P_\alpha^{st}(x)$ exists.  & state,  $P(x,t)$  &state, $P(x,t)$ \\
% [-1ex]\raisebox{1.5ex}{} for $\alpha \geq 0$
 & See Eq.~\eqref{stdy_sol}. &  with &  with \\
 & & $\mu(t)=\langle |x(t)|\rangle =0 $&$\mu(t) \underset{t \to \infty }{\sim} t$\\
\hline
&Relaxes as $e^{- bt}$ & Exact expression    & Exact expression   \\[-1ex]
\raisebox{1.5ex}{$\alpha=0$} 
& to $P_0^{st}(x).$& for $P(x,t)$ given & for $P(x,t)$ given \\ 
& & in Eq.~\eqref{main_eq14}  & in Eq.~\eqref{main_eq14} \\
& See Eq.~\eqref{main_eq15}.  & $\langle x^2(t)\rangle \underset{t \to \infty }{\sim} t$ & $\sigma_0^2(t)  \underset{t \to \infty }{\sim} t$ \\[1ex]
%& & & \\
\hline 
&Relaxes as $e^{-\zeta t}$ &Large $t$  & $P(x,t)$ obtained \\[-0.8ex]
\raisebox{1.5ex}{$\alpha=1$} 
&  to $P_{1}^{st}(x)$ with $\zeta$  & scaling form of & from ILT of  \\
&  given by the& $P(x,t)$ in Eq.\eqref{sc-sol-a-1-D-0}. & $\bar{P}(x,s)$ in Eq.\eqref{bar_P_alph-1-Del<0}, \\
& solution of Eq.\eqref{D-zero}  & $\langle x^2(t)\rangle  \underset{t \to \infty }{\sim} t^{2/3}$.  & $ \sigma_{1}^{2}(t) \sim \log (t)$ \\[1ex]
%&  & & \\[1ex]
\hline 
&Exponential  & Large $t$ & Large $t$  \\[-0.5ex]
 
&relaxation & scaling form of &  scaling form of ,\\[0.5ex]
\raisebox{1.2ex}{general }
& verified  & $P(x,t)$ in Eq.\eqref{main_eq21}, &$P(x,t)$ in Eq. \eqref{scaling-P_alpha}.\\[0.5ex]
\raisebox{1.2ex}{$\alpha \geq 0$}
&numerically&  with   & for $\alpha \leq 1$, with \\
&in Fig.~\ref{gen-alp-delgt0}b& $\langle x^2(t) \rangle \underset{t \to \infty }{\sim} t^{\frac{2}{2+\alpha}}$. &  $ \sigma_{\alpha}^{2}(t) \simeq D_{\alpha}t^{1-\alpha}$. \\[1ex]
\hline
\end{tabular}
\caption{Table summarising $P(x,t)$ for various $\alpha$ and $\Delta$. Here \\ $\sigma_\alpha^2(t) =\langle x^2(t)\rangle - \langle |x(t)|\rangle^2$ and ILT stands for 'Inverse Laplace Transform'.}
\label{table prob}
\end{table}

For $\gamma_1\leq \gamma_2$ the RTP particle does not reach a stationary state. While for $\gamma_1 <\gamma_2$ the average absolute position $|x(t)|$ of the particle grows linearly with time, for $\gamma_1=\gamma_2$,  $\langle |x| \rangle=0$. Note that the mean position $\langle x \rangle =0$ in all cases. This suggests that the distribution $P(x,t)$ has two symmetric peaks moving with equal speed in the opposite direction for $\gamma_1 <\gamma_2$ and for $\gamma_1 =\gamma_2$ there is a single non-moving peak at $x=0$.  In this case for $\alpha =0$ the distribution $P(x,t)$ was computed in \cite{Kanaya_2018} which was shown to be Gaussian at large $t$ with variance growing linearly with time. In this paper we have extended this result for general $\alpha>0$ for which we have found that $\langle x^2(t) \rangle \sim t^{\frac{2}{2+\alpha}}$. We also have proved that for large $t$, the distribution  function $P(x,t)$ follows a scaling form $f_\alpha(y)$ with scaling variable $y=|x|/t^{1/(2+\alpha)}$ for $\gamma_1=\gamma_2$. We have  obtained an explicit expression of this scaling function for all $\alpha>0$ in Eq.~\eqref{main_eq21}. 
%Quite remarkably these results are also true for $-2 <\alpha <0$. 
On the other hand for $\gamma_1 < \gamma_2$, the distribution $P(x,t)$ does not satisfy this scaling form. In this case 
 we have found that,  the dynamics of the particle at large time can effectively be  described by an Ito-Langevin equation with position dependent drift and diffusion constant (Note that such effective Ito-Langevin dynamics also holds for $\gamma_=\gamma_2$).  While for  $\alpha =0$ and $\alpha=1$, it is possible to solve the master equation exactly to find $P(x,t)$, performing the same task for general $\alpha >0$ is difficult.  In such cases the Ito-Langevin description is particularly useful to obtain  the scaling behaviour of  distribution $P(x,t)$ at large time for $\gamma_1< \gamma_2$ case (for which particle is drifted away from the origin).  In particular, using this description we have shown that   $\langle |x(t)| \rangle \sim t$ and $\sigma _{\alpha} ^2(t)=\langle x^2(t) \rangle -\langle |x|(t) \rangle^2 \sim t^{1-\alpha}$ at large $t$  for general $\alpha$. In addition we have shown that scaling form of the distribution $P(x,t)$ is in fact Gaussian as also verified through direct numerical simulation of the actual  RTP dynamics in Eq.~\eqref{langevin}.

We also have studied the survival probability of the inhomegenous RTP dynamics on semi-infinite line from an absorbing boundary at $x=0$. For $\gamma_1=\gamma_2$ the survival probability, at large $t$, decays as a power law with a persistent exponent $\theta$ \emph{i.e.} $S(t) \sim t^{-\theta}$. We have shown that the persistent exponent is given by $\theta = \frac{1+\alpha}{2+\alpha}$ which generalises the result $\theta=1/2$ for $\alpha=0$ derived in \cite{Kanaya_2018}. For $\gamma_1< \gamma_2$ the particle has non-zero probability to survive at $t \to \infty$ as it is effectively drifted away from the origin. We explicitly computed this non-zero survival probability for all $\alpha>0$. On the other hand for $\gamma_1 >\gamma_2$, the probability decays to zero at large $t$. In both cases, we have found that the approach to the value at $t \to \infty$ is exponential. We have also looked at the  exit probabilities of the RTP from a finite interval. Finally, we provide  summary of the results presented in the paper in the tables \ref{table prob} and \ref{table surv}.

\begin{table}[h]
\centering
\begin{tabular}{|c|c|c|c|}
\hline

~&$\Delta >0$ & $\Delta =0$ & $\Delta <0$
\\ [0.5ex]
\hline 
& $S(x_0,t) \xrightarrow[]{ t \to \infty} 0$ & $S(x_0,t) \xrightarrow[]{ t \to \infty} 0$ & $S(x_0,t) \xrightarrow[]{ t \to \infty} \mathcal{S}(x_0)$ \\
& & & See Eqs. \eqref{surv_genalph_dell0_eq1} and \eqref{surv_genalph_dell0_eq2} \\
\hline
&Exact expression  &Exact expression &Exact expression  \\[-1ex]
\raisebox{1.5ex}{$\alpha=0$} 
& of $S(x_0,t)$ given & of $S(x_0,t)$ given  & of $S(x_0,t)$ given\\
& in Eq. \eqref{surv_a-1}.  & in Eq.\eqref{surv_a-1}. &in Eq.\eqref{surv_a-1}. \\[1ex]
& Decays to $0$ & Decays to $0$ as &Decays to $\mathcal{S}(x_0)$  \\
&  exponentially. & $\sim \frac{1}{t^{1/2}}$ for large $t$.& exponentially.\\ 
\hline 
&Decays to zero  &Decays to zero  &Relaxes to $\mathcal{S}_{\pm}(x_0)$\\[-0.8ex]
\raisebox{1.5ex}{$\alpha=1$} 
&  as $e^{-|s^*| t}$ at large $t$, &as $\frac{1}{t^{2/3}}$  at large $t $,  &  as $e^{-|s^*|t}$ at large $t$,  \\[1ex]
&  shown in Eq. \eqref{surv_eqqq7} &shown in Eqs. \eqref{surv_alph_eqq2}&shown in Eq.~\eqref{surv_eqqq7}\\
%& for large $t$ & for large $t$ & \\

\hline 
&For large $t$,  decays to & For large $t$, decays to& For large $t$, decays to \\[-1ex]
\raisebox{1.5ex}{general} 
& $0$ exponentially. &$0$ as $t^{-\frac{1+\alpha}{2+\alpha}}$, see Eq.\eqref{surv_genalph_eq5}. &$\mathcal{S}_{\pm}(x_0)$ exponentially. \\[-1ex]
\raisebox{1.5ex}{$\alpha \geq 0$} 
& Verified numerically& Verified numerically& Verified numerically\\
& in Fig.~\ref{surv_gen_alph}c&  in Fig.~\ref{surv_gen_alph}a& in Fig.~\ref{surv_gen_alph}b\\
\hline
\end{tabular}
\caption{Table summarising $S_{\pm}(x_0,t)$ for various $\alpha$ and $\Delta$.}
\label{table surv}
\end{table}
%surv_gen_alph

 We note that in this paper we have focused on the range $\alpha \geq 0$. However, we find that some of our results remain valid for $\alpha<0$. For example the steady state distribution in case of $\gamma_1>\gamma_2$, as given in Eq.~\eqref{stdy_sol} is also valid for $-1 < \alpha < 0$. In Fig.~\ref{gen-alp-delgt0}a we show a numerical verification of this fact for $\alpha=-0.3$. The scaling form $f_\alpha(y)$ of the probability distribution $P(x,t)$ for $\gamma_1=\gamma_2$, given in Eq.~\eqref{main_eq21}. Quite remarkably it turns out that this scaling distribution holds true  for $-2 <\alpha <0$ also, which we have verified numerically (not shown here) as well. While these results remain valid for $\alpha<0$, many results, for example the scaling form of $P(x,t)$ 
given in Eq.~\eqref{scaling-P_alpha} for $\gamma_1< \gamma_2$ is not valid for $\alpha <0$. Extending these results for $\alpha <0$ remains an interesting future direction. All our results are valid in one dimension. Extending our results to higher dimension would be nice to explore. Recently it was shown that the survival probability for RTP in $d-$dimension has some universal features \cite{Mori2019}. 
It would be interesting to see what happens to the universality when rates become position dependent. It would also be interesting to study the situation in which the rates $R_1,~R_2$ 
 become time dependent where the time dependence may come from the coupling of the RTP motion to the evolution of the inhomogeneous media. 
Finally, to study more realistic situations where individual active agents like bacteria or micro-robots or Janus particle interacts among themselves, one needs to look at interacting particle dynamics which is another important future direction.

\section{acknowledgement}
The authors acknowledge fruitful discussions with Satya N Majumdar, Urna Basu and Varun Dubey. A.K would like to acknowledge support from the SERB Early Career Research Award ECR$/2017/000634$ from the Science and Engineering Research Board, Department of Science and Technology and the support of the Department of Atomic Energy, Government of India, under project no. $12\text{-R}\&\text{D-TFR-}5.10\text{-}1100$.

\appendix
\section{Derivation of $G(x,s)$ and $\bar{P}(x,s)$ for $\alpha =0$}
\label{new_stdy_appen}
In this appendix, we solve Eq.~\eqref{main_eq6} explicitly to get $G(x,s)$. We will then insert this solution in Eqs. \eqref{main_eq2} and \eqref{main_eq4} to get $\bar{Q}(x,s)$ and $\bar{P}(x,s)$. Turning to Eq.~\eqref{main_eq6}, it is straightforward to solve it, and using the boundary conditions $G(x \to \pm \infty,s)=0$, one gets  
\begin{align} 
 G(x,s)=
    \begin{cases}
    A_{+} e^{-\lambda (s) x}, & \text{if}\ x>0 \\
     A_{-} e^{\lambda (s) x}, & \text{if}\ x<0
    \end{cases}
\end{align}
where $A_{\pm}$ are position independent constants. Inserting this solution in Eq.~\eqref{main_eq4}, we get $\bar{Q}(x,s)$ which can again be substituted in Eq.~\eqref{main_eq2} to get $\bar{P}(x,s)$.
\begin{align} 
 \bar{Q}(x,s)=
    \begin{cases}
    A_{+} e^{-\left(\lambda (s)+\frac{\Delta}{v}\right) x}, & \text{if}\ x>0 \\
     A_{-} e^{\left(\lambda (s)+\frac{\Delta}{v}\right) x}, & \text{if}\ x<0
    \end{cases}
    \label{main_eq7}
\end{align}
\begin{align} 
 \bar{P}(x,s)=\frac{v}{s}\left(\lambda(s)+\frac{\Delta}{v} \right)
    \begin{cases}
    A_{+} e^{-\left(\lambda (s)+\frac{\Delta}{v}\right) x}, & \text{if}\ x>0 \\
    - A_{-} e^{\left(\lambda (s)+\frac{\Delta}{v}\right) x}, & \text{if}\ x<0
    \end{cases}
    \label{main_eq8}
\end{align}
The task now is to evaluate the constants $A_{\pm}$ which demand two conditions. One condition comes by integrating Eq.~\eqref{main_eq2} from $-\epsilon$ to $+\epsilon$ and take $\epsilon \to 0$. This will result in the following discontinuity equation
\begin{align}
\bar{Q}(x \to 0^+,s)-\bar{Q}(x \to 0^-,s)=\frac{1}{v}. \label{main_eq9}
\end{align}
The other condition comes by noting that for symmetric initial condition, the probability distribution $P(x,t)$ is symmetric about $x=0$ which gives
\begin{align}
\bar{P}(x \to 0^+,s)=\bar{P}(x \to 0^-,s). \label{main_eq10}
\end{align}
Inserting the solutions of $Q(x,s)$ and $P(x,s)$ in Eqs. \eqref{main_eq9} and \eqref{main_eq10}, we get two linear equations for $A_+$ and $A_-$ solving which one finally gets complete expressions for $P(x,s)$ as,
\begin{align}
\bar{P}(x,s)=\frac{1}{2s}\left(\lambda(s)+\frac{\Delta}{v} \right)e^{-\left(\lambda (s)+\frac{\Delta}{v}\right) \mid x \mid}. \label{main_eq11}
\end{align}
This expression for $\bar{P}(x,s)$ is also written in Eq.~\eqref{main_eq12}.

\section{Derivation of $P(x,t)$ for $\alpha = 0$ in Eq. \eqref{main_eq14}}
\label{laplace-appen}
In this appendix we will derive the expression for $P(x,t)$ for $\alpha=0$ as written in Eq.~\eqref{main_eq14}. 
As we will see later, to prove this result the following inverse Laplace transform will be useful
\begin{align}
L_{s\to t} \left[e^{- \lambda(s) y}\right]&=-v \frac{d}{dy} \left[ e^{-\gamma t}   I_0\left(\sqrt{\gamma_1 \gamma_2 \left(t^2-\frac{y^2}{v^2}\right)} \right)\Theta(vt-y)\right], 
\label{laplace-eq1} \\
\text{where} &~\lambda(s)=\frac{1}{v}\sqrt{\Delta ^2+2 \gamma s+s^2}.
\end{align}
So we first provide a derivation of this equation and then provide the derivation of Eq.~\eqref{main_eq14}.
\subsection{Derivation of Eq. \eqref{laplace-eq1}}
%We aim to prove,
%\begin{align}
%L_{s\to t} \left[e^{- \lambda(s) y}\right]\stackrel{?}{=}-v \frac{d}{dy} \left[ e^{-\gamma t}   I_0\left(\sqrt{\gamma_1 \gamma_2 \left(t^2-\frac{y^2}{v^2}\right)} \right)\Theta(vt-y)\right],
%\label{appen-laplace-1}
%\end{align}
%with $\lambda(s)=\frac{1}{v}\sqrt{\Delta ^2+2 \gamma s+s^2}$ and $y>0$. 
We begin with the following inversion.
\begin{align}
&L_{s\to t} \left[\frac{e^{- \lambda(s) y}}{\lambda(s)}\right] = \frac{1}{2 \pi i} \int_{-i \infty} ^{i \infty} ds~e^{s t}~ \frac{e^{-\lambda(s) y}}{\lambda (s)}.
\label{appen-laplace-2}
\end{align}
Proceeding further, we can rewrite $\lambda (s)$ as,
\begin{align}
\lambda (s)=\frac{1}{v} \sqrt{(s+\gamma+\sqrt{\gamma_1 \gamma_2}) (s+\gamma-\sqrt{\gamma_1 \gamma_2})}.
\label{appen-laplace-3}
\end{align}
Substituting this in Eq. \eqref{appen-laplace-2}, we get
\begin{align}
L_{s\to t} \left[\frac{e^{- \lambda(s) y}}{\lambda(s)}\right] = \frac{v}{2 \pi i} \int_{-i \infty} ^{i \infty} ds~e^{s t}~ \frac{e^{-\frac{y}{v}\sqrt{(s+\gamma+\sqrt{\gamma_1 \gamma_2}) (s+\gamma-\sqrt{\gamma_1 \gamma_2})}}}{\sqrt{(s+\gamma+\sqrt{\gamma_1 \gamma_2}) (s+\gamma-\sqrt{\gamma_1 \gamma_2})}}.
\label{appen-laplace-4}
\end{align}
Changing the variable $s+\gamma - \sqrt{\gamma_1 \gamma_2}=\sqrt{\gamma_1 \gamma_2} z$ in Eq. \eqref{appen-laplace-4}, we have
\begin{align}
L_{s\to t} \left[\frac{e^{- \lambda(s) y}}{\lambda(s)}\right] &= \frac{v}{2 \pi i} e^{-(\gamma-\sqrt{\gamma_1 \gamma_2})t} \int_{-i \infty} ^{i \infty} dz~e^{z t \sqrt{\gamma_1 \gamma_2}}~ \frac{e^{-\frac{y \sqrt{\gamma_1 \gamma_2}}{v}\sqrt{z(z+2)}}}{\sqrt{z(z+2)}}, \nonumber \\
&= v e^{-(\gamma-\sqrt{\gamma_1 \gamma_2})t} L_{z \to \sqrt{\gamma_1 \gamma_2}t}\left[ \frac{e^{-\frac{y \sqrt{\gamma_1 \gamma_2}}{v}\sqrt{z(2+z)}}}{z(2+z)}\right].
\label{appen-laplace-5}
\end{align}
The inverse Laplace transform in the right hand side has been obtained in \cite{Doussal2019} (see Eq. (C2) there). Using this, we obtain,
\begin{align}
L_{s\to t} \left[\frac{e^{- \lambda(s) y}}{\lambda(s)}\right]= v e^{-\gamma t} I_{0} \left(\sqrt{\gamma_1 \gamma_2 (t^2-\frac{y^2}{v^2})} \right) \theta(v t -y),
\label{appen-laplace-6}
\end{align}
where $I_{0}$ is the modified Bessel function of first kind and $\theta(x)$ is the Heaviside step function. To prove \eqref{laplace-eq1}, we note that 
\begin{align}
L_{s\to t} \left[e^{- \lambda(s) y}\right]=-\frac{d}{dy}\left[L_{s\to t} \left(\frac{e^{- \lambda(s) y}}{\lambda(s)}\right)\right].
\label{appen-laplace-7}
\end{align} 
Substituting Eq. \eqref{appen-laplace-6} in \eqref{appen-laplace-7}, we establish the equality in \eqref{laplace-eq1}.

\subsection{Derivation of $P(x,t)$ in Eq.~\eqref{main_eq12}}
Let us begin by rewriting $\bar{P}(x,s)$ in Eq. \eqref{main_eq12} as,
\begin{align}
\bar{P}(x,s)=-\frac{1}{2 s} \frac{d}{d |x|}\left[ e^{-\left(\lambda (s)+\frac{\Delta}{v} \right)|x|}\right].
\label{appen-laplace-8}
\end{align}
The inverse Laplace transform $P(x,t)$ read as,
\begin{align}
%\begin{split}
P(x,t)&=-\frac{1}{2} \frac{d}{d |x|} \left[e^{-\frac{\Delta}{v} |x|} L_{s \to t} \left( \frac{e^{-\lambda(s)|x|}}{s}\right) \right], \nonumber \\
&=-\frac{1}{2} \frac{d}{d |x|} \left[e^{-\frac{\Delta}{v} |x|}\int_{0}^{t} d \tau~ L_{s \to t-\tau}\left(\frac{1}{s}\right) L_{s \to \tau} \left( e^{-\lambda(s)|x|}\right) \right]  \nonumber  \\ 
&=-\frac{1}{2} \frac{d}{d |x|} \left[e^{-\frac{\Delta}{v} |x|}\int_{0}^{t} d \tau~ L_{s \to \tau} \left( e^{-\lambda(s)|x|}\right) \right]. 
%\end{split} 
\label{appen-laplace-9}
\end{align}
In going from first line to second line, we have used the convolution property of Laplace transform.
%that gives $L_{s \to t} \left( \frac{e^{-\lambda(s)|x|}}{s}\right)=\int_{0}^{t} d \tau L_{s \to \tau} \left( e^{-\lambda(s)|x|}\right)$. 
Inserting Eq. \eqref{laplace-eq1} in \eqref{appen-laplace-9}, one recovers the expression of $P(x,t)$ written in Eq.~\eqref{main_eq14}.

\section{Derivation of the approximate expression of $P(x,t)$ given in Eq.~\eqref{main_eq15} for  $\alpha =0$ at large $t$}
\subsection{$\Delta \geq 0$}
\label{assy-del-gt-0}
Here we will provide a derivation of the large $t$ behaviour of $P(x,t)$ for $\Delta > 0$ as written in Eq. \eqref{main_eq15}. We begin with the exact expression of $P(x,t)$ in Eq. \eqref{main_eq14}. We first  rewrite Eq. \eqref{main_eq14} by replacing the time integral by $\int_{0}^{t} = \int_{0}^{\infty}-\int_{t}^{\infty}$ as 
\begin{align}
P(x,t) &= \frac{1}{2} e^{-\gamma_1 t} \delta (\mid x \mid-vt)+\frac{\gamma_1}{2 v}\left( 1+\frac{\gamma_2 \mid x \mid}{2 v}\right)e^{-\frac{\gamma_1 \mid x \mid}{2 v}}\Theta \left(v t -\mid x \mid \right)\nonumber \\
&-\frac{\sqrt{\gamma_1 \gamma_2}}{2v} \int_{0}^{\infty} d\tau~e^{-\gamma \tau}~ \frac{d ~\mathcal{I}(\mid x \mid, \tau)}{d \mid x \mid}\Theta \left(v \tau -\mid x \mid \right)\nonumber \\
&+ \frac{\sqrt{\gamma_1 \gamma_2}}{2v} \int_{t}^{\infty} d\tau~e^{-\gamma \tau}~ \frac{d ~\mathcal{I}(\mid x \mid, \tau)}{d \mid x \mid}\Theta \left(v \tau -\mid x \mid \right),
\label{assygt-eq-1}
\end{align}
we note that as $t \to \infty$, $P(x,t)$ goes to the stationary distribution $P^{st}_0(x)$. Also at large $t$, the coefficient of $\delta$-functions becomes very small. Hence for large $t$, we get
\begin{align}
P(x,t)-P_0^{st}(x) \simeq \frac{\sqrt{\gamma_1 \gamma_2}}{2v} \int_{t}^{\infty} d\tau~e^{-\gamma \tau}~ \frac{d ~\mathcal{I}(\mid x \mid, \tau)}{d \mid x \mid}.
\label{assygt-eq-2}
\end{align}
Note that $\mathcal{I}(x,t)=\frac{x e^{-\frac{\Delta x}{v}}}{v}\frac{I_{1}\left( \sqrt{\gamma_1 \gamma_2(t^2-\frac{x^2}{v^2})}\right)}{\sqrt{t^2-\frac{x^2}{v^2}}}$ with $I_{1}$ being the modified Bessel function of first kind. One can easily perfom the differentiation $\frac{d ~\mathcal{I}(\mid x \mid, \tau)}{d \mid x \mid}$ in Eq. \eqref{assygt-eq-2}. Also for large $t$, $\tau$ is also very large which means we can use the large $\tau$ form of $\frac{d ~\mathcal{I}(\mid x \mid, \tau)}{d \mid x \mid}$. It turns out that we need to make use of the asymptotic form of $I_{\nu}(z) \simeq \frac{e^z}{\sqrt{2 \pi z}}$ for large $z$. For $v t >>|x|$ alongwith the aforementioned approximations, we change the variable $\tau = t w$ to get
\begin{align}
P(x,t)-P_0^{st}(x) \simeq \mathcal{C}_1 \int_{1}^{\infty} d w \frac{e^{-t\left(\mathcal{C}_2 w +\frac{\mathcal{C}_3}{w}\right)}}{w^{3/2}} \left(1-\frac{\Delta |x|}{v} - \frac{x^2 \sqrt{\gamma_1 \gamma_2}}{v^2 t w} \right),
\label{assygt-eq-3}
\end{align}
where $\mathcal{C}_1 = \frac{\left(\gamma_1 \gamma_2 \right)^{1/4}}{2 v \sqrt{2 \pi t}} e^{-\frac{\Delta |x|}{v}}$, $\mathcal{C}_2 = \gamma - \sqrt{\gamma_1 \gamma_2}$ and $\mathcal{C}_3 = \frac{x^2 \sqrt{\gamma_1 \gamma_2}}{2 v^2 t^2}$. Interestingly this integration can be performed exactly. First we write, 
\begin{align}
\int_{1}^{\infty} dw \frac{e^{-t\left(\mathcal{C}_2 w +\frac{\mathcal{C}_3}{w}\right)}}{w^{3/2}} & = \int_{0}^{\infty} dw \frac{e^{-t\left(\mathcal{C}_2 w +\frac{\mathcal{C}_3}{w}\right)}}{w^{3/2}}-\int_{0}^{1} dw \frac{e^{-t\left(\mathcal{C}_2 w +\frac{\mathcal{C}_3}{w}\right)}}{w^{3/2}},  \nonumber \\
& = \sqrt{\frac{\pi}{\mathcal{C}_3t} }e^{-2 t \sqrt{\mathcal{C}_2 \mathcal{C}_3}} - Z_{3/2}, \\
\label{assygt-eq-4}
\end{align}
where $Z_{\mu}$ is defined as,
\begin{align}
Z_{\mu} = \int_{0}^{1} dw \frac{e^{-t\left(\mathcal{C}_2 w +\frac{\mathcal{C}_3}{w}\right)}}{w^{\mu}}
\label{assygt-eq-5}
\end{align}
Similarly,
\begin{align}
\int_{1}^{\infty} dw \frac{e^{-t\left(\mathcal{C}_2 w +\frac{\mathcal{C}_3}{w}\right)}}{w^{5/2}} & =\frac{1}{2} \sqrt{\frac{\pi}{(\mathcal{C}_3 t)^3}} \left(1+2 t\sqrt{\mathcal{C}_2 \mathcal{C}_3} \right)e^{-2 t \sqrt{\mathcal{C}_2 \mathcal{C}_3}} - Z_{5/2}.
\label{assygt-eq-6}
\end{align}
Let us start by evaluating $Z_{1/2}$ which can be easily shown to be,
\begin{align}
Z_{1/2} = \frac{1}{2}\sqrt{\frac{\pi}{\mathcal{C}_2 t} }e^{-2 t \sqrt{\mathcal{C}_2 \mathcal{C}_3}} \left[ 1+\text{Erf}\left(\sqrt{t \mathcal{C}_2}-\sqrt{t \mathcal{C}_3} \right)-e^{4 t \sqrt{\mathcal{C}_2 \mathcal{C}_3}}~\text{Erfc}\left(\sqrt{t \mathcal{C}_2}+\sqrt{t \mathcal{C}_3} \right)\right].
\label{assygt-eq-7}
\end{align}
The integrals  $Z_{3/2}$ and $Z_{5/2}$ can now be obtained from $Z_{3/2} = -\frac{d}{d \mathcal{C}_3}Z_{1/2}$ and $Z_{5/2} = -\frac{d}{d \mathcal{C}_3}Z_{3/2}$.  We get
 \begin{align}
& Z_{3/2}= \frac{1}{2}\sqrt{\frac{\pi}{\mathcal{C}_3 t} }e^{-2 t \sqrt{\mathcal{C}_2 \mathcal{C}_3}} \left[ 2-\text{Erfc}\left(\sqrt{t \mathcal{C}_2}-\sqrt{t \mathcal{C}_3} \right)+e^{4 t \sqrt{\mathcal{C}_2 \mathcal{C}_3}}~\text{Erfc}\left(\sqrt{t \mathcal{C}_2}+\sqrt{t \mathcal{C}_3} \right)\right], \nonumber \\
& Z_{5/2} =  \frac{1}{2} \sqrt{\frac{\pi}{(\mathcal{C}_3 t)^3}} e^{-2 t \sqrt{\mathcal{C}_2 \mathcal{C}_3}} \left[ 1+2 t \sqrt{\mathcal{C}_2 \mathcal{C}_3} -\frac{1}{2}\left(1+2 t \sqrt{\mathcal{C}_2 \mathcal{C}_3}\right)\text{Erfc}\left(\sqrt{t \mathcal{C}_2}-\sqrt{t \mathcal{C}_3} \right)\right. \nonumber \\
& ~~~~~~~~~~~\left.+ \frac{1}{2}\left(1-2 t \sqrt{\mathcal{C}_2 \mathcal{C}_3}\right)e^{4 t \sqrt{\mathcal{C}_2 \mathcal{C}_3}}~\text{Erfc}\left(\sqrt{t \mathcal{C}_2}+\sqrt{t \mathcal{C}_3} \right) \right]  +  \frac{1}{\mathcal{C}_3 t} e^{-t(\mathcal{C}_2+\mathcal{C}_3)}.
\label{assygt-eq-8}
\end{align}
Using these explicit expressions of $Z_{3/2}$ and $Z_{5/2}$, one gets explicit expression fo the integrals in  Eqs. \eqref{assygt-eq-4} and  \eqref{assygt-eq-6}. Finally, substituting these results  in Eq. \eqref{assygt-eq-3} we get the result written in Eq. \eqref{main_eq15} valid for large $t$. This technique can also be used to evaluate the asymptotic form for $\gamma_1 =\gamma_2$.

%\subsection{Derivation of the approximate expression of $P(x,t)$ given in Eq. \eqref{main_eq15} for $\Delta < 0$ and $\alpha =0$ at large $t$}
\subsection{$\Delta<0$}
\label{assy-del-lt-0}
Here we will provide a derivation of the large $t$ behaviour of $P(x,t)$ for $\Delta \leq 0$ case using saddle point approximation. Using the expression of $\bar{P}(x,s)$ in Eq. \eqref{main_eq12}, we write $\bar{P}(x,t)$ as Bromwich integral as
\begin{align}
P(x,t) = \frac{e^{-t \Delta \bar{x} }}{2\pi i} \int_{\gamma_0-i \infty}^{\gamma_0+i \infty} ds \left(\frac{\lambda(s)+\Delta }{2 s}\right) e^{t \phi (s)},
\label{assy-eq-1}
\end{align}
where $\bar{x}=\frac{|x|}{v t}$, $\lambda(s)=\frac{1}{v}\sqrt{\Delta ^2+2 \gamma s+s^2}$ and $\phi (s)$ is given by
\begin{align}
\phi(s)=s- \bar{x} v \lambda (s).
\label{assy-eq-2}
\end{align}
From Eq. \eqref{assy-eq-1}, we see that for $t \to \infty$ the integral will be dominated by the saddle point of $\phi(s)$. The saddle points are given by the solution of $\frac{d \phi}{ds}=0$. It may look like that this equation has two solutions $s_{\pm}$ 
\begin{align}
s_{\pm}=-\gamma \pm \sqrt{\frac{\gamma ^2 -\Delta ^2}{1-\bar{x}^2}},
\label{assy-eq-3}
\end{align}
however only $s_+$ satisfies $\frac{d \phi}{ds}=0$.
%Proceeding further, we note that \bluew{$\phi (s_+),~\phi (s_-)$ are real and $\phi (s_+)>\phi (s_-)$, which means that for large $t$,  $e^{t\phi (s_+)} >>e^{t\phi (s_-)}$.} Hence the leading contribution to Eq. \eqref{assy-eq-1} will come from $s_+$. 
Expanding $\phi(s)$ about $s_+$ and substituting in Eq. \eqref{assy-eq-1}, we get
\begin{align}
P(x,t) &\simeq  \frac{e^{t\left[ \phi(s_+)-\Delta \bar{x}\right]}}{2\pi i} \left(\frac{\lambda(s_+)+\Delta }{2 s_+}\right)\int_{\gamma_0-i \infty}^{\gamma_0+i \infty} ds ~ e^{t \frac{\phi ''(s_+)}{2} (s-s_+)^2}, \nonumber \\
& \simeq  \frac{e^{t\left[ \phi(s_+)-\Delta \bar{x}\right]}}{2\pi } \left(\frac{\lambda(s_+)+\Delta }{2 s_+}\right)\int_{- \infty+i(s_++\gamma_0)}^{ \infty+i(s_++\gamma_0)} dy ~ e^{-t \frac{\phi ''(s_+)}{2} y^2} \nonumber \\
& \simeq  \frac{e^{t\left[ \phi(s_+)-\Delta \bar{x}\right]}}{2\pi } \left(\frac{\lambda(s_+)+\Delta }{2 s_+}\right)\int_{- \infty}^{ \infty} dz ~ e^{-t \frac{\phi ''(s_+)}{2} z^2},
\label{assy-eq-4}
\end{align}
In going from first to second line, we have changed the integration from complex domain to real line by substituting $s-s_+=i y$ and from second to third line we have used the fact that $\phi ''(s_+)=\frac{d \phi}{d s}|_{s_+}=\frac{(1-\bar{x})^{3/2}}{\bar{x}^2 \sqrt{\gamma ^2-\Delta ^2}}$ is greater than zero. This also implies that the integral in Eq. \eqref{assy-eq-4} is always convergent. Performing the integration gives the asymptotic behaviour of $P(x,t)$ as written in Eq. \eqref{main_eq15}. 
%In Eq. \eqref{main_eq15} we have replaced $s_+$ by $s^*$.

\section{Derivation of $G(x,s)$ and $\bar{P}(x,s)$ for $\alpha =1$}
\label{appen_alph_1}
In this appendix, we will provide the solution of $G(x,s)$ for $\alpha =1$ as given in Eq.~\eqref{main_eqG}. We consider $\Delta =0$ and $\Delta \neq 0$ cases separately. 
%As discussed in the main section, we will denote the quantities as $G^{*}(x,s)$, $\bar{Q}^{*}(x,s)$ and $\bar{P}^{*}(x,s)$ with $*$ being $0$ and $\Delta$ for $\Delta =0$ and $\Delta \neq 0$ cases respectively.

\subsection{ Case I: $\Delta =0$}
\label{al-1-D-eq-0}
We make the change of variable $x = \left(\frac{v^2 l}{2 \gamma s}\right)^{\frac{1}{3}}y$ and then write Eq.~\eqref{G-alpha-1-Del-0} in terms of $y$ as,
\begin{align}
\partial_y^{2} G-\left(\mid y \mid + d ~s^{\frac{4}{3}} \right)G=0,
\label{appen_alph_1eqqq1}
\end{align}
where $d=\left(\frac{2 \gamma}{v^2 l}\right)^{\frac{1}{3}} \frac{l}{2 \gamma}$. Solving this equation gives $G(y,s)$ in terms of Airy functions $\text{Ai}\left(\mid y \mid+d ~s^{\frac{4}{3}}\right)$ and $\text{Bi}\left(\mid y \mid+d ~s^{\frac{4}{3}}\right)$. However $\text{Bi}(y \to \infty)$ diverges while $G(y,s)$ should remain finite. We thus get
\begin{align} 
 G(y,s)=
    \begin{cases}
    C_{+}~ \text{Ai}\left(c ~s^{\frac{1}{3}}x+d~s^{\frac{4}{3}} \right), & \text{if}\ x>0 \\
    C_{-}~ \text{Ai}\left(-c ~s^{\frac{1}{3}}x+d~s^{\frac{4}{3}} \right), & \text{if}\ x<0
\end{cases}
\label{appen_alph_1_eqqq2}
\end{align}
\begin{align} 
 \bar{P}(y,s)=-\frac{1}{2s} \frac{d}{d~x}
    \begin{cases}
    C_{+}~ \text{Ai}\left(c ~s^{\frac{1}{3}}x+d~s^{\frac{4}{3}} \right), & \text{if}\ x>0 \\
    -C_{-}~ \text{Ai}\left(-c ~s^{\frac{1}{3}}x+d~s^{\frac{4}{3}} \right), & \text{if}\ x<0
\end{cases}
\label{appen_alph_1_eqqq3}
\end{align}
Next we evaluate the constants $C_+$ and $C_-$. Before that,following Eq.~\eqref{main_eq4} one sees that $\bar{Q}_(x,s)$ and $G(x,s)$ are same for this case. We now integrate Eq.~\eqref{main_eq2} from $-\epsilon$ to $\epsilon$ and take $\epsilon \to 0$ limit which gives discontinuity equation in $G(x,s)$. Also for symmetric initial condition, the probability distribution will be symmetric about $x=0$ which means $\bar{P}(x,s)$ is continous about $x=0$. Therefore, one finally gets
\begin{align}
&G(x \to 0^+,s)-G(x \to 0^-,s)=\frac{1}{v}, \nonumber\\
&~~~~~~\bar{P}(x \to 0^+,s)=\bar{P}(x \to 0^-,s),
\end{align}
These equations give rise to two linear equations for $C_{\pm}$ which can be easily solved to get them as function of $s$. Next inserting $C_{\pm}(s)$ in Eqs.\eqref{appen_alph_1_eqqq2}, one finally gets Eq.~\eqref{main_eqqq2} for $\bar{P}(x,s)$.

\subsection{ Case II: $\Delta \neq 0$} 
\label{al-1-D-neq-0}
At first, we make the transformation $\frac{z}{\sqrt{2}}= \frac{\mid \Delta \mid}{v l} x+\text{sgn}(x) \frac{\gamma s}{\mid \Delta \mid} \sqrt{\frac{l}{v \mid \Delta \mid}}$, which reduces Eq.~\eqref{main_eqG} to
\begin{align}
\partial_z^{2} G-\left(\frac{z^2}{4}-\beta s^2 + \frac{\text{sgn}(\Delta)}{2} \right) G=0
\label{appen_alph_1_eq1}
\end{align}
where $\beta=\frac{l(\gamma^2-\Delta^2)}{2 v \mid \Delta \mid^3}$.
%$\mu (s)=\frac{s^2 l(\gamma^2-\Delta^2)}{2 v \mid \Delta \mid^3}-\frac{\text{sgn}(\Delta)}{2}$. 
This equation is a standard one in the literature whose solution is given by parabolic cylinder functions $D_{\beta s^2-\frac{1+\text{sgn}(\Delta)}{2}}(\pm z)$. Recalling $D_{\mu}(- \mid z \mid)$ diverges for some $\mu (s)$ as $z \to \pm \infty$, one gets the solution for $G(x,s)$ as
\begin{align} 
 G(x,s)=
    \begin{cases}
    B_{+}~ D_{\beta s^2-\frac{1+\text{sgn}(\Delta)}{2}}\left( \sqrt{\frac{2 \mid \Delta \mid}{v l}}\left(x+\frac{\gamma s l}{\Delta^2} \right)\right), & \text{if}\ x>0 \\
    B_{-}~ D_{\beta s^2-\frac{1+\text{sgn}(\Delta)}{2}}\left( \sqrt{\frac{2 \mid \Delta \mid}{v l}}\left(-x+\frac{\gamma s l}{\Delta^2} \right)\right), & \text{if}\ x<0
\end{cases}
\label{appen_alph_1_eq2}
\end{align}
where $B_{\pm}$ are the position independent constants. Next we use Eq.~\eqref{main_eq4} to write $\bar{Q}(x,s)$ and Eq.~\eqref{main_eq2} to write $\bar{P}(x,s)$ as
\begin{align} 
 \bar{Q}(x,s)= e^{-\frac{\Delta}{2 v l} x^2}
    \begin{cases}
    B_{+}~ D_{\beta s^2-\frac{1+\text{sgn}(\Delta)}{2}}\left( \sqrt{\frac{2 \mid \Delta \mid}{v l}}\left(x+\frac{\gamma s l}{\Delta^2} \right)\right), & \text{if}\ x>0 \\
    B_{-}~ D_{\beta s^2-\frac{1+\text{sgn}(\Delta)}{2}}\left( \sqrt{\frac{2 \mid \Delta \mid}{v l}}\left(-x+\frac{\gamma s l}{\Delta^2} \right)\right), & \text{if}\ x<0
\end{cases}
\label{appen_alph_1_eq3}
\end{align}
\begin{align} 
 \bar{P}(x,s)=-\frac{1}{2s}
    \begin{cases}
    B_{+}~ \frac{d}{dx}\left[e^{-\frac{\Delta}{2 v l} x^2} D_{\beta s^2-\frac{1+\text{sgn}(\Delta)}{2}}\left( \sqrt{\frac{2 \mid \Delta \mid}{v l}}\left(x+\frac{\gamma s l}{\Delta^2} \right)\right)\right], & \text{if}\ x>0 \\
    -B_{-}~\frac{d}{dx}\left[ e^{-\frac{\Delta}{2 v l} x^2} D_{\beta s^2-\frac{1+\text{sgn}(\Delta)}{2}}\left( \sqrt{\frac{2 \mid \Delta \mid}{v l}}\left(-x+\frac{\gamma s l}{\Delta^2} \right)\right)\right], & \text{if}\ x<0
\end{cases}
\label{appen_alph_1_eq4}
\end{align}
To evaluate constants $B_{\pm}$ we need two conditions. The first one comes by integrating Eq.~\eqref{main_eq2} from $-\epsilon$ to $+\epsilon$ and taking $\epsilon \to 0$ limit which gives discontinuity equation for $\bar{Q}(x,s)$. The other condition comes by noting that for symmetric initial condition, $\bar{P}(x,s)$ is symmetric about $x=0$. These two conditions can be summarised as
\begin{align}
&\bar{Q}(x \to 0^+,s)-\bar{Q}(x \to 0^-,s)=\frac{1}{v}, \nonumber\\
&~~~~~~\bar{P}(x \to 0^+,s)=\bar{P}(x \to 0^-,s),
\end{align}
and they give rise to two linear equations for $B_{+}$ and $B_{-}$ solving which we get $B_{\pm}$ as function of $s$. Inserting the solution for $B_{\pm}(s)$ in Eq.~\eqref{appen_alph_1_eq3}, we get the final expression for $\bar{P}(x,s)$ which is written in Eq.~\eqref{main_eq17}

\section{Derivation of the asymptotic forms of $S_{\pm}(x_0,t)$ for $\alpha =0$}
\label{appen-asy-surv-alph-0}
Here we provide the derivation of $S_{\pm}(x_0,t)$ for large $t$ for $\alpha =0$ as given in Eqs. \eqref{surv-a-001} and \eqref{surv-a-002}. Let us begin with the exact expression of $S_{-}(x_0,t)$ in Eq. \eqref{surv_a-1}. In this expression, writing $\int_{0}^{t}=\int_{0}^{\infty}-\int_{t}^{\infty}$, one gets
\begin{align}
S_-(x_0,t)=&1+v e^{\frac{\Delta x_0}{v}} \frac{d}{dx_0} \left[\int_0^{\infty} d \tau e^{-\gamma \tau}   I_0\left(\sqrt{\gamma_1 \gamma_2 \left(\tau^2-\frac{x_0^2}{v^2}\right)} \right) \Theta(v \tau-x_0)\right]- \nonumber \\
& v e^{\frac{\Delta x_0}{v}} \frac{d}{dx_0} \left[\int_t^{\infty} d \tau e^{-\gamma \tau}   I_0\left(\sqrt{\gamma_1 \gamma_2 \left(\tau^2-\frac{x_0^2}{v^2}\right)} \right) \Theta(v \tau-x_0)\right].
\label{appen-asy-surv-alph-0-eq-1}
\end{align}  
Note that in the limit $t \to \infty$, the third term becomes zero. Hence we can rewrite Eq. \eqref{appen-asy-surv-alph-0-eq-1} as,
\begin{align}
S_-(x_0,t)= S_-(x_0, t \to \infty)-v e^{\frac{\Delta x_0}{v}} \frac{d}{dx_0} \left[\int_t^{\infty} d \tau e^{-\gamma \tau}   I_0\left(\sqrt{\gamma_1 \gamma_2 \left(\tau^2-\frac{x_0^2}{v^2}\right)} \right) \Theta(v \tau-x_0)\right],
\label{appen-asy-surv-alph-0-eq-2}
\end{align}
where $S_-(x_0, t \to \infty)$ is $0$ for $\Delta >0$ and $\mathcal{S}_-(x_0)$ in Eq. \eqref{surv_eq101} for $\Delta <0$. The Heaviside step function inside integration becomes redundant when $v t >x_0$. Performing the differentiation over $x_0$ and changing the variable $\tau = u t$, we get 
\begin{align}
L_-(x_0,t)= \frac{x_0}{v} \sqrt{\gamma_1 \gamma_2} e^{\frac{\Delta x_0}{v}} \int_1^{\infty}  d u \frac{e^{-\gamma u t}}{\sqrt{u^2-\frac{x^2}{v^2 t^2}}}   I_1\left(\sqrt{\gamma_1 \gamma_2 \left(u^2 t^2-\frac{x_0^2}{v^2}\right)} \right),
\label{appen-asy-surv-alph-0-eq-3}
\end{align}
where $L_-(x_0,t)=S_-(x_0,t)-S_-(x_0,t \to \infty)$. Note that in the integration, $u$ is greater than or equal to $1$. Hence for large $t$, we can use the asymptotic form of modified Bessel functions $I_{\nu}(z) \simeq \frac{e^{z}}{\sqrt{2 \pi z}}$ for large $z$.  Therefore for large $t$, we have
\begin{align}
L_-(x_0,t)\simeq e^{\frac{\Delta x_0}{v}}\frac{x_0}{v} \frac{(\gamma_1 \gamma_2)^{\frac{1}{4}}}{\sqrt{2 \pi t}} \int _{1}^{\infty} \frac{du}{u^{3/2}}e^{-u t (\gamma -\sqrt{\gamma_1 \gamma_2})}.
\label{appen-asy-surv-alph-0-eq-4}
\end{align}
 This integration can be now easily performed as,
 \begin{align}
  \int _{1}^{\infty} \frac{du}{u^{3/2}}e^{-u t y}&=2 e^{-y t}-2 \sqrt{\pi y t } ~\text{Erfc}\left[ \sqrt{y t }\right], \nonumber \\
  &\simeq \frac{e^{-y t}}{y t}.
  \label{appen-asy-surv-alph-0-eq-5}
 \end{align}
In going from first line to second line we have used that for large $t$, $\text{Erfc}\left[ \sqrt{y t }\right] \simeq e^{- y t} \left(\frac{1}{\sqrt{\pi y t}}-\frac{1}{2 \sqrt{\pi y^3 t^3}} \right)$. Substituting Eq. \eqref{appen-asy-surv-alph-0-eq-5} in \eqref{appen-asy-surv-alph-0-eq-4}, one recovers the expressions in Eq. \eqref{surv-a-002}. Proceeding similarly for $S_+(x_0,t)$ one gets Eq. \eqref{surv-a-001} for large $t$ from Eq. \eqref{surv_a-1}.
It is worth remarkin that the analysis so far assumed that $\gamma_1 \neq \gamma_2$. However for $\gamma_1 = \gamma_2$ also, the same technique gives the correct asymptotic form as obtained in \cite{Kanaya_2018, Doussal2019} and written in Eqs. \eqref{surv-a-001} and \eqref{surv-a-002}.

%\eqref{surv-a-001} and \eqref{surv-a-002} with the exact expression in Eq. \eqref{surv_a-1} 

\section{Effective equation for survival probability for general $\alpha$ at large $t$ and large $x_0$}
\label{appen_efff_surv}
In this appendix, we will derive effective differential equations for $\bar{U}(x_0,s)$ which is related to $\bar{S}_{\pm}(x_0,s)$ using Eqs. \eqref{surv_eq3} and \eqref{surv_U-H-def}.We start with Eqs.\eqref{surv_U_pm} for $\bar{U}_{\pm}(x_0,s)$ 
\begin{align}
&\left[-v \partial_{x_0} +R_1(x_0)+s \right] \bar{U}_+=R_1(x_0) \bar{U}_-, \nonumber \\
& \left[~v ~\partial_{x_0} +R_2(x_0)+s \right] \bar{U}_-=R_2(x_0) \bar{U}_+,
\label{appen_surv_gen_eq1}
\end{align}
where $R_1(x_0)$ and $R_2(x_0)$ are defined in Eqs.\eqref{rates}. One may rewrite them as,
\begin{align}
&\frac{1}{x^{\alpha}} \mathcal{O}_{+}\bar{U}_{+}=\gamma_1 \bar{U}_{-}, ~~~~~~~\text{with } \mathcal{O}_{+}=-v\partial_x +\gamma_1 \frac{x^{\alpha}}{l^{\alpha}}+s, \label{appen_surv_gen_eq2}\\
&\frac{1}{x^{\alpha}} \mathcal{O}_{-} \bar{U}_{-}=\gamma_2 \bar{U}_{+}, ~~~~~~~\text{with } \mathcal{O}_{-}=v\partial_x +\gamma_2 \frac{x^{\alpha}}{l^{\alpha}}+s.
\label{appen_surv_gen_eq3}
\end{align}
Operating both sides of Eq.~\eqref{appen_surv_gen_eq2} by $\mathcal{O}_{-}$ and Eq.~\eqref{appen_surv_gen_eq3} by $\mathcal{O}_{+}$, we can decouple these two equations and simplify them further to get
\begin{align}
&-v^2 l^{\alpha} \partial_{x_0} \left(x_{0}^{-\alpha} \partial_{x_0} \bar{U}_{+} \right)+v (\gamma_1-\gamma_2) \partial _{x_0} \bar{U}_+-\left[-\frac{v s \alpha l^{\alpha}}{x_{0}^{\alpha+1}}+\frac{s^2 l^{\alpha}}{ x_0^{\alpha}}+(\gamma_1+\gamma_2)s\right] \bar{U}_+=0, \label{appen_surv_gen_eq4}\\
&-v^2 l^{\alpha}\partial_{x_0} \left(x_{0}^{-\alpha} \partial_{x_0} \bar{U}_{-} \right)+v (\gamma_1-\gamma_2) \partial _{x_0} \bar{U}_--\left[\frac{v s \alpha l^{\alpha} }{x_{0}^{\alpha+1}}+\frac{s^2 l^{\alpha}}{ x_0^{\alpha}}+(\gamma_1+\gamma_2)s\right] \bar{U}_-=0. \label{appen_surv_gen_eq5}
\end{align}
Adding these two equations and recalling the definition in Eqs.\eqref{surv_U-H-def}, one gets
\begin{align}
 -v^2 l^{\alpha} \partial_{x_0} \left(\frac{1}{x_{0}^{\alpha}} \partial_{x_0} \bar{U}\right)+ 2 v \Delta \partial _{x_0} \bar{U}+\left(2 \gamma s +\frac{s^2 l^{\alpha}}{x_0^{\alpha}} \right)\bar{U}-\frac{\alpha v s l^{\alpha}}{x_0^{\alpha+1}} \bar{H}=0
\label{appen_surv_gen_eq4}
\end{align}
For large $t$ (equivalantly small $s$) behaviour, we neglect the $O(s^2)$ term. Also for large $x_0$, we neglect the term containing $H$ because this is sub leading with respect to $x_0^{-\alpha}\partial_{x_0}\bar{U}$, which itself is of order $\bar{H}$ (see Eqs.~\eqref{surv_H-diff} and \eqref{surv_U-diff}). The final equation thus reduces to Eq.~\eqref{surv_genalph_eq1} of the main text.

\section{Solution of Effective equation for survival probability for general $\alpha$ and $\Delta=0$}
\label{appen_efff_surv_sol}
Here we provide solution of Eq.~\eqref{surv_genalph_eq2} for general $\alpha$. Changing variable $y=x_0^{\frac{2+\alpha}{2}}$ and writing Eq.~\eqref{surv_genalph_eq2} in terms of $y$, we get
\begin{align}
y \partial _y^2  \bar{U}-\frac{\alpha}{2+\alpha} \partial_y  \bar{U} = \frac{4sy}{D_{\alpha}(2+\alpha)^2} \bar{U}.
\label{appen_surv_gen_sol_eq1}
\end{align}
The solutions of this equation are given in terms of modified Bessel functions of first kind and second kind $y^{\frac{1+\alpha}{2+\alpha}}I_{\frac{1+\alpha}{2+\alpha}}\left(\frac{2}{2+\alpha}\sqrt{\frac{s}{D_{\alpha}}} y \right)$ and $y^{\frac{1+\alpha}{2+\alpha}}K_{\frac{1+\alpha}{2+\alpha}}\left(\frac{2}{2+\alpha}\sqrt{\frac{s}{D_{\alpha}}} y \right)$. However the first solution diverges as $x \to \infty$ which leaves us with only the second solution. Writing the solution in terms of $x_0$, we get
\begin{align}
\bar{U}(x_0,s)=B~ x_0^{\frac{1+\alpha}{2}}K_{\frac{1+\alpha}{2+\alpha}}\left(\frac{2}{2+\alpha}\sqrt{\frac{s}{D_{\alpha}}} x_0^{\frac{2+\alpha}{2}} \right).
\label{appen_surv_gen_sol_eq2}
\end{align}
To get $\bar{H}(x_0,s)$ we add the Eqs.\eqref{appen_surv_gen_eq1} which gives
\begin{align}
\bar{H}(x_0,s)=v\left(s+ \frac{2 \gamma x_0^{\alpha}}{l^{\alpha}}\right)\partial_{x_0}U.
\label{appen_surv_gen_sol_eqqqqq2}
\end{align}

To evaluate the constant $B$, we use the other boundary condition $S_-(0,s)=\frac{1}{s}+\bar{U}_-(0,s)=0$ which gives $\bar{U}_-(0,s)=-\frac{1}{s}$. Next to translate this in terms of $\bar{U}(x_0,s)$, we note that $\bar{U}_-(x_0,s)=\frac{\bar{U}(x_0,s)-\bar{H}(x_0,s)}{2}$ which gives the boundary condition in terms of $\bar{U}(x_0,s)$ and $\bar{H}(x_0,s)$ as,
\begin{align}
U(0,s)-H(0,s)=-\frac{2}{s}.
\label{appen_surv_gen_sol_eq3}
\end{align}
To find $U(0,s)$ and $H(0,s)$, we substitute $K_{\nu}(z)\approx2^{\nu-1} z^{-\nu} \Gamma(\nu)$ as $z \to 0$ in Eqs.\eqref{appen_surv_gen_sol_eq2} and \eqref{appen_surv_gen_sol_eqqqqq2} which gives $H(0,s)=0$ and non-zero $U(0,s)$. Finally substituting this in Eq.~\eqref{appen_surv_gen_sol_eq3} gives $B(s)$ and the expression for $\bar{U}(x_0,s)$ which is written in Eq.~\eqref{surv_genalph_eq3} 
\section{$S_{\pm}(x_0,t)$ as $t\to \infty$ for general $\alpha$ and $\Delta <0$}
\label{surv_large_t}
In this appendix, we will solve Eqs.~\eqref{surv_eq1} for general $\alpha$ and $\Delta <0$ in the limit $t \to \infty$. For $\Delta <0$, the particle is effectively drifted away from the absorbing wall at origin which gives rise to non-zero $S_{\pm}$ as $t \to \infty$. For this case Eqs.\eqref{surv_eq1} can be rewritten as,

\begin{align}
\left(-v \frac{d}{dx_0}+\frac{\gamma_1 x_0^{\alpha}}{l^{\alpha}}\right)S_+=&\frac{\gamma_1 x_0^{\alpha}}{l^{\alpha}}S_-,\label{appen_surv_large_t_eq0} \\
\left(~v \frac{d}{dx_0}+\frac{\gamma_2x_0^{\alpha}}{l^{\alpha}}\right)S_-=&\frac{\gamma_2 x_0^{\alpha}}{l^{\alpha}}S_+.
\label{appen_surv_large_t_eq1}
\end{align}
These two equations can be recasted as,
\begin{align}
\frac{l^{\alpha}}{x_0^{\alpha}} \mathcal{L}_+ S_+&=\gamma_1 S_-, ~~~~~~~~\text{with }\mathcal{L}_+=-v \frac{d}{d x_0}+\frac{\gamma_1 x_0^{\alpha}}{l^{\alpha}}\label{appen_surv_large_t_eq2} \\
\frac{l^{\alpha}}{x_0^{\alpha}} \mathcal{L}_- S_-&=\gamma_2 S_+, ~~~~~~~~\text{with }\mathcal{L}_-=~v \frac{d}{d x_0}+\frac{\gamma_2 x_0^{\alpha}}{l^{\alpha}}.\label{appen_surv_large_t_eq3}
\end{align} 
These coupled equations can be easily decoupled by multiplying both sides of Eq\eqref{appen_surv_large_t_eq2} by $\mathcal{L}_-$ which gives ordinary differential equation for $S_{+}(x_0)$ as,
\begin{align}
\frac{d}{dx_0}\left(\frac{v l^{\alpha}}{x_0^{\alpha}} \frac{d}{dx_0}-2 \Delta \right) S_+=0.
\label{appen_surv_large_t_eq4}
\end{align}
One can now solve Eq.\eqref{appen_surv_large_t_eq4} to get $S_+(x_0)$ which could be substituted in Eq.\eqref{appen_surv_large_t_eq0} to get $S_-(x_0)$. The solved expressions for $S_{\pm}(x_0)$ read as,
\begin{align}
S_{+}(x_0)=& C_1 + C_2~ e^{-\frac{2 \mid \Delta \mid}{v l^{\alpha} (\alpha+1)}x_0^{\alpha+1}}, \label{appen_surv_large_t_eq5}\\
S_{-}(x_0)=& C_1 + C_2\frac{\gamma + \mid \Delta \mid}{\gamma - \mid \Delta \mid}~ e^{-\frac{2 \mid \Delta \mid}{v l^{\alpha} (\alpha+1)}x_0^{\alpha+1}}, \label{appen_surv_large_t_eq6}
\end{align}
where $C_1$ and $C_2$ are constants that remain to be evaluated. To evaluate them, we use the boundary conditions $S_{\pm}(x_0 \to \infty)=0$ and $S_-(x_0 \to 0)=0$ which give $C_1=1$ and $C_2=-\frac{\gamma - \mid \Delta \mid}{\gamma + \mid \Delta \mid}$. Inserting them in Eqs. \eqref{appen_surv_large_t_eq5} and \eqref{appen_surv_large_t_eq6}, one gets the final expression for $S_{\pm}(x_0)$ written in Eqs.\eqref{surv_genalph_dell0_eq1} and \eqref{surv_genalph_dell0_eq2}.

\end{document}